\documentclass[fleqn,usenatbib]{mnras}

\usepackage{newtxtext,newtxmath}

\usepackage[T1]{fontenc}

\DeclareRobustCommand{\VAN}[3]{#2}
\let\VANthebibliography\thebibliography
\def\thebibliography{\DeclareRobustCommand{\VAN}[3]{##3}\VANthebibliography}

\usepackage{graphicx}	
\usepackage{amsmath}	
\usepackage{float}
\usepackage[dvipsnames]{xcolor}
\usepackage{hyperref}
\usepackage{subcaption}
\usepackage{rotating}
\usepackage{footnote}
\usepackage{pdflscape}

\usepackage[export]{adjustbox} 
\usepackage{enumitem} 

\definecolor{light-gray}{gray}{0.95}

\newcommand{\kms}{km\,s\textsuperscript{-1}}

\title[\textbf{HARPS exocomet occurrence in Ca\,{\sc ii}}]{Quantifying spectroscopic Ca II exocomet transit occurrence in two decades of HARPS data}

\author[R. Bendahan-West et al.]{
Rapha\"el Bendahan-West$^{1,2}$\thanks{E-mail: rb941@exeter.ac.uk},
Grant M. Kennedy$^{2,3}$,
David  J. A. Brown$^{2,3}$,
Paul A. Str\o{}m$^{2}$
\\
$^{1}$Department of Physics and Astronomy, University of Exeter, Stocker Road, Exeter EX4 4QL, UK\\
$^{2}$Department of Physics, University of Warwick, Gibbet Hill Road, Coventry, CV4 7AL, UK\\
$^{3}$Centre for Exoplanets and Habitability, University of Warwick, Gibbet Hill Road, Coventry CV4 7AL, UK\\
\\
}
\date{Accepted XXX. Received YYY; in original form ZZZ}
\pubyear{2024}

\begin{document}
\label{firstpage}
\pagerange{\pageref{firstpage}--\pageref{lastpage}}
\maketitle

\begin{abstract}
The field of exocomets has been built around the unmatched number of detections made in the circumstellar disc of the archetypal star Beta Pictoris. An exocomet detection in spectroscopy is identified by variable atomic absorption features in a stellar spectrum, associated with transiting gas in and trailing an exocomet coma. This paper presents the largest spectroscopic search for exocomet transits to date, which overcomes the limitations of biased samples of stars with debris discs, and instead looks through the $\approx7500$ stars in the HARPS archive for signs of exocomets in the Ca\,{\sc ii}~doublet (H:$396.847$~nm and K:$393.366$~nm). The search resulted in 155 candidate stars, which after filtering for false positives (e.g. binaries, stellar activity, etc.), were cut down to 22 stars. These 22 stars are classified into Tier~1, 2, and 3 exocomet candidates, reflecting the confidence level of their exocomet detection. Our two best candidates (Tier~1: Beta Pictoris, HD~172555) and four lower confidence candidates (Tier~2: Gl~1, HIP~5158, HD~94771, HR~1996) are discussed, yielding a detection rate of 0.03\% (Tier~1 only) and 0.1\% (Tier~1 \& 2) in the HARPS sample. Both Tier~1 stars are known exocomet host stars. These two young A-type stars correspond to 0.4\% of all A-types in the sample, suggesting that detecting signs of exocomet transits using Ca\,{\sc ii} is more likely around young A-type stars. Reanalysing a past HARPS study, we found no evidence to support the previously claimed four exocomet detections, indicating either that those detections are not robust or that we are only sensitive to the strongest signals.
\end{abstract}




\section{Introduction}
\label{intro}
Comets are commonly observed in the Solar System and can be pictured as the remnant bodies of planet formation. Given the presence of planets around other stars ("exoplanets"), it is then expected that "exocomets" should be found in other planetary systems. In the Solar System, individual comets can be visible to the naked eye once they come within a few au of the sun, showing signs of cometary activity such as comae and tails. We expect exocomets to show similar signs of activity. However, to be detected, we rely on an exocomet to transit which typically occurs when the exocomet is much closer to the star, allowing a high enough abundance of exocometary gas and dust for spectroscopic and photometric detections.

Exocomets are thought to have the same formation process as Solar System comets, and hence are expected to have analogous physical properties and composition. Comets remain stable and inactive in their reservoirs until they are gravitationally scattered towards the Sun by a combination of mechanisms generated by planet-comet interactions, e.g. interactions with giant planets \citep{fernandez-94, weissman-96, wiegert-99}, mean-motion resonances \citep{moons-95, morbidelli-95, beust-morbidelli-96}, and secular resonances \citep{farinella-94, levison-94, beust-morbidelli-96}. It is only when these comets get close to the star that they become visible, as the comet nuclei are usually too small (a few kilometres in size) to be identified \citep{blum+17, moulane+18}. The close distance allows gas expansion through sublimation of the condensed volatiles in the nucleus which also drags along dust particles, forming an envelope of gas and dust around the star referred to as a coma, which can reach sizes comparable to the size of the star. The evolution of this gas and dust throughout the comet's orbit generates visible gas \citep{cravens-91} and dust \citep{finson-68} tails from the various forces the different particles are sensitive to.

However, the similarity between Solar System comets and exocomets remains poorly understood. Despite two interstellar objects entering the Solar System, neither has fully matched our expectations of what an exocomet should look like. On one hand, 1I/'Oumuamua exhibited cometary orbital effects with non-gravitational acceleration without typical signs of cometary activity \citep{Meech+17, Micheli+18}, while 2I/Borisov showed many similar gas, dust, and nuclear properties to Solar System comets, but was unusually enriched in CO compared to what is observed in our comets \citep{guzik+20, bodewits+20, cordiner+20}. The sample of interstellar objects is still too small to understand the true distribution of what a typical exocomet could look like,
but the differences seen so far suggest that we should expect the unexpected.

Exocometary reservoirs are thought to reside primarily in the outer regions of planetary systems. Stars exhibiting mid/far infrared or sub-mm excesses caused by thermal emission from small mm or $\mu$m sized dust particles \citep{backman+paresce-93} are likely to harbour these reservoirs known as debris discs. Methods to probe exocometary material include detecting excess thermal and scattered light from sub-$\mu$m dust grains near the star \citep{absil+13, ertel+14, nunez+17, faramaz+17, marino+18} or observing CO gas via UV spectroscopy \citep{vidal-madjar-94, jolly+98, roberge+2000} and cold molecular gas emission \citep{marino+16, matra+17b}. Since cold CO gas is short-lived, its presence suggests continuous replenishment from cometary collisions in debris discs, and can yield an estimate of the CO + CO$_2$ ice fraction \citep{matra+17b}. While these methods provide valuable insights into exocometary material, they do not offer specific details about individual cometary bodies. For such detailed information, we expect that planets are necessary to scatter exocomets into the inner regions of planetary systems, where the sublimated material transits in front of the host star, producing detectable features in UV and optical spectroscopy, as well as in photometry.

\subsection{Exocomet detections in the literature}

\begin{figure}
    \centering
    \includegraphics[width=\linewidth]{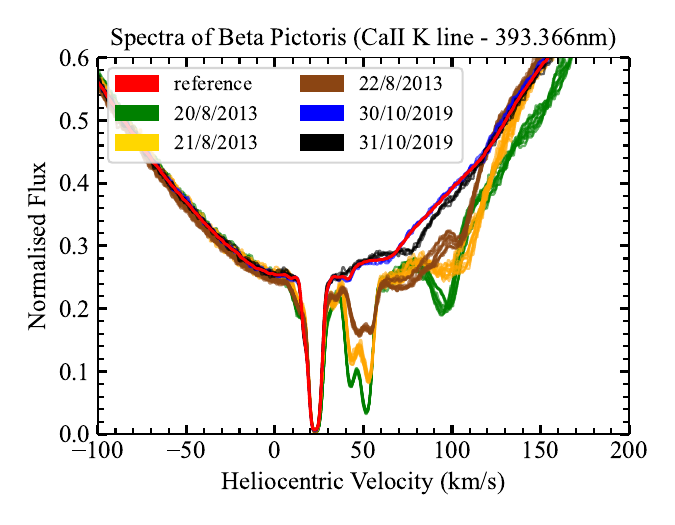}
    \vspace{-17pt}
    \caption{Exocomet signatures around $\beta$~Pic in the Ca\,{\sc ii}~K line. The red spectrum shows the reference spectrum with most exocomet features removed. The absorption feature around 20~\kms ($\beta$~Pic systemic velocity) corresponds to constant circumstellar gas absorption. All other spectra (different colours for different nights) show variable absorption features interpreted as signs of exocomet transits. Note that a reference spectrum for $\beta$~Pic will almost certainly still contain signs of exocomet absorption features unless produced by theoretical photosphere modelling. Here the reference spectrum is generated using spectra showing the fewest exocometary features.}
    \label{fig:intro/bpic-var}
\end{figure}

The first mention of exocomet detections is made around the archetypal A6V star Beta Pictoris \citep[or $\beta$~Pic;][]{paperv}, to explain variable absorption features observed in the Ca\,{\sc ii}~H \& K lines. Although these sporadic events were first observed over 30 years ago, the rate of detections in both the H \& K lines remains high, with each absorption event differing from others \citep{paperxiv, kiefer+14b}. The strength and variability of the observed absorption features around $\beta$~Pic were consistently modelled \citep{paperix, paperx, paperxi} by the presence of dense enough gas in the coma and tail of a comet transiting very close to the star \citep[i.e. a few tens of stellar radii][]{paperx, paperxi, paperxxi, paperxxii, paperxxv}). The most popular absorption line used to detect exocomet transits is the calcium doublet observed with optical spectroscopy (see example in Fig.~\ref{fig:intro/bpic-var}), however, exocometary absorption features can also be seen around other species in UV spectroscopy \citep[e.g. Fe\,{\sc ii}, Mg\,{\sc ii}, Al\,{\sc iii}][]{paperiv, papervi}. In all cases, the velocity distribution of the absorption features encodes information about specific exocomet orbital properties \citep{paperix, paperx, paperxi, paperxiii, paperxxi, paperxxii, paperxxv, kiefer+14b}. Monitoring $\beta$~Pic intensively over the years has revealed a trend where most transient absorption events are redshifted from the stellar systemic velocity \citep{paperv,paperxiii, kiefer+14b}, hinting at a specific orbital configuration.

Detecting exocomet transits tends to be a rare event, with no star matching the frequent and clear features observed around $\beta$ Pic. Most of the detections are made using Ca\,{\sc ii}, however, the level of confidence in an exocomet detection is linked to the multiplicity of exocomet detection techniques used. Table~1 and Table~2 in \citet{strom+20} as well as \citet{rebollido+20} provide a good summary regarding multi-wavelength detections as well as single-wavelength detections. Models have also predicted that exocomet transits should be detectable in photometric measurements \citep{lecavelier-99}, which is true with Kepler \citep[e.g. in ][]{strom+20} and with the Transiting Exoplanet Survey Satellite (TESS) \citep[$\beta$~Pic;][]{zieba+19, pavlenko+22, lecavelier+22}. With the addition of a hint of a photometric exocomet transit around HD~172555 using the CHaracterising ExOPlanet Satellite (CHEOPS) \citep{kiefer+23}, HD~172555 and $\beta$~Pic are the only two stars that have detections in both multi-wavelength spectroscopy and photometry.

\subsection{Exocomet surveys and statistics}

The main limitations from exocomet surveys in the literature \citep{WandM+18, iglesias+18, rebollido+20} arise from biased samples of stars in addition to "eye-balling" detection methods made possible with small samples. The stars observed tend to have previously shown specific features such as circumstellar absorption or a detected debris disc, indicating that there would be a higher chance of observing an exocomet transit around these stars. That is, previous samples are strongly biased so it is hard to draw conclusions about exocomet occurrence rates, which is an aim of our work. Currently, exocomets seem to be predominantly found around A-type stars, with some around F and B-type stars. However, observing a wider range of spectral types can test the validity of this trend as later-type stars do not yet contribute to any spectroscopic exocomet detections in the literature, although a few have been detected with photometry \citep{kiefer_2017, rappaport+18, kennedy+19}. We use the entire publicly available HARPS archive in this work with the purpose of spanning a wide range of spectral types. This is a large dataset which in turn maximises our chances of observing an exocomet transit.

We attempt to automate our search method to allow for a uniform detection technique in our large sample of stars but inevitably require interactive input to filter candidates. There might be other biases implied with HARPS as its main research focus is looking at targets that facilitate the detection of exoplanets using radial velocity. However, the sample of stars used is not biased towards stars with debris discs, as there is no significant correlation between the presence of radial velocity planets and debris discs \citep{yelverton+20}.

We can estimate the current detection rate of exocomet transits from the exocomet survey in \citet{iglesias+18}. This survey looked at a subset of 23 stars that show multiple narrow gas absorption features, and therefore biases their sample towards stars with circumstellar gas and close to edge-on inclinations, though this absorption can also be interstellar. They concluded that two stars out of the 23 showed exocometary activity, indicating that exocomet detections in such samples of stars should not be considered rare. In a broader context, their exocomet detection rate of 8.7\% is for a sample of stars that have detected debris discs, and are biased towards both edge-on orbits and the presence of circumstellar gas. Estimating the overall rate of detectable exocomets for a randomly selected sample of stars requires several corrections. If the probability of being sufficiently edge-on to detect circumstellar gas absorption is 10\% \citep{iglesias+18}, then the rate drops to 0.87\%. If all debris disc systems host detectable circumstellar gas, given the debris disc detection rate of 20\% \citep{sibthorpe+18} the exocomet detection rate for all stars is roughly 0.17\%. To date, most debris disc systems are not seen to have circumstellar gas, even those that are known to be edge-on, and most stars have later spectral types than those considered by \citet{iglesias+18}, so the exocomet detection rate for all stars is probably significantly less than 0.17\%.

Here, we aim to evaluate this value for the spectroscopic occurrence rate of exocomet transits in Ca\,{\sc ii} by firstly, using all the HARPS archive to ensure a wide range of spectral types, and secondly, using automated candidate detections to ensure consistency in our results. In Section \ref{sample} we describe the HARPS dataset used in this work. Section \ref{search} describes our search algorithm, its parameters, and its application. In Section \ref{chap:exo sec:final-vet} we explain the filtering process of false positive detections leading to our main results. We discuss our results in Section \ref{discussion} and conclude in Section \ref{conclusions}.

\section{Sample}
\label{sample}
We use data from the High Accuracy Radial velocity Planet Searcher (HARPS) spectrograph mounted on the 3.6 meter telescope at the La Silla Observatory. HARPS allows for exocomet detections as it produces both, high-resolution spectra that are necessary to observe the small widths of transient exocomet features, and observes optical wavelengths permitting the observation of the ionised calcium doublet (Ca\,{\sc ii}~H \& K lines - 396.847~nm and 393.366~nm respectively).

The entire publicly available HARPS archive was downloaded from the Phase 3 ESO HARPS platform\footnote{\url{http://archive.eso.org/wdb/wdb/adp/phase3_main/form}}. The data covers observations from 2003 to 2022, totalling more than 300,000 one-dimensional spectra (downloaded on 21/03/2022). We only extracted the relevant metadata information from the fits file header and subsets of spectra near the Ca\,{\sc ii}~H and K lines $\pm$ 1~nm ($\pm 760$~\kms).

\begin{figure}
\centering
    \includegraphics[width=\linewidth]{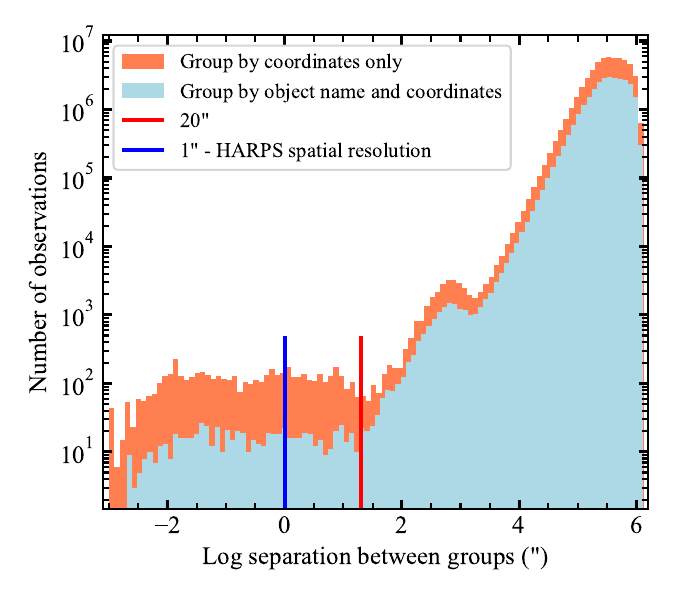}
    \vspace{-17pt}
    \caption{Histogram of separations in arcseconds between every group generated by both grouping methods - (\textit{orange}) grouping by commanded coordinates, (\textit{blue}) grouping with our "reduced" OBJECT name and commanded coordinates. Note the distribution at smaller separations ($<20"$ - marked by the red line), where the grouping by our "reduced" OBJECT name and commanded coordinates is more efficient. The blue line marks the spatial resolution of the HARPS instrument.}
    \label{fig:sample/grouping}
\end{figure}

A key step of this work is grouping all the data for each star as we want to compare all the spectra of a specific star to observe variations over time. Although this seems like a trivial task, where SIMBAD could be used to match the observations to the different stars, the challenge came from the lack of naming convention (e.g. $\beta$~Pic is known as beta~pic, HD~39060, etc.) which hinders matches by object name, and changes in coordinates due to proper motion, which can be significant for nearby stars. Two different grouping methods have been attempted; (1) grouping by the commanded coordinates which yield 10485 groups, and (2) grouping by our "reduced" OBJECT name (where all the characters from the OBJECT name in the fits file header are converted to lower case and unnecessary spaces/dashes are removed), and then by the commanded coordinates which yields 7539 groups. Incorrect grouping or naming errors would be flagged by the search algorithm (Section~\ref{search}) by causing large apparent spectral variations, however, only a few cases were found and this scenario is considered to be rare. The potential issue is therefore individual targets' spectra being spread across multiple groups.

To evaluate the efficiency of each grouping method, we calculate the separation distance between every group that both methods have created and show the result in Fig.~\ref{fig:sample/grouping}. At large separations ($>20"$), the two distributions are fairly similar, only varying by a factor of 1.9, linked to the difference in the number of groups generated by the two methods, i.e. $(10485/7539)^2 \approx 1.9$.

At smaller separations ($<20"$), the difference between both methods is approximately 1 order of magnitude, with far fewer distinct groups that are very close together in the sky using the "reduced" OBJECT name and the coordinates. This method is used in this work.

\begin{figure}
\centering
    \includegraphics[width=\linewidth]{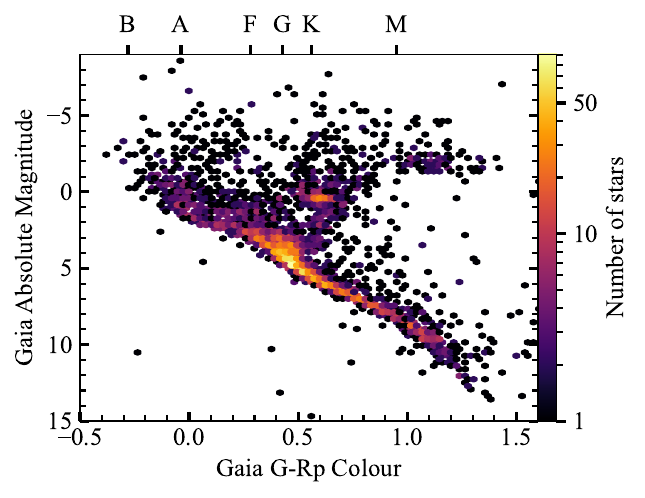}
    \caption{Hertzsprung-Russel (HR) diagram of our sample of stars using the Gaia eDR3 catalogue. The density of stars at each point in the HR diagram is represented by the colour scale on the right of the plot. The labels on top of the plot present a rough guide to the distribution of spectral types of stars across the HR diagram.}
    \label{fig:sample/HR}
\end{figure}

This step resulted in a total of 7539 unique stars in our sample. This number can be considered as approximate, as some targets remain in different groups, e.g. if their "reduced" OBJECT name and coordinates were not exactly the same even though they are the same star. For example, this was seen for $\beta$~Pic, where after the grouping process spectra were grouped into 4 groups: `betapic', `hd39060', `hd039060', and `hip27321'. For Beta Pictoris only, these were manually merged. However, this tends to happen for targets with many spectra, not affecting the ability to search for variations.

The effect of this imperfection in the grouping can be estimated by looking at Fig.~\ref{fig:sample/grouping}. There are $\approx 300$ groups (in the blue distribution) that have separations below the HARPS spatial resolution of 1 arcsecond. Assuming that stars appearing in 3 or more groups are rare, this number of groups can be treated as an upper limit of 150 stars that have been placed in two different groups when they should be merged. From the result presented in this work, a difference of $\approx 150$ stars in the denominator of the occurrence rate for exocomet transits calculation only affects the value on the $10^{-3}$\% level, which can be considered negligible.

Information from the Gaia Early Data Release 3 catalogue was used for all the stars in our sample where available (2199 stars had no match), enabling us to place them on a Hertzsprung-Russell diagram. Fig.~\ref{fig:sample/HR} therefore illustrates the demographics in our sample. There is an apparent diversity in the ages of the stars, where the Main Sequence is less well-defined for A-type stars and we also have several giant stars. It is important to note that the HARPS instrument's primary focus is exoplanet detection using the radial velocity method. This implies that the sample of stars used in this search is dominated by late-type stars (FGKM), simply because they are easier targets to achieve the required radial velocity precision for exoplanet detections. Thus, if exocomets, and our sensitivity to them, were independent of spectral type we would expect most detections to be around later-type stars.

There may nonetheless be biases that impact the probability of exocomet detection in our sample, with implications for interpreting our statistical results. Given that most known exocomet hosts are earlier type stars, we later quote detection rates for both A-type stars and the whole sample. Planet hunters commonly avoid binaries, which makes the HARPS sample different from a randomly chosen set of field stars. Dynamical mechanisms such as Kozai might increase exocomet occurrence in binary systems \citep{young+24}, but close to medium separation binaries are also less likely to host debris disks \citep{yelverton+20}. Additionally, young stars, which seem likely to be the most probable exocomet hosts, are typically more active and often more distant, making them less frequently observed. However, our aim is to derive statistics representative of ``typical'' stars, which are predominantly located in the field. Some bias could also emerge from systems closer to edge-on as HARPS has followed up candidates found by transit surveys, but given that Solar system comets have a wide range of inclinations, the magnitude of any effect on exocomet detection is unclear. Thus, there are potential biases in the HARPS sample, though whether the overall effect favours, disfavours, or does not change the exocomet detection rate, is unclear.

\section{The Exocomet Search}
\label{search}
The main idea behind this work is to search for the characteristic sporadic exocomet absorption features uniformly and automate this detection process as much as possible, for a maximum number of stars. This search functions by comparing a spectrum of a star against a reference spectrum, which is a typical spectrum of the star that is assumed to be free of any exocomet features. This comparison should allow the identification of any variable absorption features in a spectrum to then quantify the significance of the exocomet signal. Of course, some post-search work to classify candidates inevitably needs to be done by hand.

\begin{figure}
\centering
    \includegraphics[width=\linewidth]{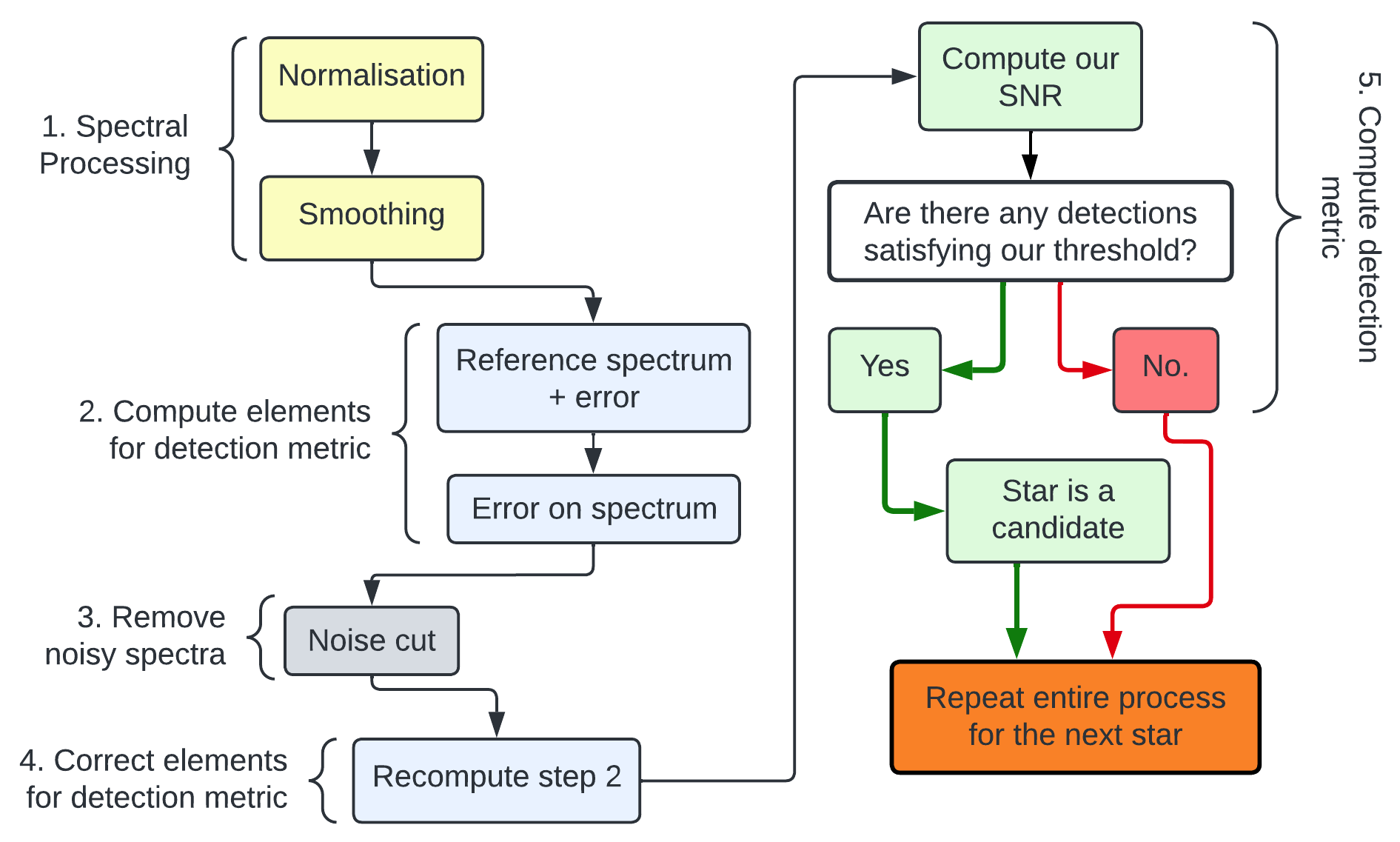}
    \caption{Systematic five step spectroscopic search for exocomet transits, independent of the line choice. Colours define the different tasks done in each step. The details of each step are given in the text.}
    \label{fig:exo/flowchart}
\end{figure}

\subsection{Systematic Five-Step Search for Exocomet Transits}
The spectra used in this search are 1D HARPS spectra that cover $\pm 1$~nm ($\pm 760$~\kms) from the line of interest, the ionised calcium H \& K lines. No correction for the systemic velocity of each star, or any radial velocity variation due to binarity/planets, is attempted. No interpolation is needed as all spectra have the same wavelengths and bin widths.

A brief overview of this search method is shown in Fig.~\ref{fig:exo/flowchart}. All spectra of a star are first normalised and smoothed (Step 1), then used to calculate the reference spectrum, its corresponding uncertainty, and the noise for each spectrum (Step 2). The next step consists of filtering out the noisy spectra that would otherwise interfere with our search process (Step 3), and Step 2 is then repeated (Step 4). The final step consists of calculating our own Signal-to-Noise Ratio (SNR) detection metric, which compares each spectrum to the reference spectrum individually while taking into consideration the noise in both elements (Step 5).

\begin{figure}
        \centering
        \includegraphics[width=\linewidth]{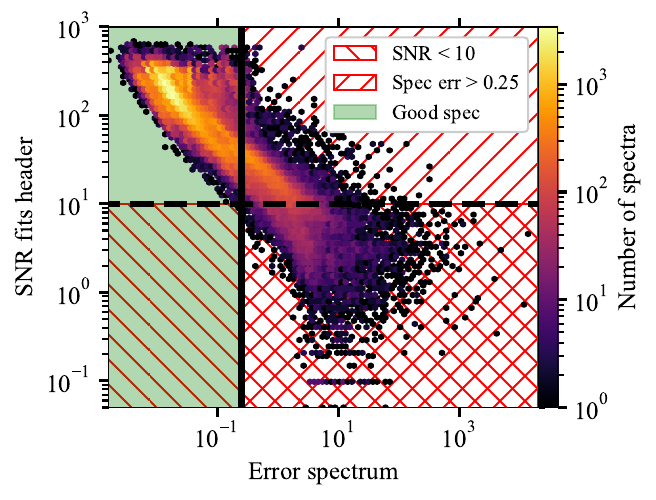}
        \caption{Comparison between the fits header signal-to-noise value and our calculated value for the error of a spectrum near the Ca\,{\sc ii}~H \& K lines. The dashed line shows the error on the spectrum $= 0.25$, the cut used here. The solid line denotes when the fits header value is equal to 10, which would be the cut used if only this metadata value were used to exclude noisy spectra.}
        \label{fig:exo/SNR-accuracy}
        \vspace{-5mm}
    \end{figure}

\subsection*{Step 1}
All spectra of a star have to be processed in the same way to be compared to each other, even though spectra change a lot between spectral types. This step consists of normalising and smoothing the spectra, hence preparing the data for the following steps of the exocomet search.
\begin{itemize}[wide, labelwidth=!, labelindent=0pt]
    \item \textbf{Spectrum normalisation:} \\
The aim of this normalisation process is for all the spectra of a star to have the same flux level. There is no need for the wings of the spectra to line up perfectly at 1 in normalised flux, as the only goal here is to detect variable absorption features. After converting each spectrum into radial velocity space, we visually investigate regions that are relatively constant by superposing spectra of $\beta$~Pic, which is expected to be more variable than any other exocomet system, and find that the regions $\pm (200$ to $700)$~\kms~are sufficient as a normalisation base due to the lack of Ca\,{\sc ii} variation caused by exocomet activity. Although this normalisation region is used for the Ca\,{\sc ii}~doublet, in reality, it should be dependent on the absorption line being studied, e.g. strong variability around $\beta$~Pic have been observed up to 400~\kms~in other lines \citep{papervi}.

Normalised spectra are obtained by dividing a spectrum by the mean across our normalisation region. Even though stars with different spectral types show large variations in their spectra, this method has been seen to work correctly for all spectral types as there is no strong variation in lines in these regions (see Fig.~\ref{fig:normalisation}). It is also worth noting that due to the normalisation being somewhat arbitrary (e.g. could use other ranges, and does not provide a true continuum estimate), the fractional absorption depths derived in this work are also only approximate. This however does not impact the analysis made in this work.
    
    \item \textbf{Smoothing:} \\
Each spectrum has the same wavelength bin width of 0.001~nm which converted to radial velocity space is approximately equal to 1~\kms. The spectra used in this search are smoothed using a median filter of 3 data points in order to increase the signal-to-noise ratio and allow for clearer exocomet detections, without deteriorating the HARPS spectral resolution. This can be explained by the fact that each HARPS spectral resolution element is sampled by 3.2 CCD pixels \citep{mayor+03}, meaning that when ''smoothing" using a 3-point median filter, we actually get back a spectrum that is at the HARPS spectral resolution ($\approx 3$~\kms). 
    
This bin of 3 data points, is set as the threshold to counter the oversampling of HARPS spectra and ensures that no exocometary-like absorption features are narrower than 3~\kms, which would be below HARPS resolution \citep{kiefer+14b}. An additional benefit of smoothing spectra is that it removes single-point outliers which could cause false positive detections later on.
\end{itemize}

\subsection*{Step 2}
\begin{itemize}[wide, labelwidth=!, labelindent=0pt]
    \item \textbf{Reference spectra:} \\
    The reference spectrum $r$ represents the typical spectrum of a star, meaning that most spectra should look like $r$. A key assumption here is that in general exocomet transits are rare, which aside from $\beta$~Pic is borne out by the low detection rate of previous searches. Empirically, we define the reference spectrum as the median of all normalised spectra obtained for a star. There is also an error associated with this reference spectrum $\sigma_{r}$, which is simply a vector with the standard deviation of all normalised spectra, computed at each velocity. The reference spectrum is therefore more precise for systems with a larger number of spectra.
    
    \item \textbf{Uncertainty of target spectra:} \\
    The error $\sigma_{s}$ for each spectrum is obtained by computing the standard deviation of the difference between the spectrum and the reference spectrum. As the reference should be a robust representation of a typical spectrum, any overall deviations from this reference could either be attributed to the noise in the spectrum or an exocomet signature. However, we assume here that exocomets do not cover a wide range in velocity, meaning that the noise will overall dominate over the weak exocomet signatures. There is a signal-to-noise value recorded in each fits file header, however, this value is calculated across the whole HARPS spectral range. We did not use this value as it was not found to correlate well with our estimate of the noise more focused near our lines of interest, the Ca\,{\sc ii}~H \& K lines (see the comparison between the fits header value and our computed values in Fig.~\ref{fig:exo/SNR-accuracy}).
\end{itemize}

\subsection*{Step 3}
\label{chap:exo sec:step3}
The removal of any noisy spectra is important in the search process as this yields a more precise reference spectrum for each star. Primarily this step aims to remove spectra that are much noisier than average for a given star, which makes the search less sensitive by increasing the reference spectrum noise.

\begin{itemize}[wide, labelwidth=!, labelindent=0pt]
    \item \textbf{Noise cut} \\
    This step has some dependency on the threshold used to quantify the significance of an exocomet detection ($4\sigma$ which is explained in \hyperref[chap:exo sec:step5]{Step 5}). In the case of an arbitrarily precise reference spectrum, the error of an individual spectrum sets whether an exocomet can be detected.
    
    A transient absorption feature can be as deep as the difference between the reference spectrum of a star and zero flux. Given that we have normalised spectra to near 1, and that we require $4\sigma$ for a detection, spectra must have $\sigma_{s} < 0.25$. Spectra with $\sigma_{s} > 0.25$ are too noisy and are discarded. This process is performed on every spectrum over the same velocity range as for the normalisation, ie. $\pm (200$ to $700)$~\kms.
    
    The distribution of fits header metadata for the signal-to-noise is compared to our measurement of the error in a spectrum in Fig.~\ref{fig:exo/SNR-accuracy}. If we wanted to use the fits header value, then a cut would be necessary for values $< 10$ (dashed line). However, a cut only based on this metadata would still include some noisy spectra in the search (seen in the top right part of Fig.~\ref{fig:exo/SNR-accuracy}). On the other hand, filtering using $\sigma_{s} > 0.25$~removed $\approx 40000$ spectra and resulted in stars with a more precise reference spectrum.
\end{itemize}

\subsection*{Step 4}
\label{chap:exo sec:step4}
This is an easy and complementary step to \hyperref[chap:exo sec:step3]{Step 3}. After the filtering of noisy spectra, the reference spectrum, its error, and the noise in individual spectra are recalculated.

\subsection*{Step 5}
\label{chap:exo sec:step5}
Step 5 is the main component of this search for exocomet transits as it determines whether or not a spectrum contains any exocometary features. This detection process is controlled by our own Signal-to-Noise Ratio (SNR) metric which can be simply expressed as 
\begin{equation}
    \mathrm{SNR} = \frac{\mathrm{signal}}{\mathrm{noise}} = \frac{s - r}{\sqrt{\sigma_{s}^2 + \sigma_{r}^2}},
\label{eq:snr}
\end{equation}
with $s$ being an individual spectrum, $r$ the reference spectrum, and $\sigma_{s}$, $\sigma_{r}$, the errors on the spectrum and reference spectrum respectively.

As mentioned previously, the main distinguishable characteristic of an exocomet feature is its sporadic nature. Our SNR metric therefore calculates the difference between a spectrum and the reference spectrum, where any deviations are transient features.

Assuming noise in the spectra follows a Gaussian distribution, the SNR should have similar properties. Significant variations between a spectrum and the reference spectrum of a star would result in strong outliers in the SNR distribution. The transient absorption features caused by exocomets would therefore represent significant outliers in the SNR. Only negative SNR outliers are considered in this search as we are looking for absorption features and not emission features. As a consequence, the threshold is empirically chosen to be $-4\sigma$ as hundreds of thousands of spectra are being considered and we do not want too many statistical false positives.

The general idea of the final step of this search for exocomet transits is to calculate this SNR distribution for every spectrum of a star and detect any points that pass the $-4\sigma$ outlier threshold. If this is the case, the spectrum shows possible signs of exocomets and the star is flagged as a potential host for exocomets. On the other hand, if there are no outliers detected, the star is not considered a candidate and the search proceeds to the next star.

\begin{figure}
    \centering
    \includegraphics[width=0.8\linewidth]{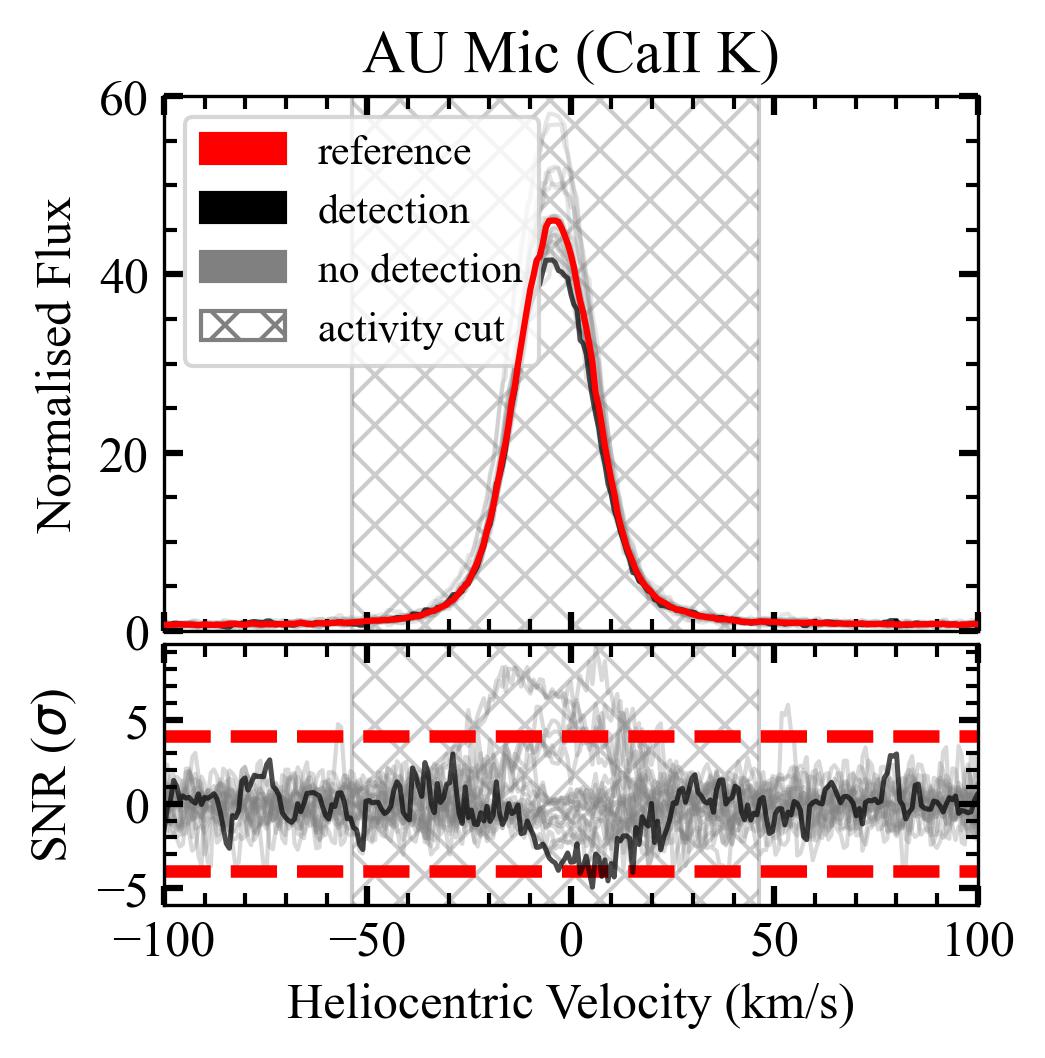}
    \caption{Example of stellar activity filtering with M1V star AU~Mic. All the spectra are focused around the Ca\,{\sc ii}~K line. The emission line is centred at $\approx -5$~\kms, and appears to be variable in strength. The black and grey lines denote spectra with a detection and no detection respectively. This false positive detection is caused by variability in the emission feature and can be avoided by masking the data points within the area of variability (denoted by the gridded area).}
    \label{fig:exo/aumic}
\end{figure}

\subsection{Additional parameters}
\label{add-filters}
Running the search systematically over a large sample will inevitably generate false positive detections. Three additional parameters were therefore applied to \hyperref[chap:exo sec:step5]{Step 5} to filter out a considerable amount of those unavoidable false positive detections: a filter on the position of the absorption feature, a filter on stellar activity, and a filter on the widths of the exocomet absorption features.

The position, in radial velocity space, of an exocomet feature depends on the atomic line that is being monitored. Typically, an absorption feature linked to an exocomet transit in the Ca\,{\sc ii} doublet is limited to a velocity of approximately $\pm 200$~\kms~beyond which the exocomet would be closer to the star and radiation pressure on Ca\,{\sc ii} ions would be too strong to allow for a detectable size coma \citep{paperx, paperxxii}. These very close exocomets, seen at velocities greater than 200~\kms, are only seen in the UV, where the effect of radiation pressure on ions such as Al\,{\sc iii} and Mg\,{\sc ii} (seen in $\beta$~Pic) is at least a factor of 10 smaller than on Ca\,{\sc ii} ions \citep{paperix, paperxxi, paperxxii}. There is therefore a filter only allowing the detection of transient absorption features for a radial velocity range of $\pm200$~\kms.

Since the HARPS archive includes late-type stars, stellar activity can be a common source of false positive detections. An example of stellar activity can be seen in Fig.~\ref{fig:exo/aumic}, where there is broad variable emission at the star's radial velocity. The strength of this emission line varies significantly from one star to another but also for spectra of the same star. For the late-type stars showing these signs of stellar activity, it would be extremely hard to confidently determine whether any absorption seen in the emission line is caused by stellar activity, an exocomet transit, or something totally unrelated. This emission feature is therefore masked out instead of discarding stars showing these features, which reduced significantly the number of false positive detections.

The masking of the emission feature was done by cross-correlating an individual spectrum with a Gaussian that has the typical width of this emission line, ie. $\pm30$~\kms~from the centre. This allows us to find the centre of the emission to then mask out $\pm 50$~\kms~around this point, as seen in Fig.~\ref{fig:exo/aumic}. This process is however only activated when a reference spectrum shows flux above a specific threshold (value determined empirically) for radial velocities between $\pm100$~\kms. Any data above the threshold of 2 in normalised flux in this specific radial velocity range is considered to indicate the presence of stellar activity, and the masking process of the central emission line is activated. We found that 398 stars have normalised flux $> 2$ and go through this masking process. 

The last filter is on the width of the transient absorption feature. It is expected that exocomet features at high radial velocity should experience a larger Doppler broadening than exocomet signatures at low velocities, due to the increased thermal velocity distribution of Ca\,{\sc ii} ions in the coma. In the lower velocity region, no exocomet features around $\beta$~Pic have been found to be narrower than 3~\kms~\citep{kiefer+14b}, ie. HARPS spectral resolution, hence being the threshold used in this work to filter out detected absorption features that are too narrow to be exocometary (and are likely not astrophysical).

\begin{table}
    \caption{Overview of the number of candidates generated by the search when focused on the Ca\,{\sc ii} K line, while varying the SNR outlier threshold, the absorption width filtering (w), and stellar activity filtering (act.). The boldface value in green highlights the final number of candidates with the parameters used in this work (summarised in \autoref{tab:params}).}
    \label{tab:overview-params}
    \centering
    \begin{adjustbox}{max width=0.98\linewidth}
    \begin{tabular}{l c c c c c c c c c c c}
        \hline
        \hline
        & \multicolumn{11}{c}{outlier threshold}\\ \cline{2-12}
        & \multicolumn{3}{c}{$-3\sigma$} & & \multicolumn{3}{c}{$-4\sigma$} & & \multicolumn{3}{c}{$-5\sigma$}\\
        & \multicolumn{3}{c}{w (\kms)} & & \multicolumn{3}{c}{w (\kms)} & & \multicolumn{3}{c}{w (\kms)} \\ \cline{2-4} \cline{6-8} \cline{10-12}
        act. & 0 & 2 & 3 & & 0 & 2 & 3 & & 0 & 2 & 3 \\ \hline
        ON & 2710& 1888& 522& & 493& 442& {\normalsize \color{OliveGreen} \textbf{155}}& & 61& 59& 48\\
        OFF & 2747& 1931& 590& & 538& 480& 195& & 84& 82& 69\\
        \hline
    \end{tabular}
    \end{adjustbox}
\end{table}
We approximate the width of the absorption feature by counting the number of consecutive data points where the signal is below $-3\sigma$. Each data point, as measured using the HARPS instrument, corresponds to a width of 1~\kms. If fewer than 3 consecutive points have an SNR below $-3\sigma$, the absorption feature is considered too narrow and is not classified as a real detection. Hereafter, this approximation will be used when referring to the width of any detected absorption features.

\subsection{Application}
Looking for exocomet transits is only possible when there is more than one spectrum taken of a star. A little bit less than 20\% of the sample of stars (or 1431 stars) have no more than one spectrum and are therefore discarded from the search. These spectra could show interesting absorption components in the Ca\,{\sc ii} doublet, however, they would only be visible if a theoretical reference spectrum was used.

The search algorithm was applied to the remaining $\approx 6100$ stars, focusing the hunt on the Ca\,{\sc ii} doublet. More precisely, we prioritise the Ca\,{\sc ii}~K line as the absorption features are deeper than in the H line. This can be explained by the oscillator strength of the Ca\,{\sc ii}~K line being twice that of the Ca\,{\sc ii}~H line \citep{paperxiii, kiefer+14a}. Up to a factor of 2 in absorption depth is visible between the H and K lines, meaning that it is more likely to observe transient absorption features in the K line. This ratio can in some cases vary to values very close to 1 according to the optical thickness of the absorbing clouds \citep{lecavelier-97}. In our case, for optically thin absorption, a $4\sigma$ detection in the Ca\,{\sc ii}~K line would only have an equivalent $2\sigma$ detection in the Ca\,{\sc ii}~H line, which will be a very marginal detection and mostly buried within the noise of the spectra. Although a detection in K that is not present in H does not rule out the initial detection, a detection in H as well as K would strengthen the exocomet scenario. Expecting that any detection in H should be at least as strong for K, we focus our search for exocometary features in H around targets that have been previously flagged in K. As a sanity check, we also conduct a search for exocomet features in H over the entire sample of stars.

\begin{table}
    \caption{Summary of all the final values for the thresholds used in the search for exocomet transits.}
    \label{tab:params}
    \centering
    \begin{tabular}{l c}
        \hline
        \hline
        Parameters & \\
        \hline
        Outlier threshold & $-4\sigma$ \\
        Smoothing & 3 data points \\
        RV range & $\pm 200$~\kms \\
        Activity threshold & 2 \\
        Absorption width & 3~\kms \\
        \hline
    \end{tabular}
\end{table}

\autoref{tab:overview-params} presents an overview of how the absorption width filter and activity filter impact the number of candidates. The absorption width filter seems to be the most effective filter, and removes false positive detections caused by narrow absorption features at a range of velocities. The activity filter removed a significant amount of false positive detections, however, many of those persist (as seen in the next Section). Using the final parameters summarised in \autoref{tab:params}, the search resulted in 155 candidate stars potentially hosting exocomets.

\begin{table}
    \caption{Distribution of false positives among the 155 search candidates.}
    \label{tab:initial-vetting}
    \centering
    \begin{tabular}{l c}
        \hline
        \hline
        Cause of variation & Number of stars\\
        \hline
        Binaries & 14 (9\%)\\
        Other variability & 27 (18\%)\\
        Stellar activity & 57 (37\%)\\
        Short timescale variations & 16 (10\%)\\
        Noisy/Instrumental& 19 (12\%)\\
        \hline 
        Total & 133 (86\%)\\
        \hline \hline 
        Remaining stars & 22 (14\%)\\
        \hline
    \end{tabular}
\end{table}

\begin{figure*}
\centering
  \begin{subfigure}[t]{0.33\linewidth}
    \centering
    \includegraphics[width=\linewidth]{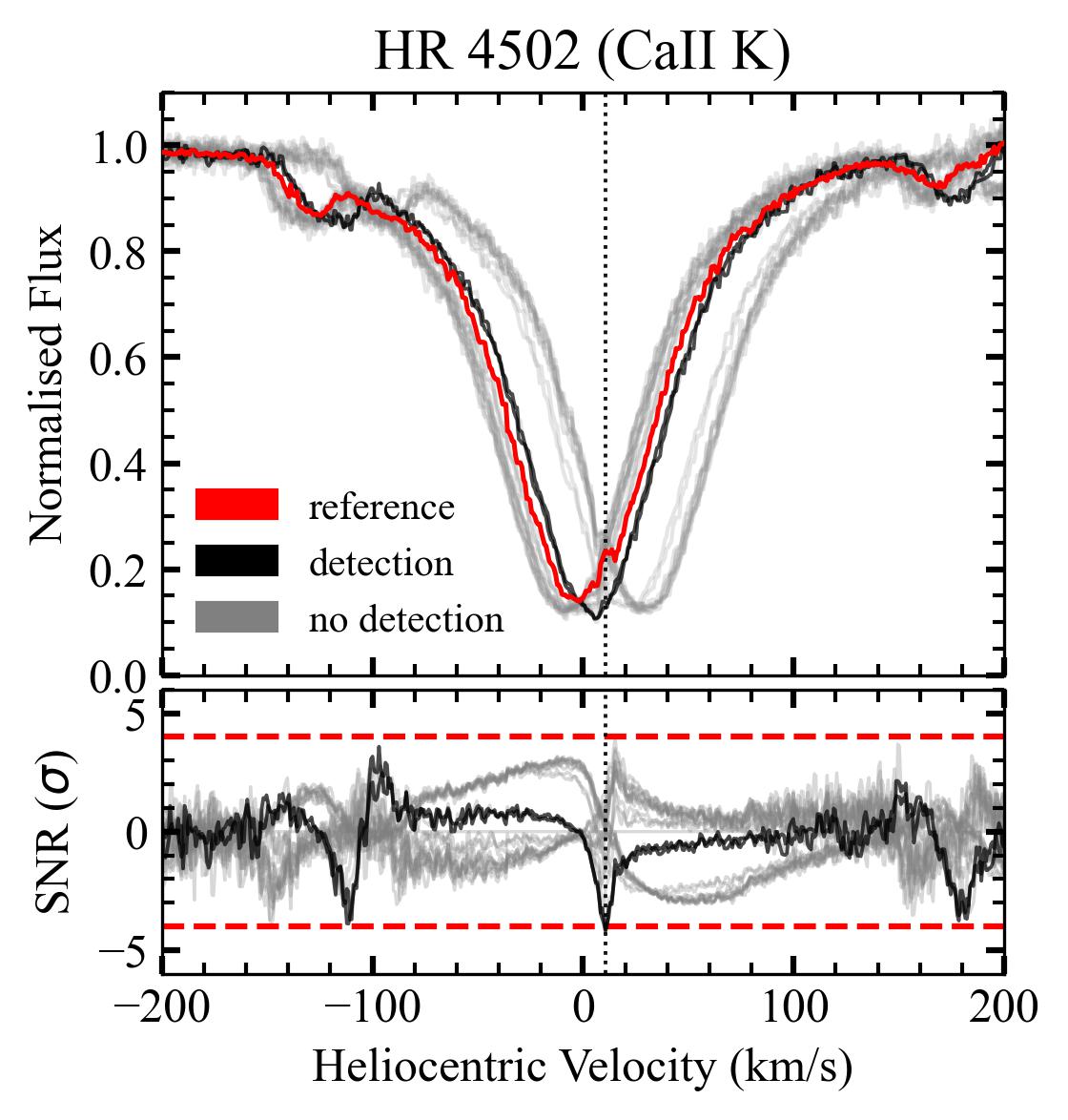}
    \captionsetup{margin={10pt,10pt}} 
    \caption{A0V star HR~4502 determined as a false positive detection due to the visible radial velocity shifts in the central absorption line.}
    \label{fig:false-positive/binary} 
  \end{subfigure} \hfill
  \begin{subfigure}[t]{0.33\linewidth}
    \centering
    \includegraphics[width=\linewidth]{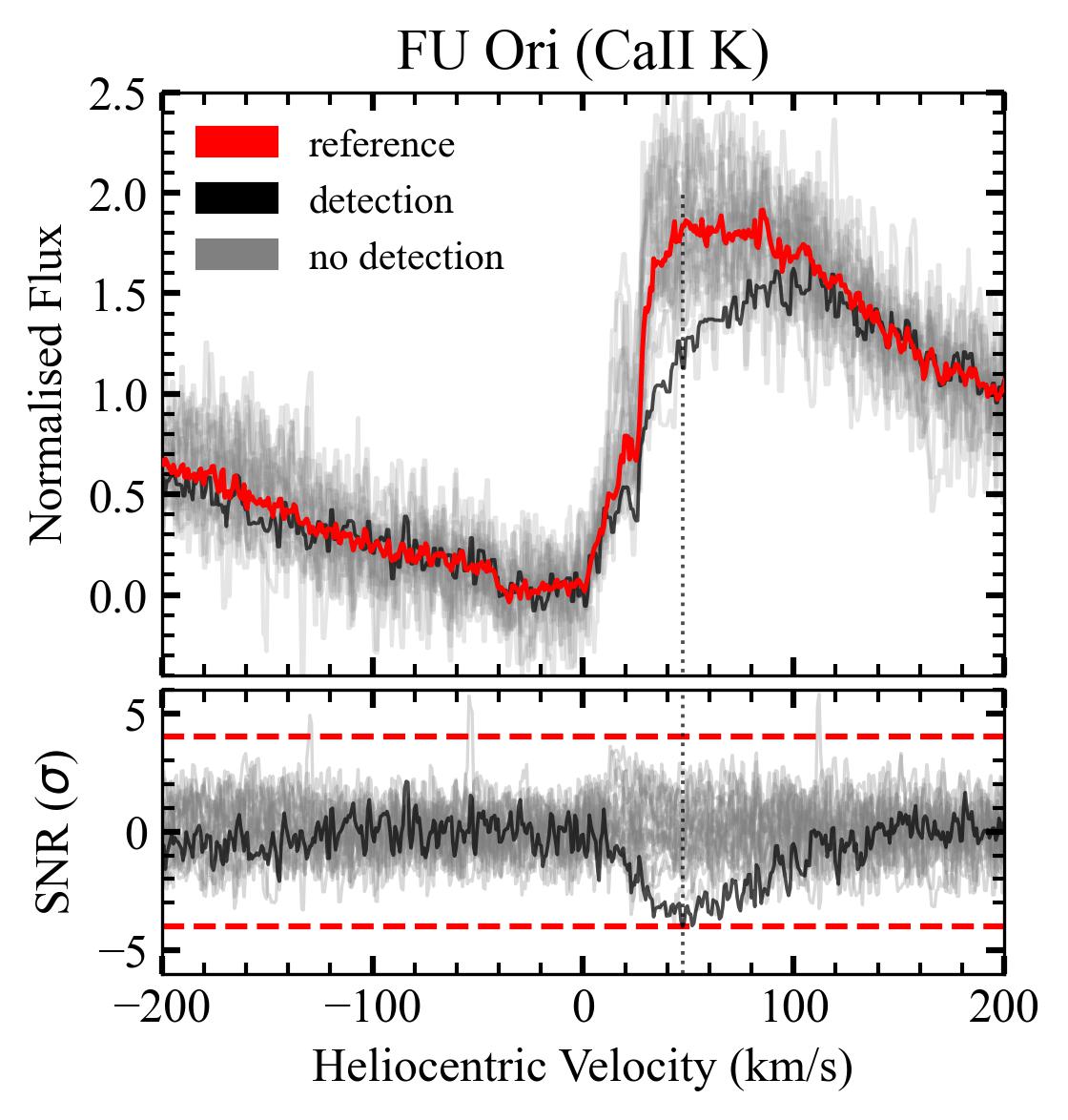}
    \captionsetup{margin={10pt,10pt}} 
    \caption{F0Iab star FU~Ori determined as a false positive detection due to the known variability caused by outbursts in disc accretion.} 
    \label{fig:false-positive/fuori} 
  \end{subfigure}  \hfill
  \begin{subfigure}[t]{0.33\linewidth}
    \centering
    \includegraphics[width=\linewidth]{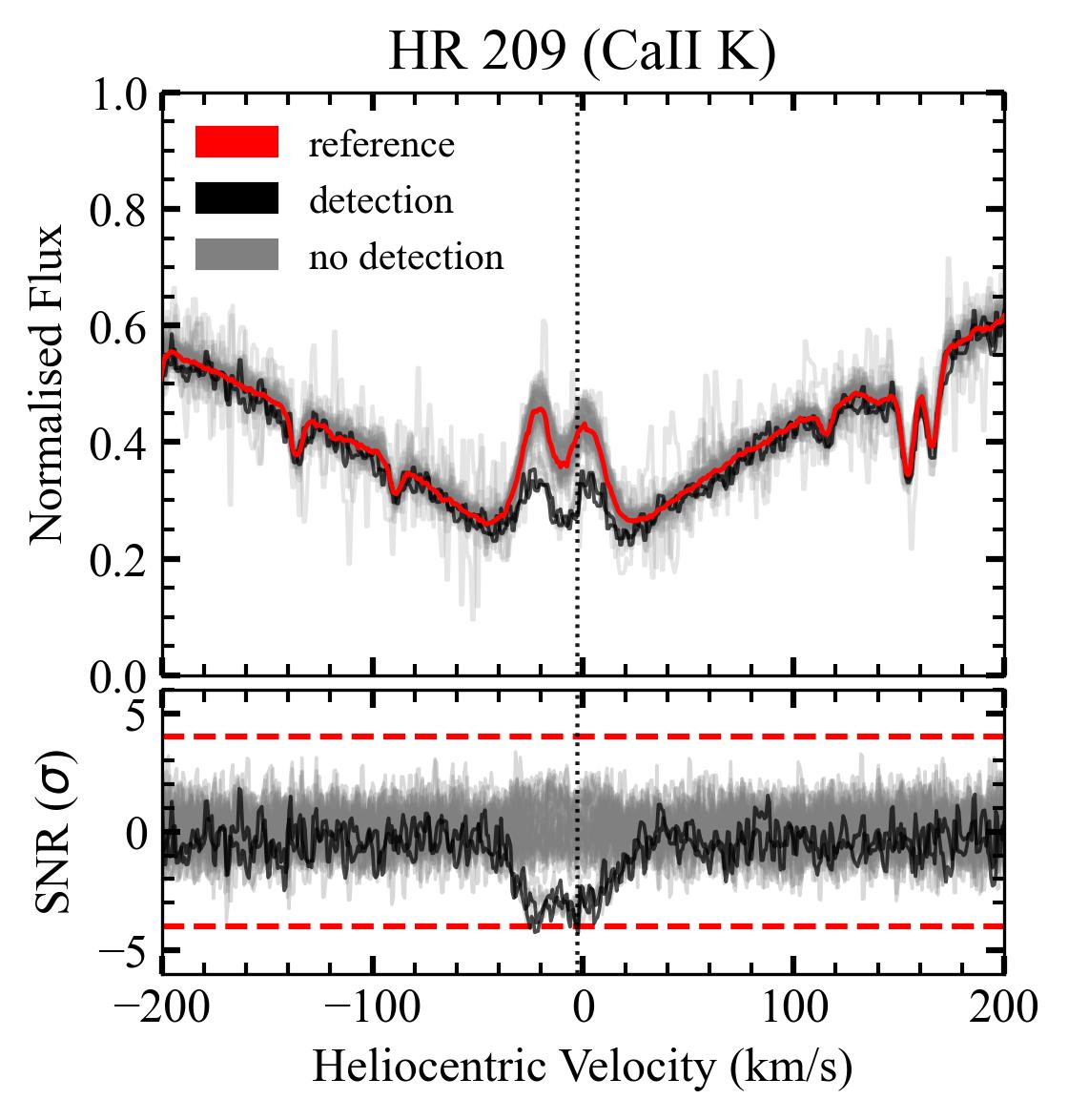}
    \captionsetup{margin={10pt,10pt}} 
    \caption{G3V star HR~209 determined as a false positive detection due to the visible variability in the stellar activity emission feature.} 
    \label{fig:false-positive/stellar-act} 
  \end{subfigure}  \hfill
  \vspace{0.5cm} 
  \begin{subfigure}[t]{0.33\linewidth}
    \centering
    \includegraphics[width=\linewidth]{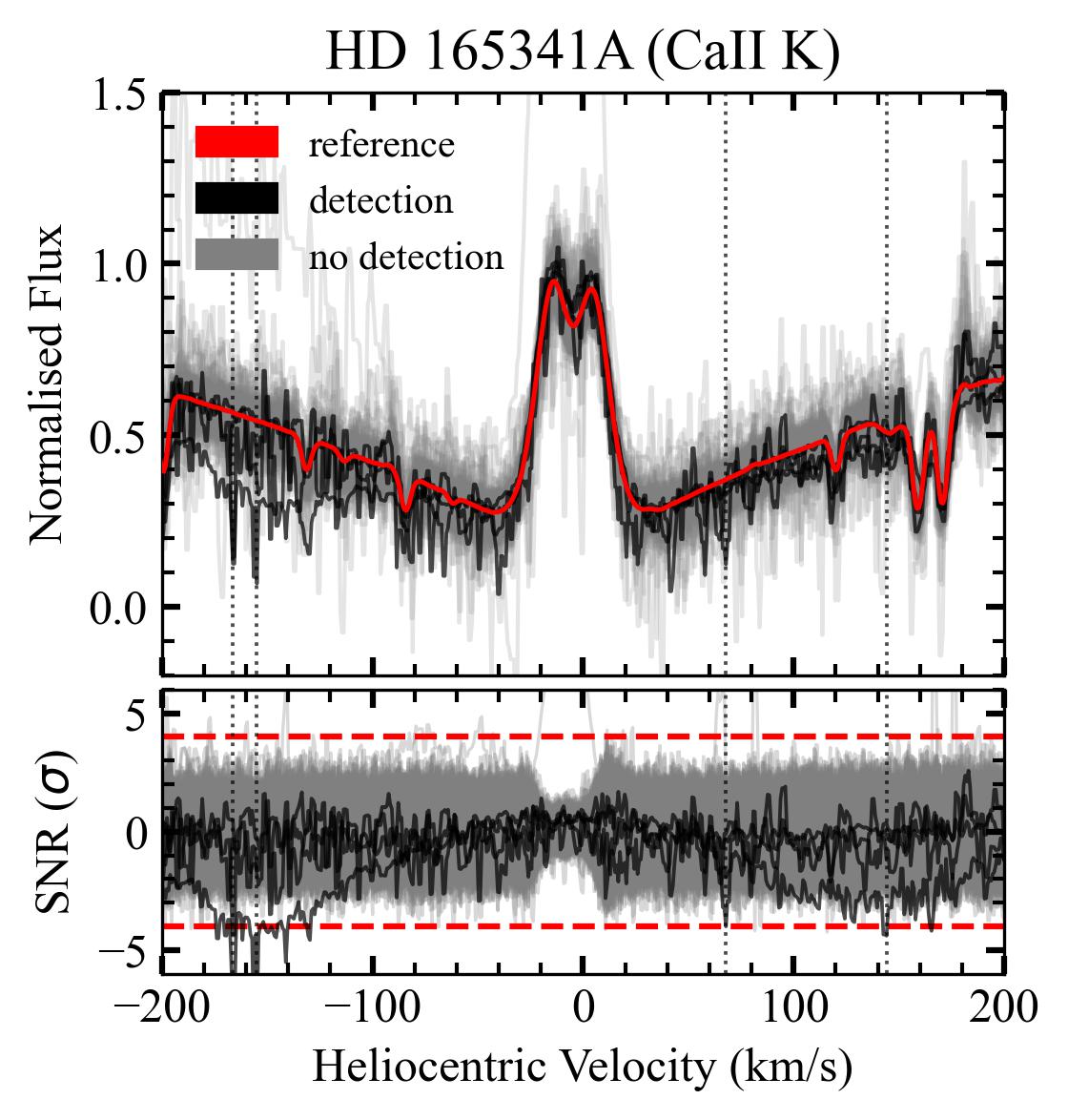} 
    \captionsetup{margin={10pt,10pt}} 
    \caption{K0V star HD~165341a determined as a false positive detection due to the short timescale of the observed transient absorption features.} 
    \label{fig:false-positive/short-var} 
  \end{subfigure}  \hfill
  \begin{subfigure}[t]{.66\linewidth}
      \includegraphics[width=0.5\linewidth]{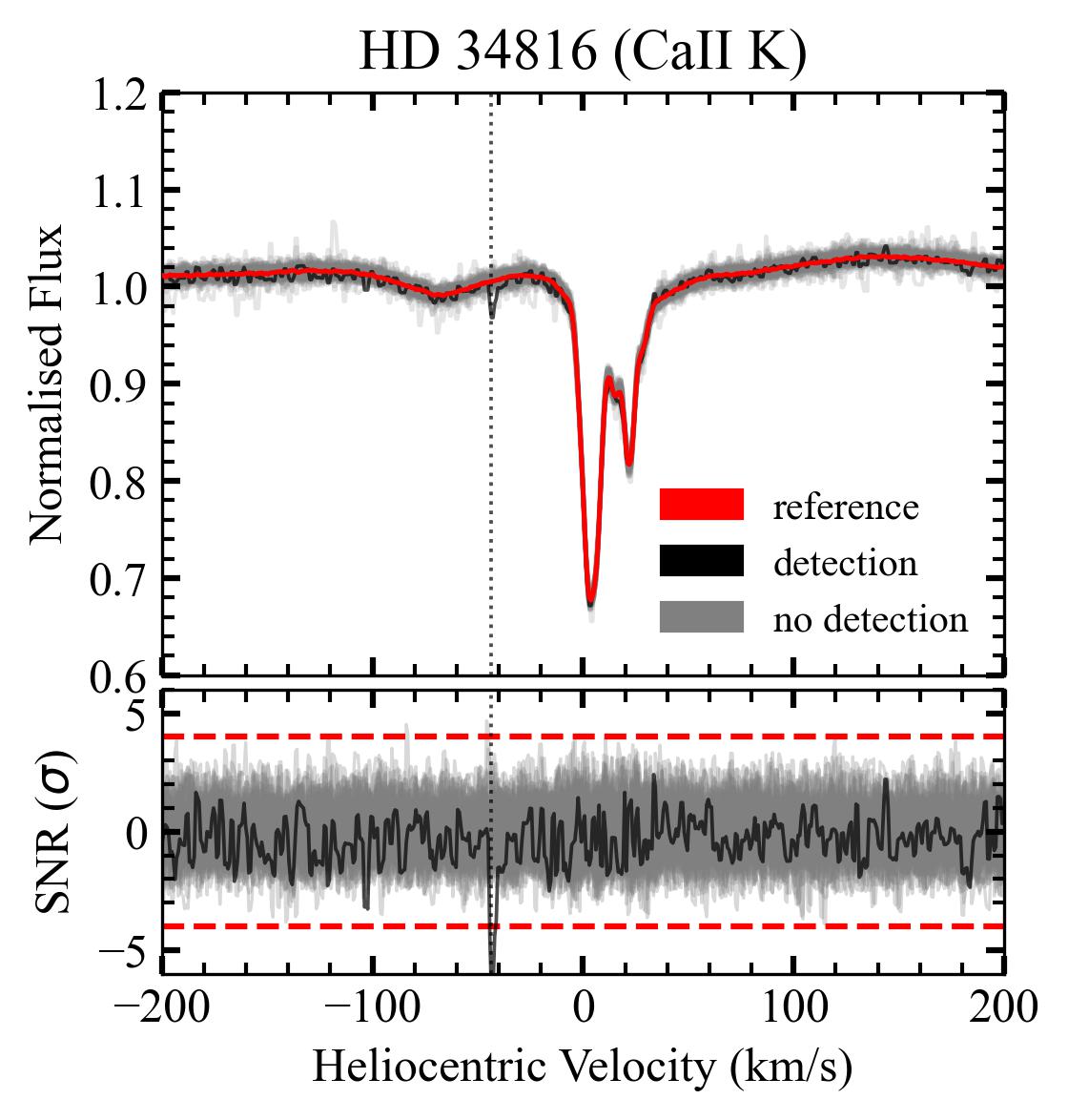}
      \includegraphics[width=0.5\linewidth]{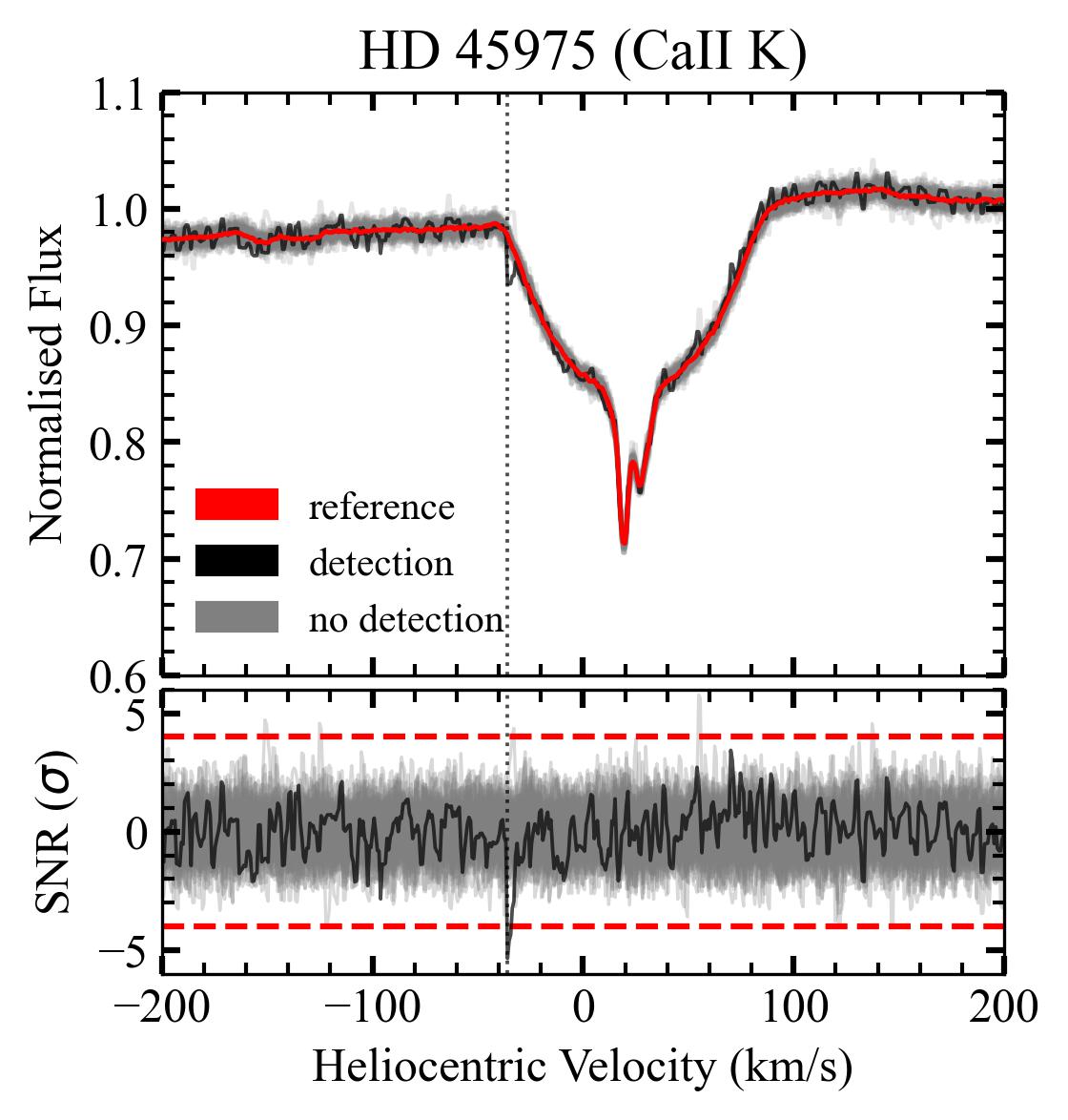}
      \captionsetup{margin={10pt,10pt}} 
      \caption{2 stars HD~34816 and HD~45975 with false positive detections due to some instrumental artefact. The detections are made on the same night and at the same radial velocity position when correcting for the Earth's Barycentric Radial Velocity.}
      \label{fig:false-positives/instr}
  \end{subfigure}
  \vspace{-17pt}
  \caption{Examples per category of false positive detections summarised in \autoref{tab:initial-vetting}. All figures have the same format. The top panel shows the spectra of the star centred on the Ca\,{\sc ii}~K line, with red, grey and black lines respectively representing the reference spectrum, spectra with no detection, and spectra with a detection. The SNR distribution is shown in the bottom panel, where the dotted red lines show the $\pm 4\sigma$ threshold, and the grey and black lines follow the same format as the top panel. While variable absorption features are harder to observe in the superimposed spectra, they are more prominent in the lower SNR panels. The dotted vertical black lines mark the radial velocity at which the transient absorption features are detected.}
  \label{fig:false-positives} 
\end{figure*}

\section{Results}
\label{chap:exo sec:man-vet}
\subsection{False positive detections}

A manual vetting process was performed to verify the validity of the potential detections, and in fact, 80\% of these initial candidates were found to be false positive exocomet detections.

These false positive detections can be classified into 5 different groups, all summarised in \autoref{tab:initial-vetting}. The majority of the initial 155 candidates are flagged because of some real astrophysical variability that is simply not characteristic of exocomet transits. However, based on their HR diagram position, these stars are actually very different to the majority of the stars observed in the HARPS archive, meaning that the search is flagging real astrophysical variability and not picking stars randomly.

One of the common false positive detections is caused by binary stars. An example can be seen in the A0V star HR~4502 shown in Fig.~\ref{fig:false-positive/binary}, where there is an apparent shift in the radial velocity of the central Ca\,{\sc ii}~K absorption line. This shift creates outliers in the SNR distribution, eventually flagging HR~4502 as a candidate. Similar shifts caused false positive detections for 9\% of the initial 155 candidates.

Other known astrophysical variability linked to the star itself has caused false positive detections in 18\% of the 155 candidates. An example can be seen with the star FU~Ori in Fig.~\ref{fig:false-positive/fuori}. The P~Cygni profile, with the blueshifted absorption and redshifted emission features, is very different to the typical spectral profiles dealt with in this work. FU~Ori stars are known to show variable outbursts of disc accretion \citep{powell+12}. These emission and absorption features are therefore caused by variability in mass accretion \citep{hartmann-96}.

Another source of false positive detections is linked to stellar activity. Even though the number of stars showing emission features is reduced by the filter, 57 active stars do not get filtered out. An example can be seen in Fig.~\ref{fig:false-positive/stellar-act}. The G3V star HR~209 shows a variable emission feature at the centre of the broad Ca\,{\sc ii}~K line. For this feature to be masked, the stellar activity threshold would need to be significantly lower than the current threshold of 2 in normalised flux (i.e. $<1$), which would result in some unwanted behaviour when applied across the full search. Such false positives caused by weaker stellar activity are therefore filtered out manually.

For all candidates, the maximum transit time ($\Delta \mathrm{t}_{\mathrm{max}}$) is calculated by adding the time to the closest spectrum with no detection taken before and after the spectrum with the detected absorption feature. No exocomet feature should vary on timescales that are shorter than 30min \citep{paperxxii}. Therefore $\Delta \mathrm{t}_{\mathrm{max}} \leq 30$min is used as a filter to identify false positives caused by variations that are too short to be exocometary. This filter forced 16 candidates to be determined as false positive detections. Those transient absorption features, as seen in Fig.~\ref{fig:false-positive/short-var} around the K0V star HD~165341A, vary on very short timescales, in this case appearing and disappearing with $\Delta \mathrm{t}_{\mathrm{max}} =$ 2min 30sec, therefore ruling out a possible exocomet transit as the cause for the detection.

The last category of false positives corresponds to unclear detections, whether they are detections in the noisiest spectrum or that the detection is an instrumental artefact. Fig.~\ref{fig:false-positives/instr} shows two examples where the detected absorption feature seems plausible until being compared to each other. These two stars, in addition to two others, have detections made just a few hours apart on the same observing night - 05-01-2013. Although these exocomet-like absorption features have different radial velocity positions, the fact that they occur on the same night is very suspicious. \\
The HARPS spectra we are using are corrected for the Barycentric Earth Radial Velocity (BERV), therefore if the detections are found at the same radial velocity pre-BERV correction, the detections would be considered not astrophysical. The BERV value is obtained from the fits file header (HIERARCH ESO DRS BERV) for each spectrum with the identified detection, and we simply subtract the BERV from the radial velocity of the exocomet-like feature. The pre-BERV correction radial velocity of the detected feature in both stars in Fig.~\ref{fig:false-positives/instr} agree with a slight 0.3~\kms~ difference, however insignificant as the HARPS wavelength bin width is only $\approx 1$~\kms. Investigating the other two stars with detections on the same night resulted in a similar conclusion, where the detections around these 4 stars are considered instrumental and not real astrophysical variation.

For the remaining 22 stars, a more in-depth analysis is required to determine the validity of the detections.

\subsection{Ca II H line vetting}
For completeness, we run the entire search over the Ca\,{\sc ii} H line to check whether detections in K have any H counterparts. We lower the outlier threshold to $-3\sigma$ as we expect the H line absorption to be shallower than what is seen in K. Overall, we recovered with H the false positive detections found with K that were flagged as real astrophysical variations (e.g. stellar activity, binaries, ...). In some cases, there are detections found in H that have no K counterparts, however, these are false positive detections linked to stellar activity or other more complicated astrophysical variations. No new candidates are flagged using the Ca\,{\sc ii} H line. More information regarding the H line counterparts to the interesting 22 candidates flagged by K is given throughout the next section.

\subsection{Final Vetting}
\label{chap:exo sec:final-vet}
The remaining 22 candidates could not be ruled out as false positive detections as easily as the other stars presented in the previous section. A more in-depth analysis is needed to discover the cause of the variability detected.

The 22 candidates are categorised into three Tiers based on the strength of evidence for exocometary activity. Tier 1 includes stars with strong and clear indications of exocometary activity, in both H \& K lines. Tier 2 includes stars exhibiting potentially interesting but ambiguous variability which might be due to exocomet activity, but could also be caused by instrumental effects or different astrophysical processes. Lastly, Tier 3 contains stars with complex variable absorption features that are most likely the result of sources of astrophysical variation other than exocomets.

The main factors used in this classification process are the strength of the SNR detection, the radial velocity position of the absorption feature, the absorption width, the absorption depth, the number of detections per star, the spectral type of the star, the duration of the absorption variability, the noise in the spectrum of the detection, and the presence of any ISM cloud absorption in the line of sight.

Although the ISM is unlikely to vary on short timescales, we still want information regarding whether an observed absorption feature is caused by ISM or not. Detecting variation around these ISM absorptions would add complexity and uncertainty regarding the true nature of the variation. When looking for those local interstellar clouds that could be crossing the line of sight and causing specific absorption features, we use the online Local Inter Stellar Medium (LISM) Kinematic Calculator\footnote{\url{http://lism.wesleyan.edu/LISMdynamics.html}} \citep{Redfield+08} which provides the radial velocity of every known cloud that is present between the observer and a queried star.

\begin{table*}
    \caption{Detailed summary of the different parameters of the Tier 1 and Tier 2 exocomet detections. Values correspond to a single detection unless stated otherwise. Columns show the name of the star, the spectral type (SpT), the Tier of the star, the date of the spectrum with the detection in UT, the significance of the detection (SNR in $\sigma$), the radial velocity position of the detection (RV in \kms), the width of the detected feature (W in \kms), the depth of the detected feature, and the maximum transit time calculated ($\Delta \mathrm{t}_{\mathrm{max}}$ in days).}
    \label{tab:final-candidates}
    \centering
    \begin{tabular}{l c l c c c c c c c}
        \hline
        \hline
        Star & SpT  & Tier & Date & SNR & RV & W & Depth & $\Delta \mathrm{t}_{\mathrm{max}}$ \\
         &  &  & (UT) & ($\sigma$) & (\kms) & (\kms) & (\%) & (days) \\
        \hline  
        $\beta$~Pic & A6V & $1$ & many & many & many & many & many & - \\
        Gl~1 & M2V & $2$ & 28-12-2008 01:05:41 & $-4.9$ & $-162$ & $3$ & $114$ & $216$ \\
        HD~94771 & G3/5V & $2$ & 12-12-2020 08:33:14 & $-4.3$ & $-36$ & $5$ & $84$ & $542$ \\
        HD~172555 & A7V & $1$ & 21-09-2004 23:54:43 & $-4.5$ & $4$ & $4$ & $10$ & 0.008\textsuperscript{\tiny a} \\ 
        HIP~5158 & K5V & $2$ & 16-09-2009 07:11:31 & $-4.1$ & $-47$ & $3$ & $154$ & $1153$ \\
        HR~1996 & O9.5V & $2$ & 19-08-2016 09:35:58 & $-5.3$ & $41$ & $4$ & $28$ & $337$ \\
         &  &  & 03-09-2016 09:35:30 & $-5.6$ & $42$ & $8$ & $7$ & $44$ \\
         & &  & 02-10-2016 07:46:30\textsuperscript{\tiny b} & $-7$ & $42$ & $7$ & $20$ & $131$ \\
         &  &  & 12-01-2017 01:42:44\textsuperscript{\tiny c} & $-6$ & $42$ & $5$ & $21$ & $101$ \\
         &  &  & 04-12-2019 01:11:51 & $-4.3$ & $38$ & $3$ & $8$ & $2.2$ \\
        \hline
    \end{tabular}
     \begin{itemize}
         \item [\textsuperscript{a}] Considering weaker detections on the same night (6 spectra over 3h20), the average values (starting from SNR) would be $-3.5\sigma$, $4$~\kms, $2$~\kms, $10$\%, and $333$ days 
         \item [\textsuperscript{b}] Two detections from the same night 02-10-2016 07:46:30/08:09:26. Other values are averaged
         \item [\textsuperscript{c}] Four detections from the same night 12-01-2017 01:42:44/01:47:20/02:19:03/02:23:36. Other values are averaged
    \end{itemize}
\end{table*}

The main reason a star is classified as a Tier 3 candidate is because the variability observed in the spectra is too complicated for any classification to be done properly. These complicated spectra could show complex features such as a combination of emission and absorption features, multiple strong absorption features, suspiciously narrow absorption, and variability detected around other known absorption lines like LISM absorption (examples can be found in Appendix \ref{appendix-vetting}). Differentiating the causes for the detected absorption features would be too complicated, therefore explaining their Tier 3 classification. Also note that none of the Tier 3 candidates has significant Ca\,{\sc ii} H absorption counterparts, which does not provide arguments to overrule the Tier 3 classification. Only Tier 1 and Tier 2 candidates will be individually presented. The final vetting for the 22 Tier 1, Tier 2, and Tier 3 candidates are described in Appendix \ref{appendix-vetting}.

\section*{Individual stars}
Six stars survive the manual vetting process as Tier 1 and Tier 2 candidates. The specifics of these stars are discussed below and summarised in \autoref{tab:final-candidates}. In what follows we quote the number of nights on which targets were observed; exocomet transits last of order hours, so the number of nights is a rough estimate of how many independent measurements were made for each star.

\subsection*{Tier 1}
\label{chap:exo sec:tier1}

Two stars are classified as Tier 1 exocomet systems: $\beta$~Pic and HD~172555. Both of these stars have previously been discussed as exocomet host stars in the literature by \citet{hobbs-85, paperv, kiefer+14b} and \citet{kiefer+14a} respectively. \\

\begin{figure*}
  \centering
    \begin{subfigure}[t]{0.33\linewidth}
      \centering
      \includegraphics[width=\linewidth]{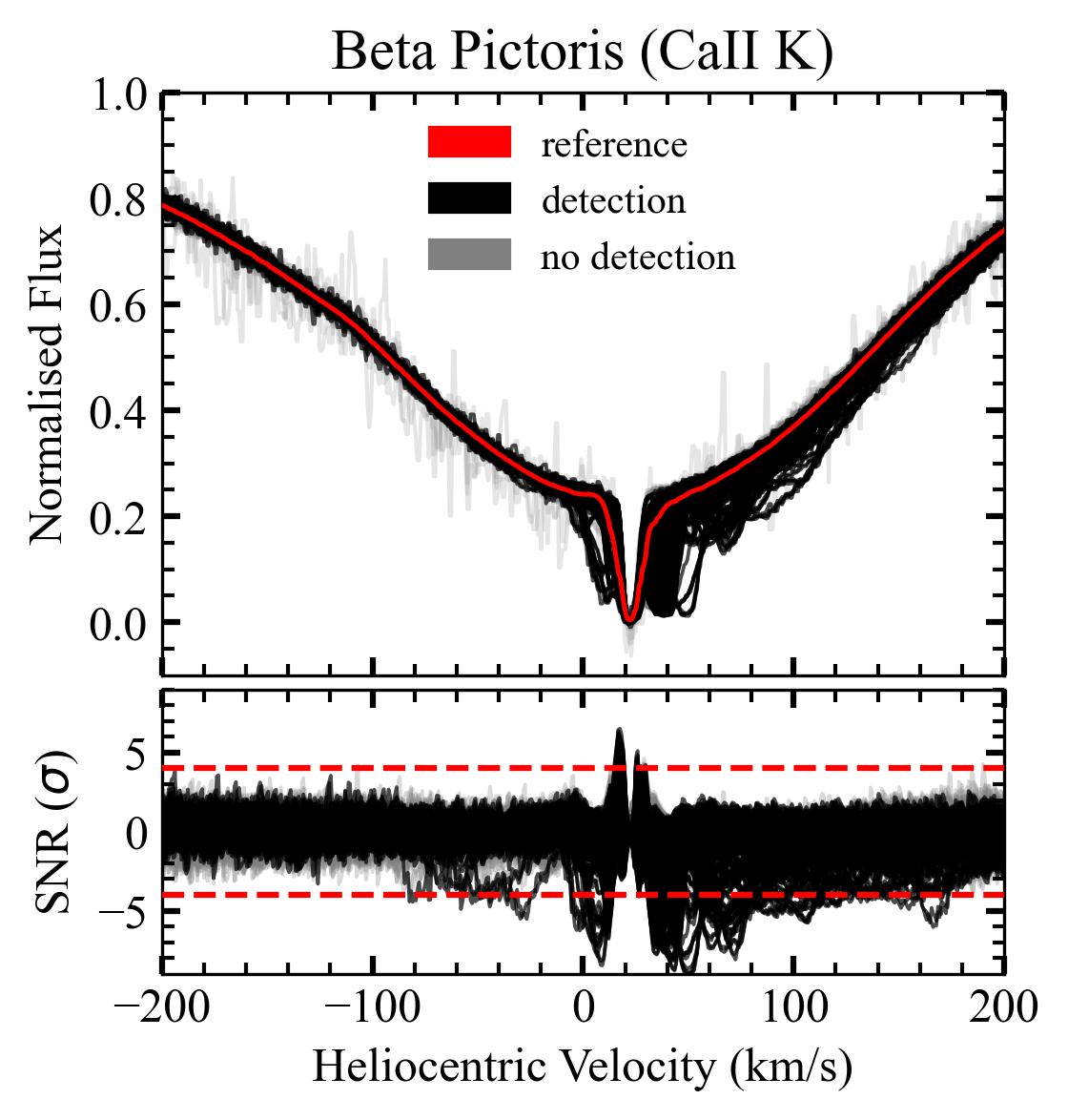}
    \end{subfigure} \hfill
    \begin{subfigure}[t]{0.33\linewidth}
      \centering
      \includegraphics[width=\linewidth]{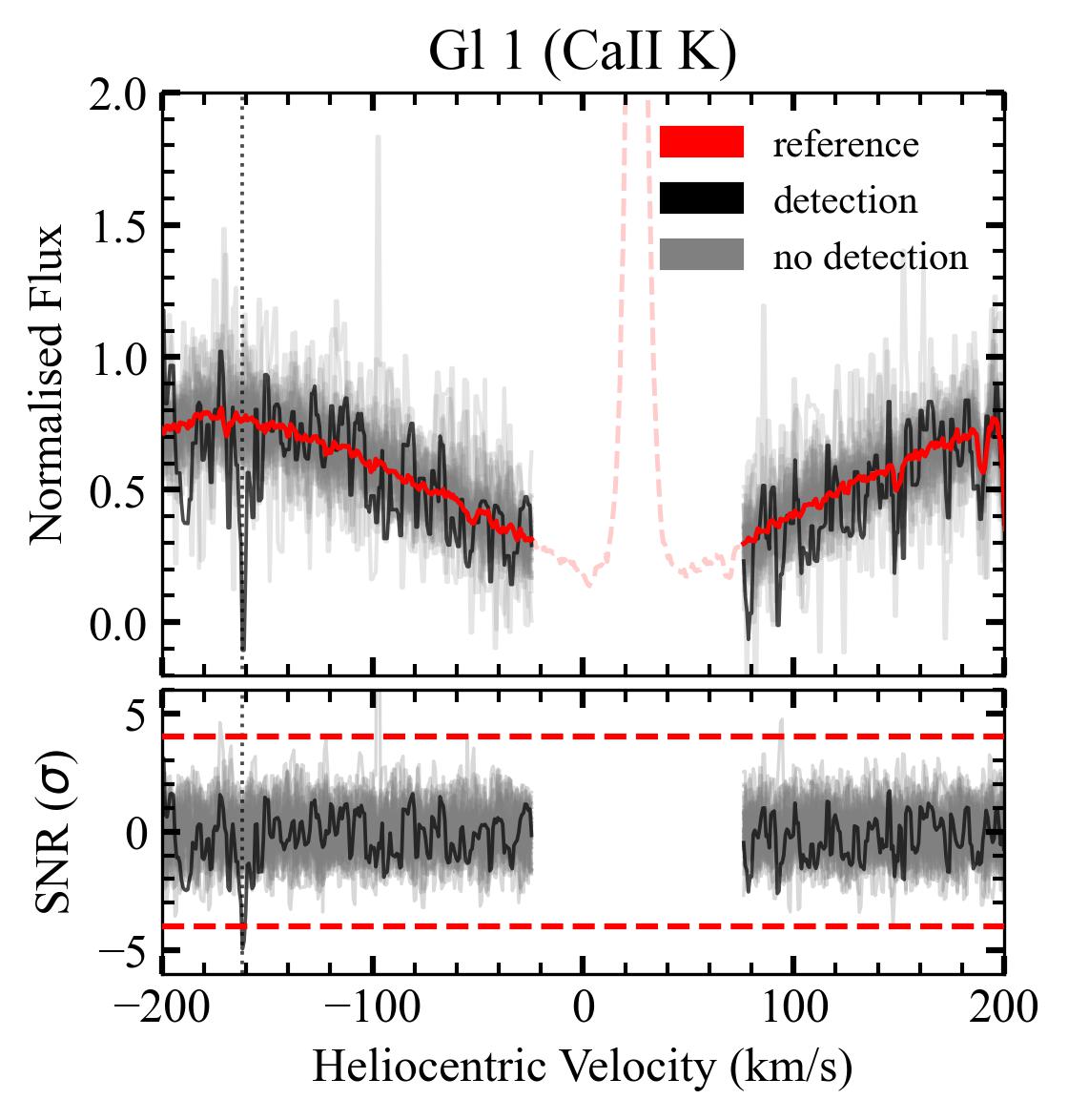}
    \end{subfigure}  \hfill
    \begin{subfigure}[t]{0.33\linewidth}
      \centering
      \includegraphics[width=\linewidth]{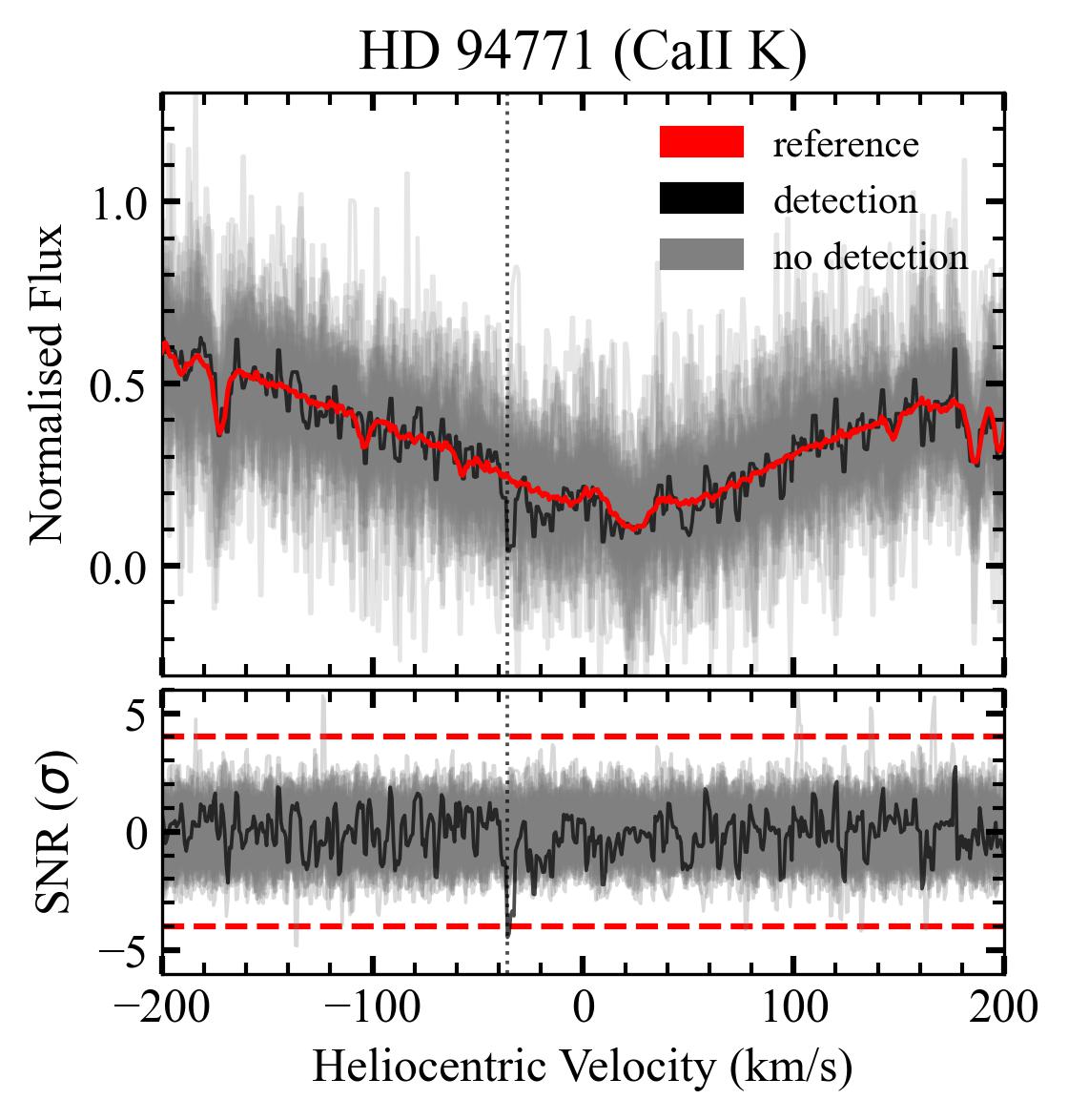}
    \end{subfigure}  \hfill
    \vspace{0.5cm} 
    \begin{subfigure}[t]{0.33\linewidth}
      \centering
      \includegraphics[width=\linewidth]{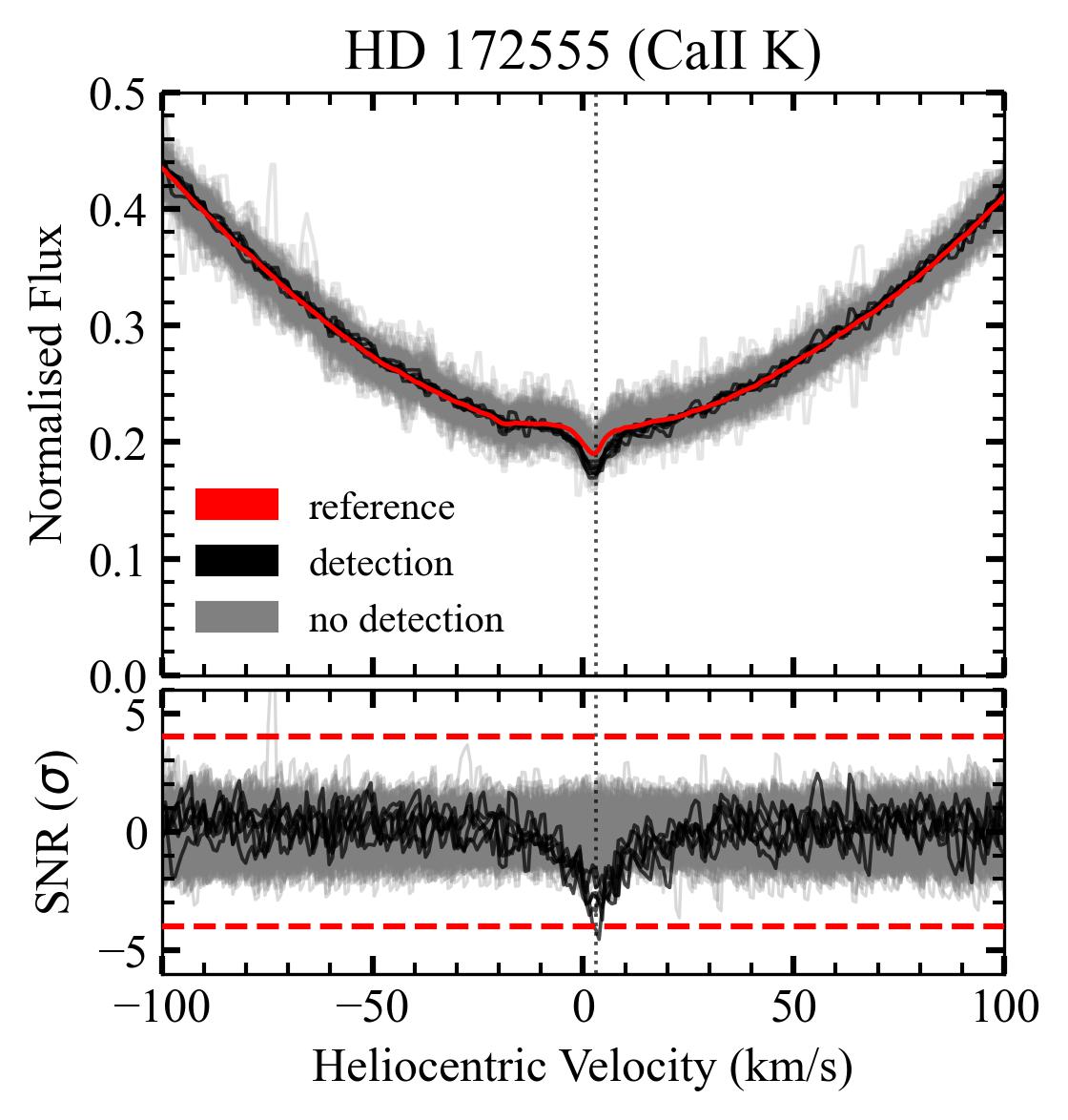} 
      \caption{Tier 1 candidates}
      \label{fig:tier1-cands}
    \end{subfigure}  \hfill
    \begin{subfigure}[t]{.66\linewidth}
        \includegraphics[width=0.5\linewidth]{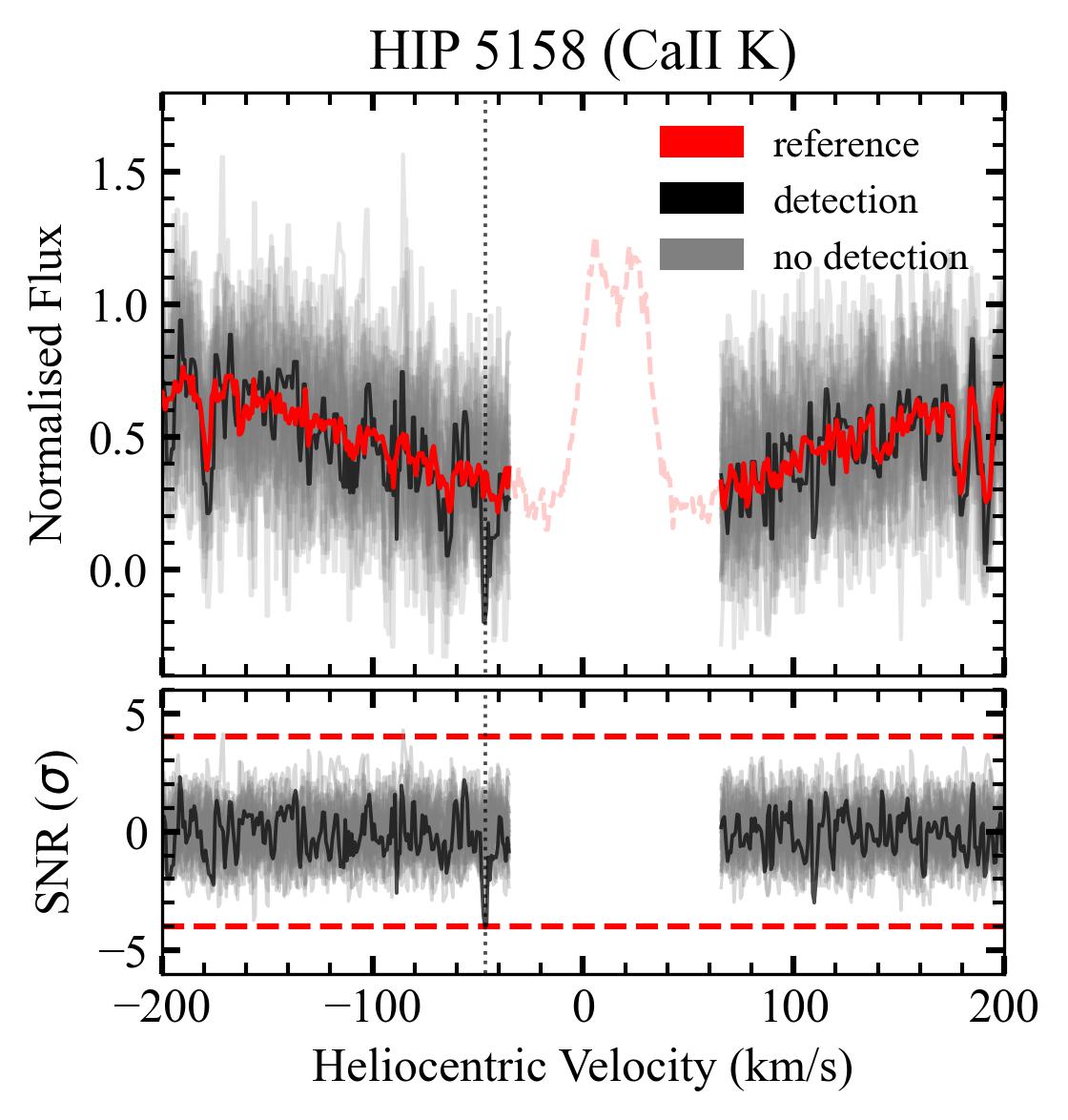}
        \includegraphics[width=0.5\linewidth]{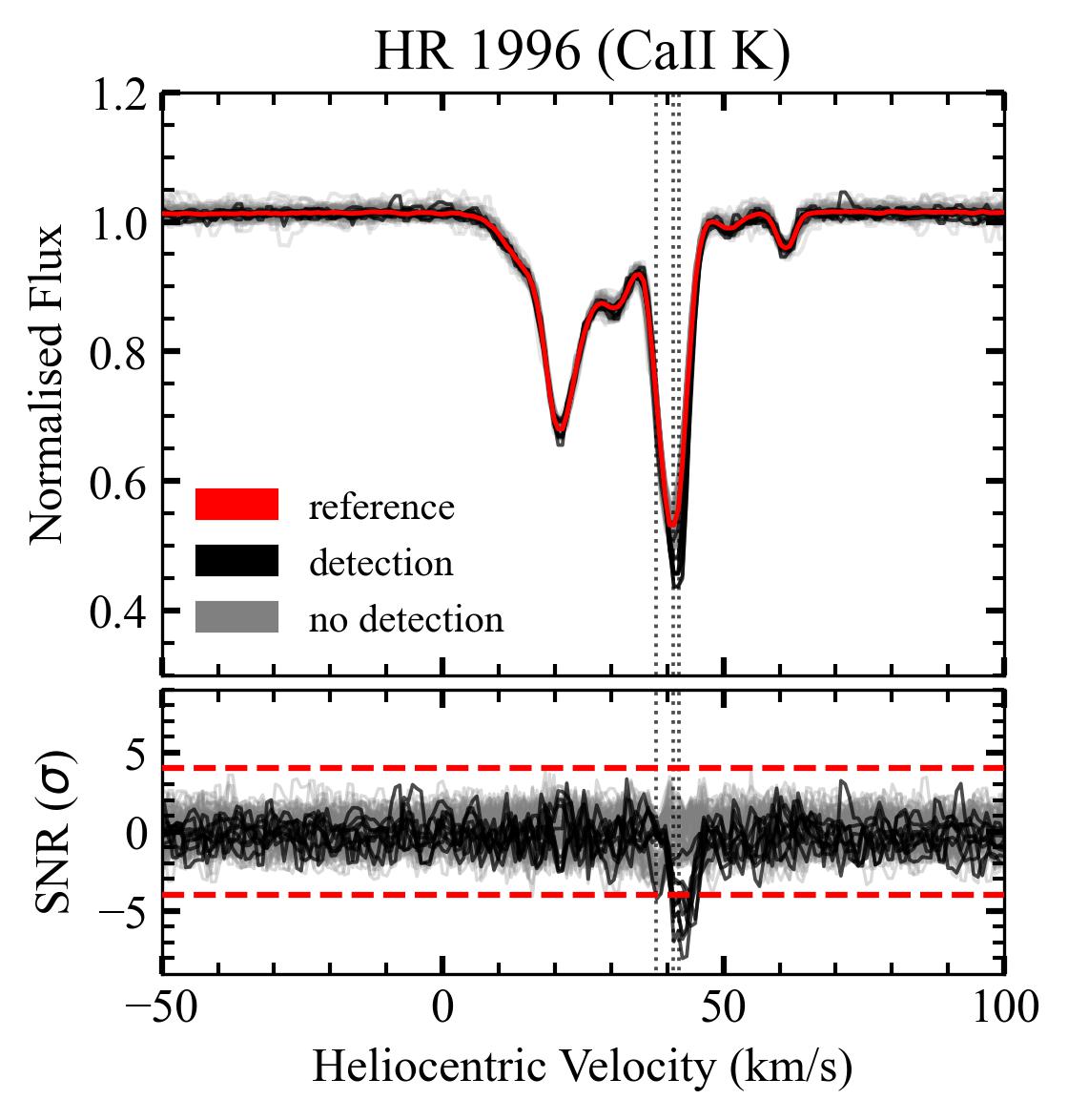}
        \caption{Tier 2 candidates}
        \label{fig:tier2-cands}
    \end{subfigure}
    \vspace{-17pt}
    \caption{Tier 1 and Tier 2 candidates. All plots have the same format. (\textit{top panel}) superimposed spectra in the Ca\,{\sc ii}~K line with the reference spectrum in red, the spectrum with a detection in black, and spectra with no detections in grey. (\textit{bottom panel}) SNR distribution with the same black and grey colour code. While variable absorption features are harder to observe in the superimposed spectra, they are more prominent in the lower SNR panels. The dotted vertical black lines indicate the radial velocity position of the detected transient absorption feature.}
    \label{fig:cands}
  \end{figure*} \hfill

\begin{figure*}
  \centering
    \begin{subfigure}[t]{0.33\linewidth}
      \centering
      \includegraphics[width=\linewidth]{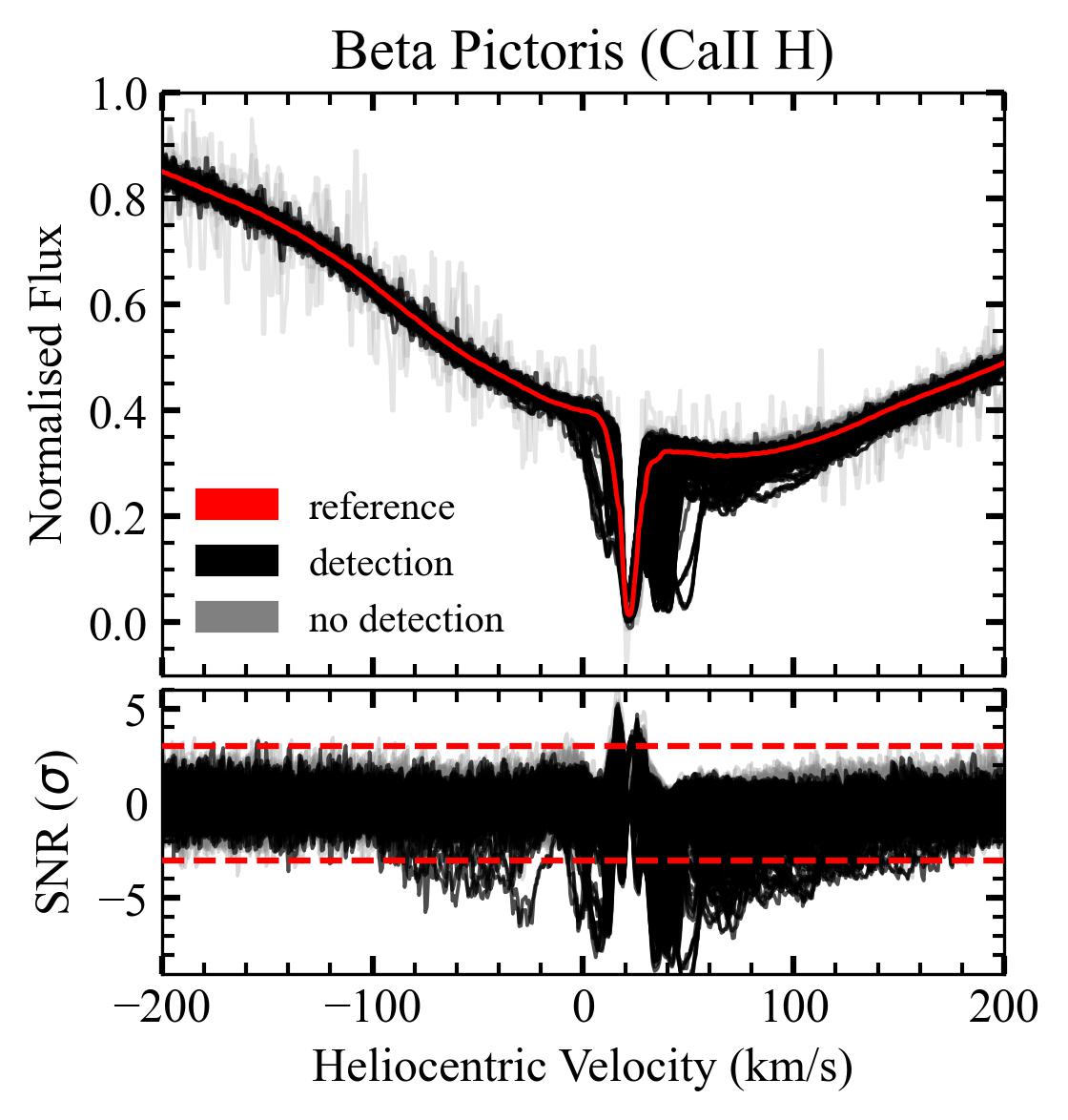}
    \end{subfigure} \hfill
    \begin{subfigure}[t]{0.33\linewidth}
      \centering
      \includegraphics[width=\linewidth]{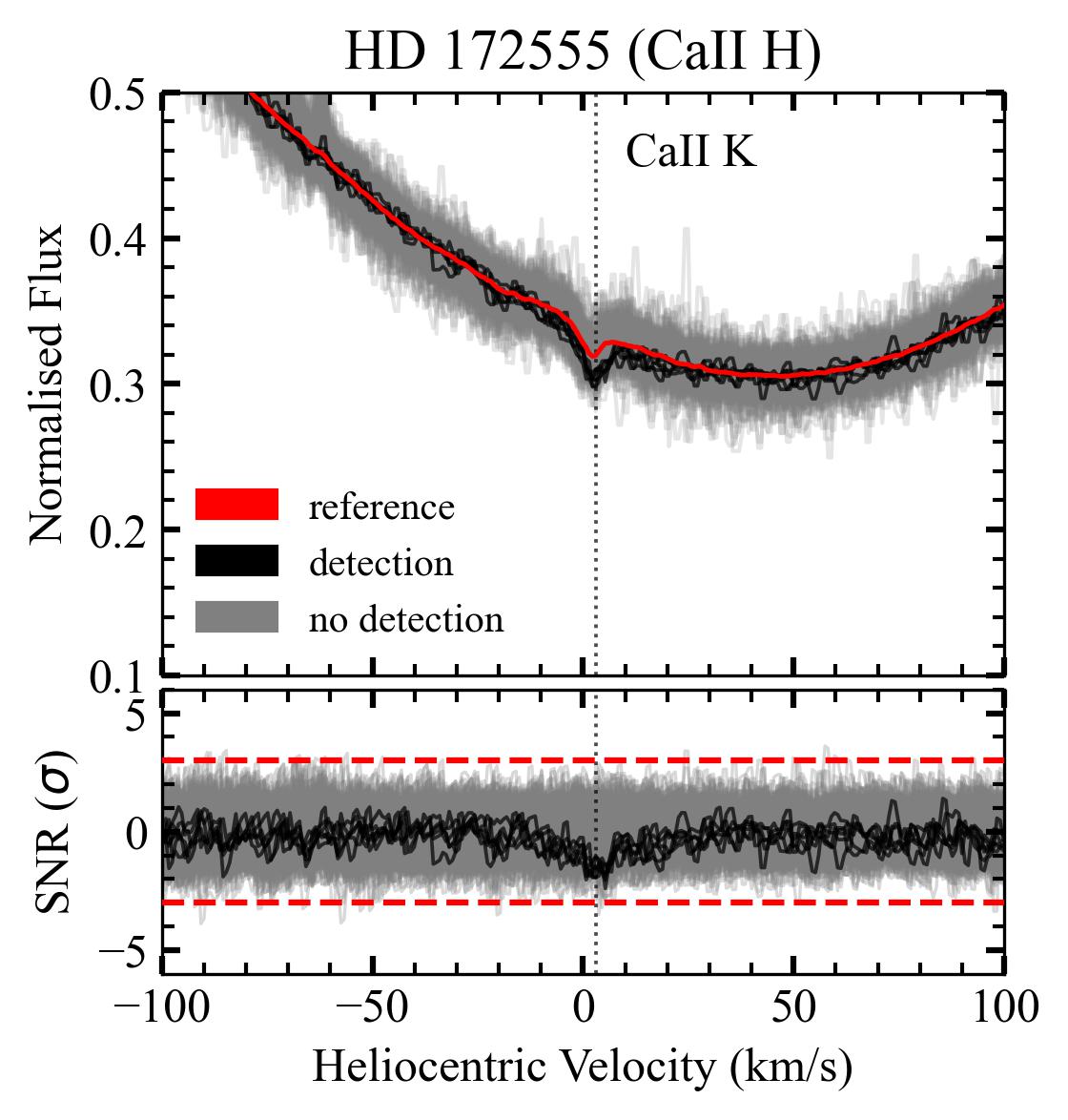}
    \end{subfigure}  \hfill
    \begin{subfigure}[t]{0.33\linewidth}
      \centering
      \includegraphics[width=\linewidth]{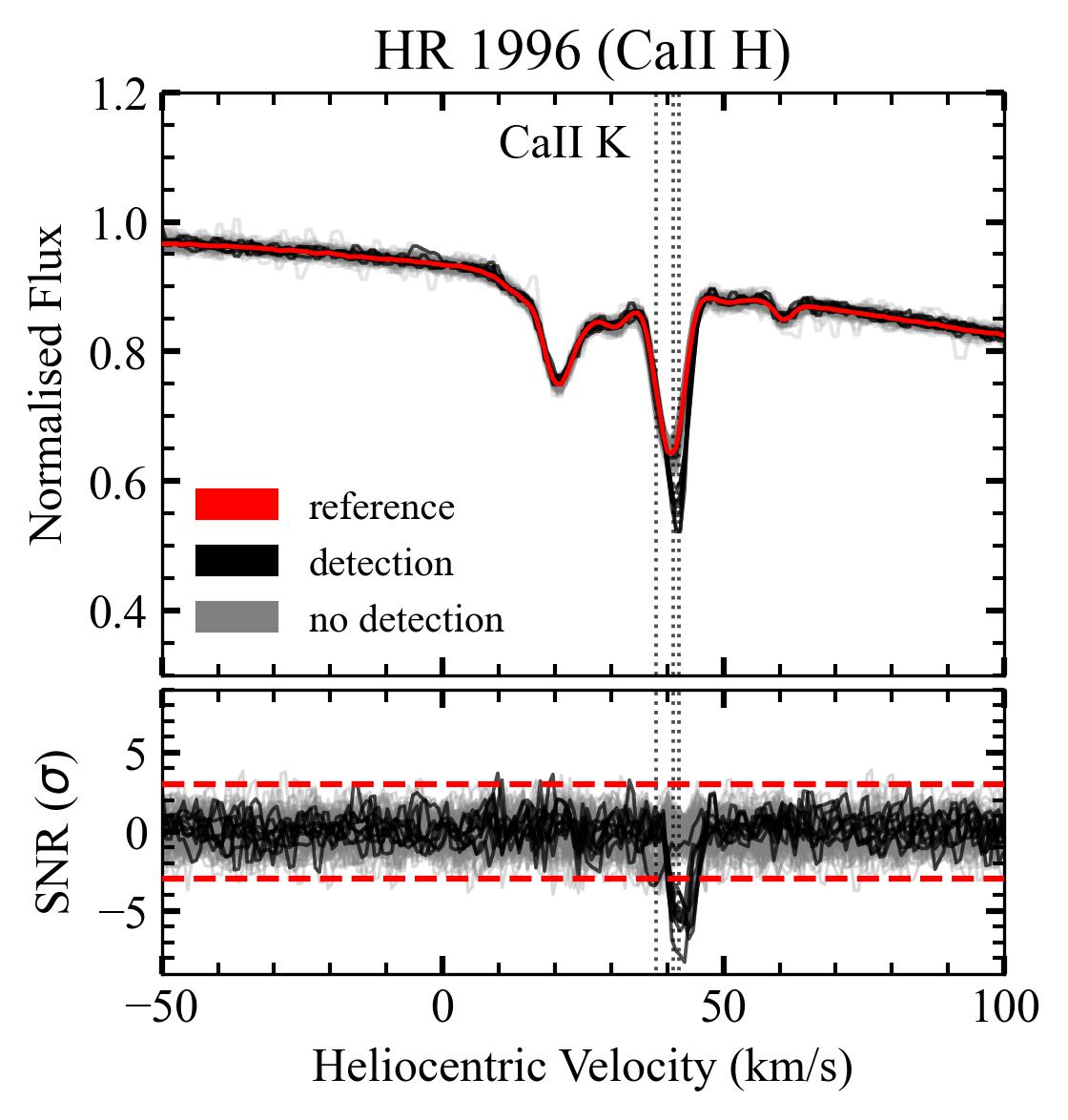}
    \end{subfigure}  \hfill
    \caption{Ca\,{\sc ii} H plots for Tier 1 candidates and HR~1996, the only Tier 2 candidate with distinguishable H counterparts. The format of the plots is the same as in Fig.~\ref{fig:cands}. The black spectra represent those with detections in the K line. The variable absorption features are easier to see in the lower SNR panels. For HD~172555 and HR~1996, the dotted vertical black lines denote the radial velocity of each detection in the Ca\,{\sc ii} K line, which are consistent with the absorption seen in the H line. Note these lines are not plotted for Beta Pictoris as there are too many detections.}
    \label{fig:T1-H}
  \end{figure*} \hfill

\noindent {\normalsize \textit{$\beta$~Pic (HD~39060)}} \\
The A6V star $\beta$~Pic is unsurprisingly classified as a Tier 1 exocomet candidate due to the very large number of exocomet detections over 210 observing nights, as can be seen in Fig.~\ref{fig:tier1-cands} and Fig.~\ref{fig:T1-H}. As expected, transient absorption features are observed every night in both the Ca\,{\sc ii} H and K lines, changing in absorption width and depth as a function of radial velocity. The exocomet scenario is further strengthened by multi-wavelength spectroscopic detections \citep[in the UV][]{papervi,paperxv} and photometric detections using TESS \citep{zieba+19, pavlenko+22, lecavelier+22}. Although many spectra of $\beta$~Pic exist, estimating transit times is complicated as multiple exocomet transits are detected in the same spectrum. A more manual approach is needed to do so by tracking specific absorption features, however, this is not the intention of this work. The feature observed at $20$~\kms~(the systemic velocity of the star) is circumstellar absorption and no obvious LISM cloud absorption is detected. Also notice that features are mostly redshifted, in agreement with previous work.\\

\noindent {\normalsize \textit{HD~172555}} \\
Fig.~\ref{fig:tier1-cands} shows the most significant exocomet detection around the A7V star HD~172555. This star is classified as a Tier 1 candidate despite only having one detected absorption feature satisfying the $-4\sigma$ threshold. This detection is from a spectrum taken on 21-09-2004 at 23:54:43, which recovers one of the detection made in \citet{kiefer+14a}. There are 5 other spectra taken on the same night, two $\pm 5$min from our detection and three 3 hours later. The most significant detections in those 5 spectra do not meet the threshold requirement, however, they all show this same low-velocity feature with a varying significance that averages to $-3.6 \sigma$. We therefore consider these spectra to also show exocometary features. All 6 spectra are taken during the same observing night, when HD~172555 has been observed on 33 nights, yielding a 3\% chance of detecting an exocomet transit per observing night.

The detected transient absorption features are found at the same radial velocity of $4$~\kms, where the star's radial velocity is 2~\kms~\citep{gontcharov+06}. No LISM cloud is found close to the $4$~\kms~region. Only one cloud seems to cause a very weak absorption visible at $-17$~\kms~in both the spectrum with the exocomet detection and the reference spectrum. However, no significant variability is detected in that radial velocity region. We note that no significant variable absorption features are recovered in the H line when using our search method, even when lowering the outlier threshold to $-3\sigma$. However, when investigating the H line in spectra where K line detections are present, weak absorption features are observed at the same radial velocity location as the K line detections, and these are not present in the other spectra (see Fig.~\ref{fig:T1-H}). While these detections align with the reported H \& K line detections in \citet{kiefer+14a}, our method only recovers the detections from 21-09-2004.

Although we do not detect strong Ca\,{\sc ii} H counterparts to the K detection using our method, the presence of exocomets in the HD~172555 system is not unlikely. In spectroscopy, significant exocomet signatures have been detected around HD~172555 in both the optical \citep[seen in this work and][]{kiefer+14a} and the UV \citep{grady+18}. In addition to the multi-wavelength spectroscopic exocomet detections around HD~172555, there is also photometric data that hints towards the identification of transiting exocomets around the star \citep{kiefer+23}. The presence of volatile-rich bodies such as comets in this system is also suggested by a spatially resolved warm almost edge-on inner disc (less than 10~au) around the star \citep{smith+12, engler+18} as well as a detected CO emission \citep{schneiderman+21}. The surprising presence of CO in this young 23~Myr system and atypical dust composition, suggest a production originating from a giant collision between planetary bodies at least 0.2~Myr ago \citep{schneiderman+21}. Such planetary collisions are thought to be common in the final stages of planet formation, which combined with the dust and gas observed in the terrestrial planet-forming region, increase the interest in observing this system. It may even be that the exocomet signature is related to this putative collision, though this would be hard to confirm.

Comparing both Tier 1 candidates, $\beta$~Pic is clearly seen as an outlier in the number of exocomet detections, with detections made every observing night. On the other hand, exocomet signatures have been detected in a single observing night in 2004 for HD~172555, i.e. 3\% of the time, when spectra for this star range from 2004 to 2017. Note that no spectra have been taken within 450 days of the 21/22-09-2004 detections, therefore restricting the transit time to 3h$30 < \Delta t < 450$days.

\subsection*{Tier 2}
\label{chap:exo sec:tier2}
Four stars are classified as Tier 2: Gl~1, HD~94771, HIP~5159, HR~1996. It is the first time these four stars are presented as potential exocomet candidates due to detected absorption variability in their spectra. The Tier 2 classification for these 4 stars originates from the uncertain origin of the variability: 3/4 are single spectrum detections that are not detected in H, so instrumental variation remains a possibility, and 1/4 is clearly a robust H \& K detection, but may be explained by other astrophysics.\\
  
\noindent {\normalsize \textit{Gl~1 (HD~225213)}} \\
The M2V star Gl~1 shows a single deep $-4.9 \sigma$ absorption feature at $-162$~\kms~seen in Fig.~\ref{fig:tier2-cands}, resulting in a Tier 2 exocomet candidate. The central region of the Ca\,{\sc ii}~K line is masked with the stellar activity filter due to the variable emission feature at the star's radial velocity of $\approx 25$~\kms. All LISM clouds have velocities in the masked region of the spectrum, therefore not impacting the detection. Note that the spectrum is noisy which is expected from the late spectral type of the star. However, the noise of the spectrum is taken into account in the detection, leaving this absorption feature at $-162$~\kms~to be significant enough for a potential exocomet detection. Exocometary-like features have only been detected in a single spectrum over 44 observing nights, yielding a 2.3\% chance of making a detection per night. However, there are no consecutive spectra within 216 days of the claimed detection, therefore contributing to the uncertainty in the exocomet nature of the feature. Searching for Ca\,{\sc ii}~H counterparts to the detection in K, the lack of significant detection of variability at the correct radial velocity location (see Fig.~\ref{fig:tier2-H}) does not give a reason for a promotion to a Tier 1 classification. \\

\noindent {\normalsize \textit{HD~94771}} \\
The G3/5V star HD~94771 shows a single $-4.3 \sigma$ transient absorption feature at $-36$~\kms, seen in Fig.~\ref{fig:tier2-cands}. No LISM absorption is detected near the claimed exocomet detection. Variable absorption features have only been detected in a single spectrum over 118 nights, yielding a 0.8\% chance of making a detection per night. The lack of spectra close to the detection, with no other spectra taken within 542 days, contributes to its Tier 2 classification. Additionally, no strong signs for Ca\,{\sc ii} H line counterparts (Fig.~\ref{fig:tier2-H}) have been detected around this star.
\\

\noindent {\normalsize \textit{HIP~5158}} \\
Fig.~\ref{fig:tier2-cands} shows the single transient absorption feature of the K5V star HIP~5158 seen around $-50$\kms, just outside the masked stellar activity region. Note that here the activity filter threshold has been manually lowered to 1.2 for this star, to mask the emission feature and remove any false positive detections in that region of the spectrum. No LISM cloud absorption has been identified near the detected exocomet-like feature. Although the spectrum seems very noisy, the SNR value for this detection satisfies our threshold ($-4.1 \sigma$) and stands out in absorption depth (150\%). The negative flux value could be explained by the absorption dip being a combination of both noise and a signal. This star only shows an exocomet-like feature in a single spectrum over 37 observing nights, corresponding to a 2.7\% chance of making a detection per night. The lack of spectra taken close to the detection ($\Delta \mathrm{t}_{\mathrm{max}}$ = 1153 days) contributes to the uncertainty in classifying this candidate higher than a Tier 2. No significant Ca\,{\sc ii} H absorption (Fig.~\ref{fig:tier2-H}) is detected around this star which does not add arguments towards a firm exocomet detection. It might also be interesting to note that this star has 2 confirmed exoplanets detected with radial velocity, HIP~5158b at 0.9~au \citep{locurto+10} and HIP 5158c at 7.7~au \citep{feroz+11}.\\

\noindent {\normalsize \textit{HR~1996 ($\mu$~Col, HD~38666)}} \\
The O9.5V star HR~1996 seen in Fig.~\ref{fig:tier2-cands} shows multiple transient absorption features around the same 40~\kms~region. Although 9 spectra show transient absorption features, these are spread over 5 different observing nights. HR~1996 has been observed for 63 nights, yielding an 8\% chance of making a detection per night.

The Tier 2 classification can be explained by multiple factors. A key component is the radial velocity position of the claimed detections, which only slightly changes even though the detections are separated by several months and years. Compared to systems like Beta Pictoris with multiple detections, the radial velocity position should be more variable. Assessing whether the signal is periodic rather than sporadic is challenging, as no spectra are collected between the widely spaced monthly observing nights. However, this consistency in the radial velocity of the absorption feature cannot be explained by any LISM clouds along the line of sight. This consistency in the radial velocity position of the detected absorption features is also noticed in the counterpart detections observed in the Ca\,{\sc ii} H line (see Fig.~\ref{fig:T1-H}). HR~1996 and $\beta$~Pic are the only two candidates showing strong H-line counterparts using our search method.

Although HR~1996 displays variability in both the H \& K lines, another reason for its Tier 2 classification can also be explained by the very early spectral type of the star. O-type stars are very bright and hot, where the presence of exoplanets has not been proven so far, questioning the presence of smaller bodies such as exocomets. O-type stars are known to show variability in their spectra that can be caused by stellar wind activity, magnetic phenomena, binarity or sometimes line-profile variability that is hard to explain \citep{Fullerton+96, Holgado+2018}, which adds uncertainty to the nature of the detection and prevents us from claiming a firm Tier 1 detection of an exocomet. If our detection is exocometary, one possible scenario involves an exocomet located at a significant distance from the star but still experiencing intense stellar insolation. This setup would explain the low-velocity feature, suggesting a considerable distance from the star. Overall, more information about this system as well as more consistently spaced spectra are needed to promote this candidate to a Tier 1 classification.

\section{Discussion}
\label{discussion}
The first part of the discussion will focus on the statistics and trends found for the Tier 1, Tier 2, and Tier 3 candidates. To achieve this, the demographics of the different detected transient absorption features are discussed through population plots, showing the distribution of absorption depths and widths as a function of radial velocity for all 22 Tier 1, Tier 2 and Tier 3 candidates. These stars will also be positioned on an HR diagram, enabling the characterisation of any trends in spectral type and age for our Tiered exocomet candidates. The second part of the discussion will evaluate how the performance of our exocomet search method compares to previous literature surveys, with a focus on the HARPS survey in \citet{WandM+18}.

\subsection{IR excesses of Tier 1 \& 2 candidates}
Other than the number of measured exocomet detections per star, another property differentiates Tier 1 candidates from Tier 2; the presence of a debris disc.

Checking for the presence of circumstellar dust can be done by searching for infrared flux excess in the star's spectral energy distribution. If circumstellar dust were present, it would be heated up by the star and the dust would re-emit at longer wavelengths where a greater measured infrared flux would be detected on top of the star's blackbody continuum. We use ALLWISE \citep{wright+10} photometry and more specifically, the W3 ($12~\mu$m) and W4 ($22~\mu$m) wavelength bands to calculate the magnitude difference and determine whether there is strong evidence for an infrared excess. There is a clear difference between Tier 1 and Tier 2 candidates, where W3$-$W4 $>1$ for Tier 1 and W3$-$W4 $<0.1$ for Tier 2, suggesting infrared excess in both Tier 1 candidates but no strong evidence in the Tier 2 candidates. While the numbers are small and considering that Tier 2 candidates are less certain compared to the confirmed Tier 1 stars, there is circumstantial evidence for a connection between confident exocomet detection and the presence of a detectable debris disc.

\begin{figure}
    \centering
    \includegraphics[width=\linewidth]{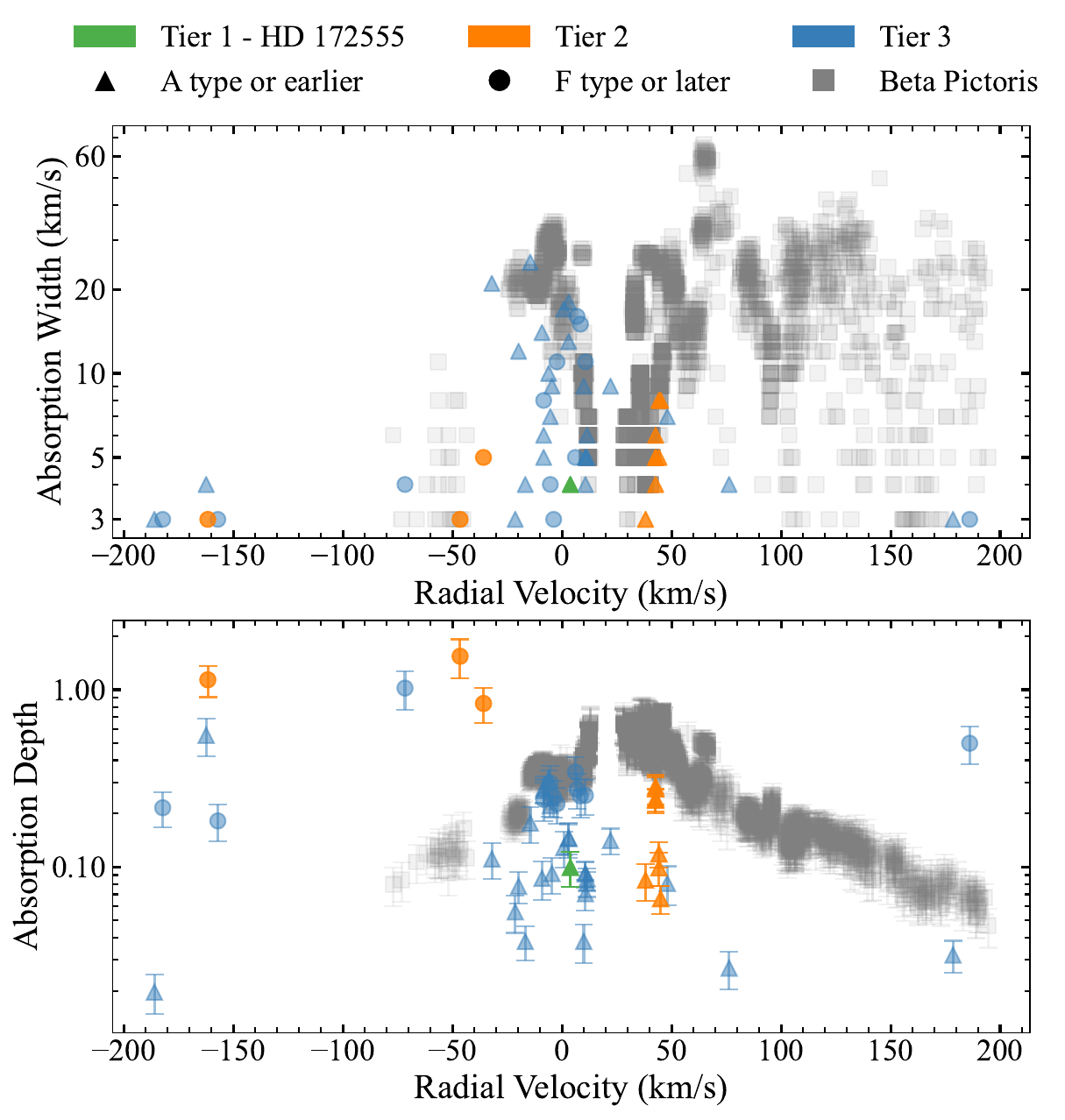}
    \vspace{-17pt}
    \caption{Absorption widths and absorption depths for all individual exocometary-like features as a function of radial velocity. Some of the data points correspond to detections made around the same star. Colours represent the different classification tiers while the shapes give information about the spectral type of the star with the detection. For comparison, the distribution of exocomet absorption features around $\beta$~Pic is plotted with the grey squares.}
    \label{fig:dis/pop}
\end{figure}

\begin{figure}
    \centering
    \includegraphics[width=\linewidth]{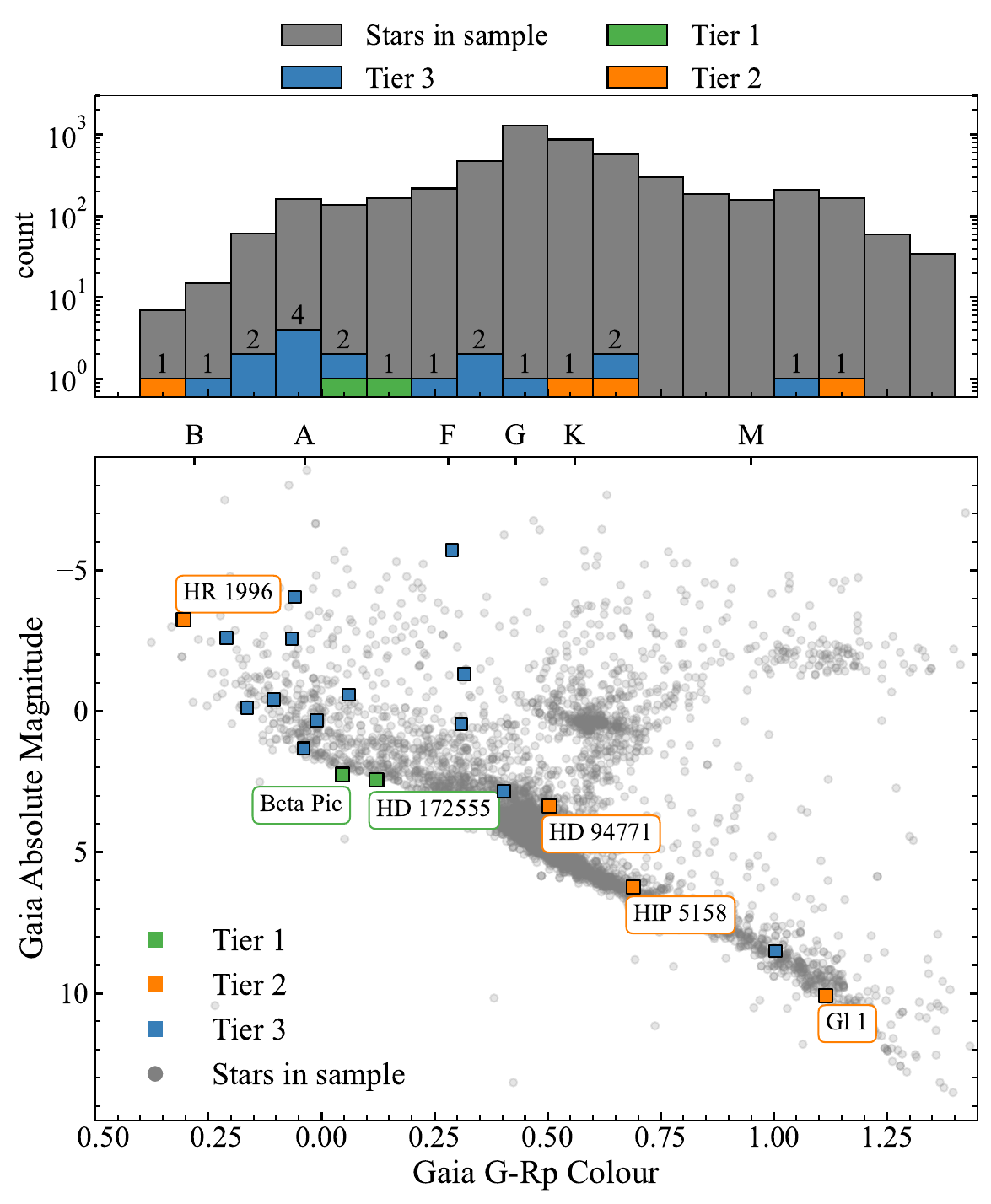}
    \vspace{-17pt}
    \caption{Hertzsprung-Russell diagram of all the 22 Tiered candidates over the entire sample of stars used in this work. The same colour code is used as for Fig.~\ref{fig:dis/pop}, where Tier 1 candidates are in green, Tier 2 candidates in orange, Tier 3 candidates in blue, and other non-classified stars are in grey. The top panel histogram shows the distribution of stars as a function of Gaia G-Rp colour as well as the relative number of Tiered candidates over the total number of stars in the sample.}
    \label{fig:dis/HR}
\end{figure}

\subsection{Population plots}
Fig.~\ref{fig:dis/pop} shows the distribution of absorption widths and depths as a function of radial velocity, for every detection of the 22 Tier 1, Tier 2 and Tier 3 candidates. There are two main components in these plots, the demographics of the detected absorption features per classification Tier, represented by the different colours, but also as a function of spectral type, seen with the different shapes.

Overall, a wide range of detected absorption features across the $\pm 200$~\kms~radial velocity region exists. Most late-type stars (F-type or later) are found to have high-velocity features (HVF). This is an expected trend as the variable activity emission features in the low-velocity features region (LVF) preclude potential exocomet detection. Most of these HVF are Tier 3 candidates and are suspiciously deep and narrow absorption, which might be expected from the noisier spectra of dimmer stars, e.g. if the noise is not strictly Gaussian or there are rare systematic effects. For most of the HVF detected in this work, the absorption features are outside of the main rotationally broadened Ca\,{\sc ii}~K absorption line, with an increased chance of being blended with some other atomic line \citep[e.g. Fe\,{\sc i} at $-257$~\kms, $-194$~\kms, or $-165$~\kms;][]{nist}. 

There is an apparent increase in the number of detections in the LVF region, for early-type stars presumably related to lower levels of stellar activity. The observed trend in Fig.~\ref{fig:dis/pop}, irrespective of Tiered classification, indicates that detections around late-type stars predominantly occur in the HVF regions, characterised by deep, narrow absorption features. In contrast, detections around earlier-type stars are more frequently found in the LVF region, which is marked by shallow, wider absorption features.

A final comparison can be made between the two Tier 1 candidates. Fig.~\ref{fig:dis/pop} previously indicates a trend where LVF are deeper and narrower, while HVF are wider and shallower. This behaviour is primarily governed by the size, density, and velocity distribution of the gas in the coma of the exocomet, influenced by radiation pressure \citep[e.g.][]{paperix}. Comparing $\beta$~Pic to HD~172555, the absorption width of the detection in HD~172555 (seen in green in Fig.~\ref{fig:dis/pop}) is slightly narrower but still consistent with what is seen from the $\beta$~Pic LVF detections (seen in grey in Fig.~\ref{fig:dis/pop}). On the other hand, the absorption depth of the HD~172555 detection is approximately $10\%$ shallower than the $\beta$~Pic LVF detections. Although there are not many exocomet detections around HD~172555, the differences between the detected features of our two Tier 1 candidates might question whether $\beta$ Pic should be a model or an outlier when it comes to typical exocometary activity.

\begin{figure*}
    \centering
    \includegraphics[width=0.25\linewidth]{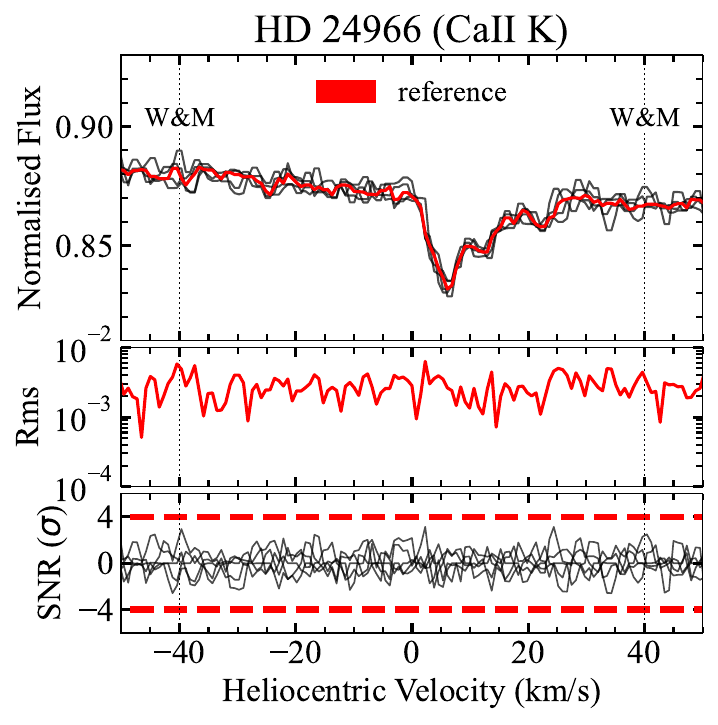}
    \includegraphics[width=0.24\linewidth]{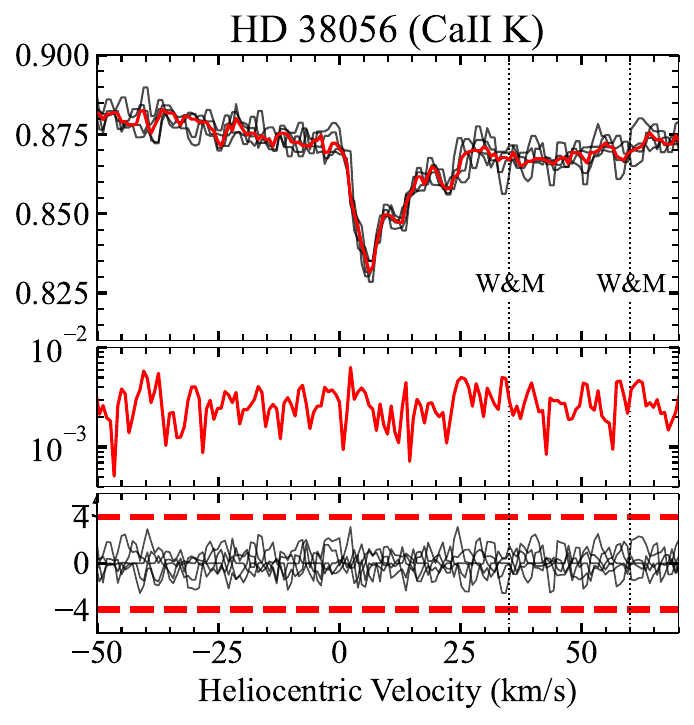}
    \includegraphics[width=0.24\linewidth]{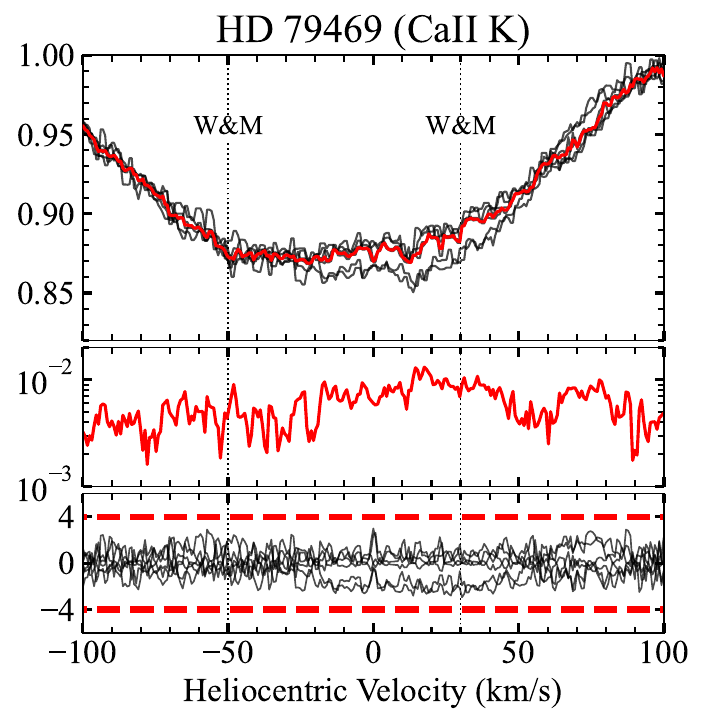}
    \includegraphics[width=0.24\linewidth]{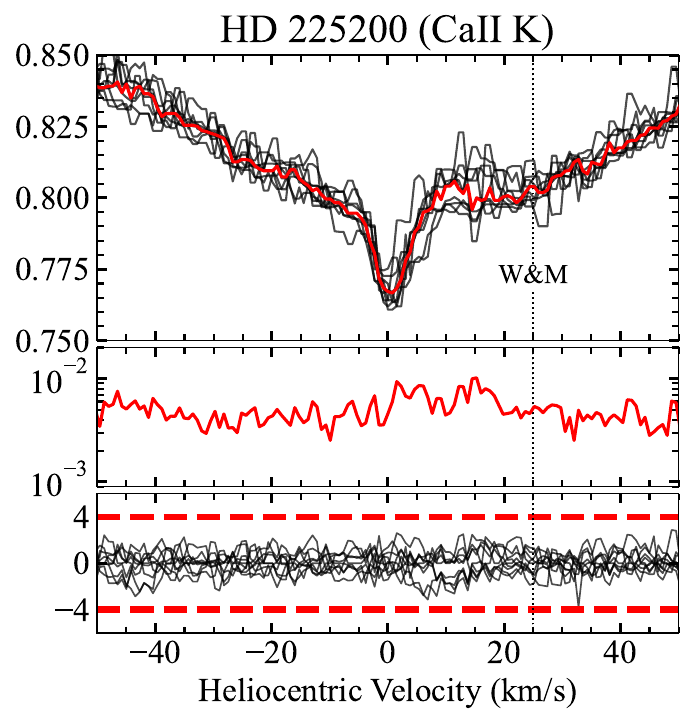}
    \caption{HD~24966, HD~38056, HD~79469, and HD~225200 compared to Fig.~4 of \citet{WandM+18}. Each plot shows three different panels: (\textit{top panel}) superimposed spectra of the star with the reference spectrum in red, (\textit{middle panel}) error on the reference spectrum as a function of heliocentric velocity, (\textit{lowest panel}) SNR distribution with the dashed red lines marking the $\pm 4\sigma$ threshold. The dotted vertical black lines point towards the position of the claimed detections in \citet{WandM+18}.}
    \label{fig:WandM} 
\end{figure*}

\subsection{HR diagram distribution}
Fig.~\ref{fig:dis/HR} shows the distribution of all 22 Tier 1, Tier 2, and Tier 3 candidates on an HR diagram on top of the sample of stars in the HARPS archive. Overall, the candidates are spread across the HR diagram with most of the stars on the Main Sequence (MS). The candidates off the MS tend to be Tier 3 candidates, due to the complexity in their spectra leading to inconclusive exocomet classification.

The top plot of Fig.~\ref{fig:dis/HR}, shows the relative number of Tiered candidates over the total number of stars in the sample. The sensitivity to exocomets depends on the distance of the star and its brightness, with stellar activity restricting the potential number of exocomet detections. From these different criteria, this work is less sensitive to late-type stars which are less bright and more prone to stellar activity, and therefore we are more sensitive to earlier-type stars which are brighter and lack stellar activity signatures.

There are no clear trends for Tier 2 candidates based on these plots, however our two Tier 1 candidates, being A-type stars, agree with the sensitivity statement. This corresponds to an exocomet detection rate of 0.4\% for A-type stars in our sample. Also based on the distribution of points above the Zero Age Main Sequence (ZAMS), there is a large variety of ages for these A-type stars, whereas both Tier 1 candidates are young $\approx 20$~Myr old stars which sit near the ZAMS. From the results observed in this work, there is a strong suggestion that it is more likely to detect exocomet transits around young A-type stars when looking for Ca\,{\sc ii} absorption variability.

\subsection{HARPS exocomet detection rate in Ca II}
Including both our Tier 1 and Tier 2 stars, the detection rate for exocomet transits determined in this large HARPS archive survey is calculated to be 0.1\%, i.e. 6 candidates over the total number of stars in the sample that have more than a single spectrum, i.e. $\sim 6100$. In the case where our Tier 2 stars are not exocometary, our detection rate would only include our Tier 1 stars and drop to 0.03\%. However, this value could also be higher as exocomet transits may have been missed due to stellar activity. Although our overall detection rate might is therefore rather uncertain, our exocomet detection rate of 0.4\% for A-type stars is likely reasonable.

Comparing to the literature estimate of 0.17\% (upper limit derived in Section \ref{intro}), our Tier 1 detection rate (0.03\%) and our Tier 1 + Tier 2 rate (0.1\%) are significantly lower. This can probably be explained by the demographics of the large sample of stars used in this search, which spans a wide range of spectral types while also being dominated by late-type stars highly affected by stellar activity.
Our rate can also be expressed as a function of observing nights. The median number of observing nights per star is 7 (5th percentile: 1, 95th percentile: 67), which means that the chance of making a detection per star per observing night is 0.004\% (only considering Tier 1) and 0.014\% (Tier 1 + Tier 2).

\subsection{Comparison to literature}
As we have only used HARPS spectra, this work can only compare its detections with previous surveys use HARPS data. Unfortunately, there is not much literature on exocomet host stars that have been observed with HARPS, but instead, the majority of detections are made using other instruments such as FEROS. The closest exocomet spectroscopic survey to our work \citep{WandM+18} looked at a sample of 20 A-type stars with HARPS, providing a good example to evaluate our exocomet search algorithm.

There are 4 claimed transient absorption features attributed to infalling exocomets in \citet{WandM+18} around the stars HD~24966, HD~38056, HD~79469, and HD~225200. Fig.~\ref{fig:WandM} illustrates our attempt to reproduce the results found in their Fig.~4. It is important to note that we are using the exact same data but it has been processed and displayed differently; we superimpose the spectra to visually guide towards the stable and variable features, we do not plot residual intensity which is obtained by fitting every spectrum with a model that removes the rotationally broad main Ca\,{\sc ii}~K line and only leaves additional absorption features. The detection method is also different, as our initial sample is much larger. To ensure we are looking at the same absorption features, the same heliocentric velocity range is used. Note that the spectra of Fig.~\ref{fig:WandM} are focused on a small region at the bottom of the much broader Ca\,{\sc ii}~K absorption line. For viewing convenience, black arrows and dashed lines were added to mark the position at which the exocomet detections have been made in \citet{WandM+18}.

The general result of this comparison is that we do not detect significant transient absorption features associated with exocomets in any of the 4 stars. Note that our detection method relies significantly on computing a strong exocomet-free reference spectrum, which means that for stars with a reduced number of spectra (as seen for some of these stars), it might be challenging to achieve this. However, in 3/4 stars (i.e. HD~38056, HD~79469, HD~225200), we might be able to identify hints about what caused a detection in \citet{WandM+18}. Multiple narrow dips are visible in the superimposed spectra at the position of the black arrows, which do not get picked up by our SNR. This suggests that some of the variations might be caused by an increased noise level. Note that for the fourth star HD~24966, we did not detect any obvious variation at the position of the black arrows that would match the claimed detection. HD~79469 is a well-known spectroscopic binary with a main sequence B9.5 star and white dwarf companion \citep{Burleigh+Barstow-99, holberg+13}, which may explain the observed variation over a wide velocity range.

\section{Conclusions}
\label{conclusions}
In this work, we have presented a search for exocomet transits in spectroscopic data, which are identified through transient absorption features in specific atomic species. This search method is used to detect variability in the ionised calcium (Ca\,{\sc ii}) doublet over a sample of 7500 stars found in the entire HARPS spectrograph archive.

After filtering out 133 false positive exocomet detections from the 155 candidates initially generated by the search in the Ca\,{\sc ii} K line, we perform the search again, lowering the outlier threshold, looking at the H line for any counterpart variability to K detections. No new candidates were found by looking at the Ca\,{\sc ii} H line. The 22 stars not considered false positive detections are classified into Tier 1, Tier 2, and Tier 3 exocomet candidates based on the confidence in assigning a detected transient absorption feature to an exocomet transit. There are no real trends in the distribution of spectral types for these 22 Tiered candidates, however, the majority are Main Sequence stars. When a star moves off the Main Sequence, there seems to be an increase in variability associated with the star itself interfering with exocomet detection.

The classification process resulted in 6 candidates, four Tier 2 (Gl~1, HIP~5158, HD~94771, HR~1996) and two Tier 1 ($\beta$~Pic, HD~172555) stars, representing 0.1\% of the entire HARPS archive. More care is needed when attributing a firm exocomet host star classification to the Tier 2 stars. Gl~1, HIP~5158, and HD~94771 show a similar single detection with a significant narrow absorption feature only found in K, which requires more continuous data to be confidently attributed to an exocomet transit. On the other hand, HR~1996, shows multiple detections, in both H and K, but the fact that it is an O-type star inserts more complexity and forces this Tier 2 classification. None of the Tier 2 candidates show strong evidence of an infrared excess.

Both Tier 1 candidates are known exocomet host stars with recorded debris disc observations. Only considering Tier 1 stars, our detection rate across the entire HARPS archive would be 0.03\%. More generally, there is a trend in the Tier 1 candidates, where both are young A-type stars and members of the Beta Pictoris moving group. Out of the diverse sample of A-type stars in the HARPS archive, these two Tier 1 candidates represent 0.4\% of all A-type stars. This suggests that it is more likely to observe these transient absorption features associated with exocomet transits around young A-type stars, when looking at the Ca\,{\sc ii}~doublet.

An additional comparison with the literature was attempted, as most of the known exocomet host stars have not been observed using the HARPS instrument. The \citet{WandM+18} survey of 20 A-type stars using HARPS was used to evaluate our results. None of the 4 claimed exocomet host stars have been recovered using our search algorithm. This may suggest that either the initial detected variability was influenced by some noisy spectra, or that our algorithm is too conservative and instead is only sensitive to the strongest exocomet detections. As a general note, ideal conditions for exocomet transit detections would be to have a large number of observations taken on consecutive nights with a time between observations of $\Delta t \approx$ 30 min. The multiple observations would allow for an accurate and precise reference spectrum, while the $\Delta t \approx$ 30 min would ensure that no potential exocomet transit would be missed.

Currently, the sample of stars already observed in optical spectroscopy is large and diverse enough to perform exocomet surveys such as the one presented in this work. However, new interesting routes might now include comparing photometric and spectroscopic data (e.g. HD~172555, \citet{kiefer+23}; 5~Vul, \citet{rebollido+23}), which could eventually lead to the first simultaneous detections in spectroscopy and photometry.

\section*{Acknowledgements}

RBW is supported by a Royal Society grant (RF-ERE-221025). GMK is supported by the Royal Society as a Royal Society University Research Fellow. DJAB and PAS are supported by the UK Space Agency. Computing facilities were provided by the Scientific Computing Research Technology Platform of the University of Warwick. We acknowledge valuable discussion with Amelia Bayo and Don Pollacco during the preparation of this work. The code used in this research is publicly available on GitHub at \url{https://github.com/raphhbw/HARPS-exocomet}.

\section*{Data Availability}
 
Data is publicly available in the HARPS archive: \url{http://archive.eso.org/wdb/wdb/adp/phase3_main/form}.



\bibliographystyle{mnras}
\bibliography{references}



\appendix
\section{Spectrum normalisation}
\begin{figure}
      \centering
      \includegraphics[width=0.49\linewidth]{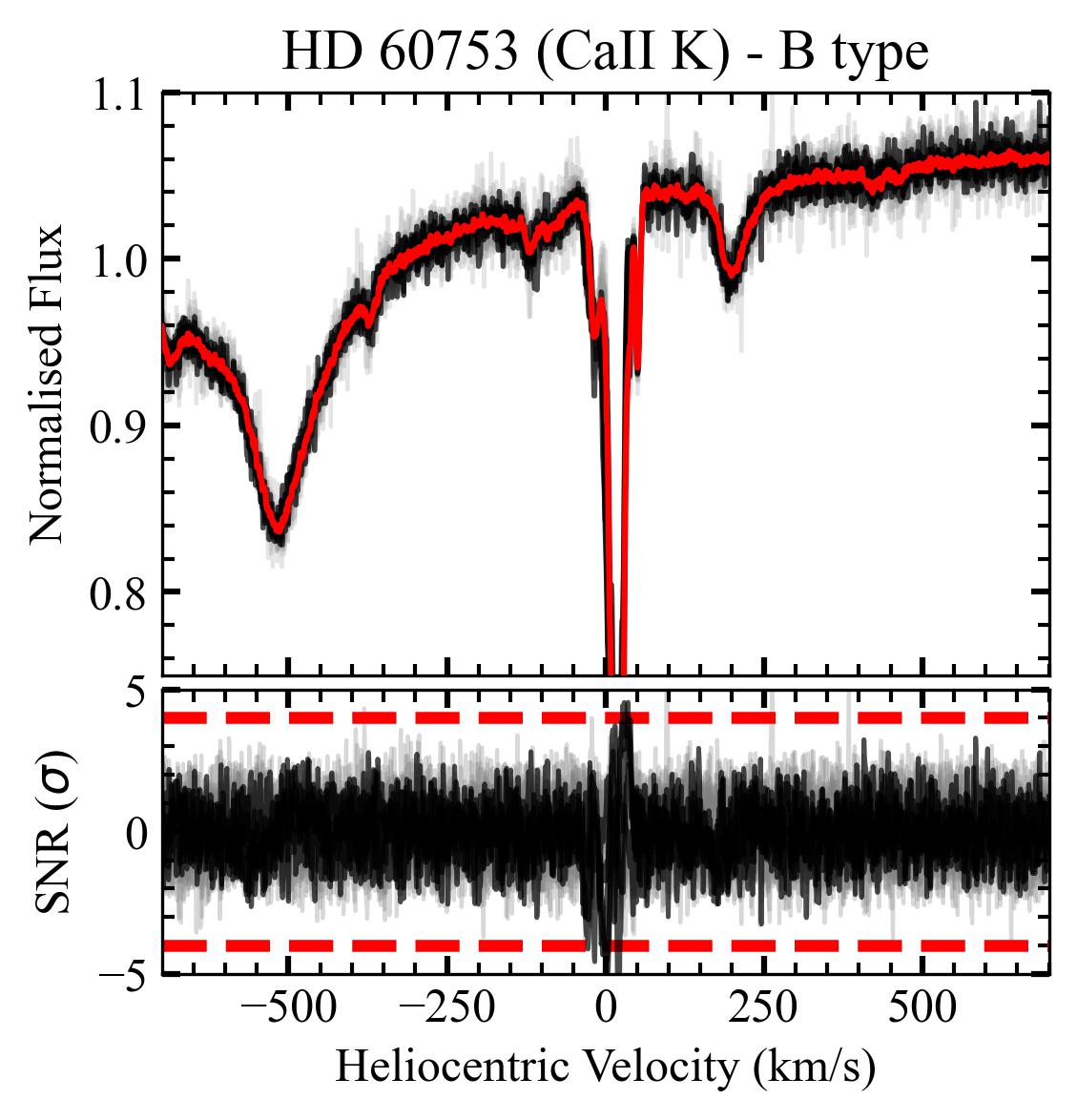}
      \includegraphics[width=0.49\linewidth]{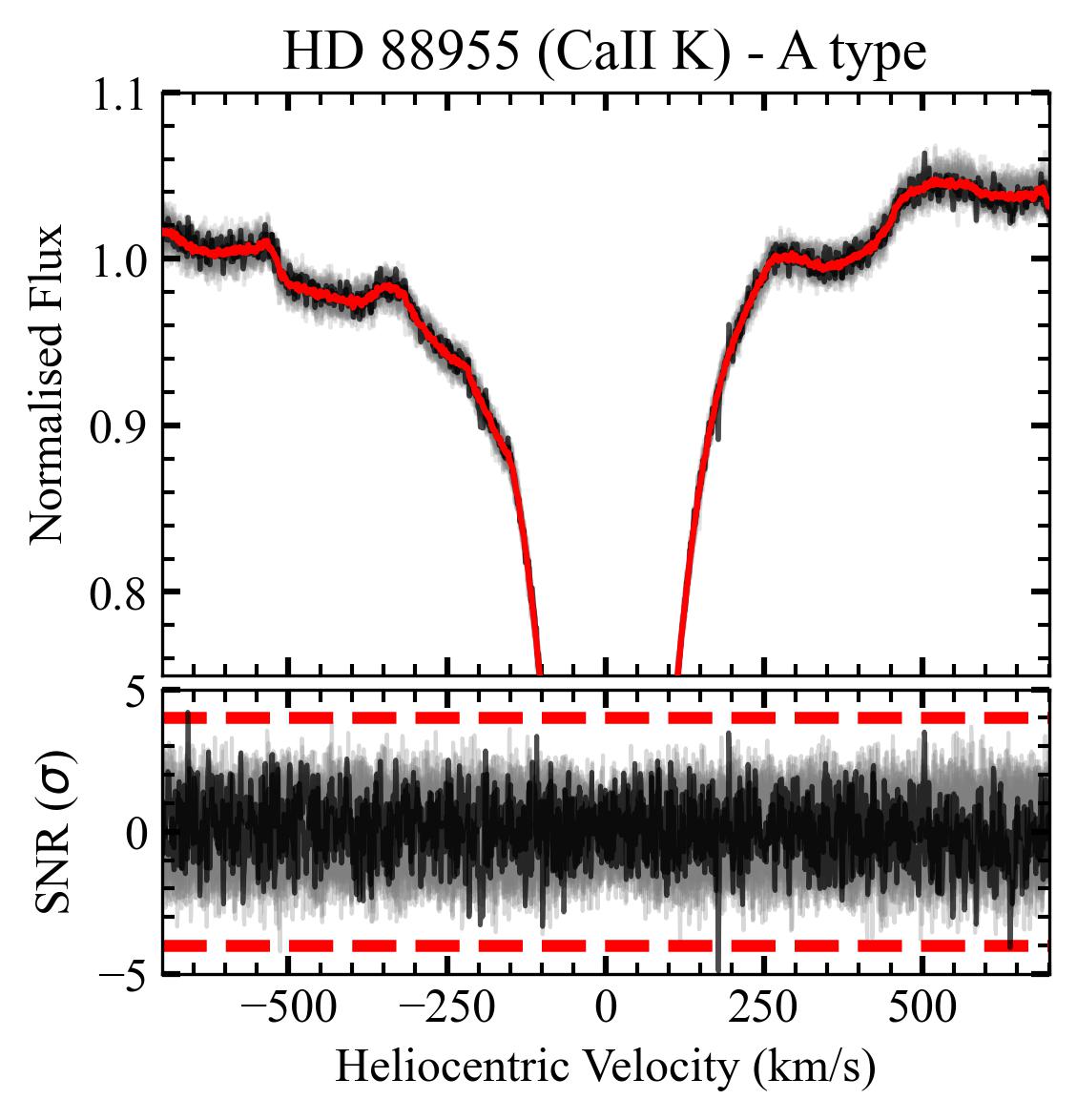} \hfill

      \includegraphics[width=0.49\linewidth]{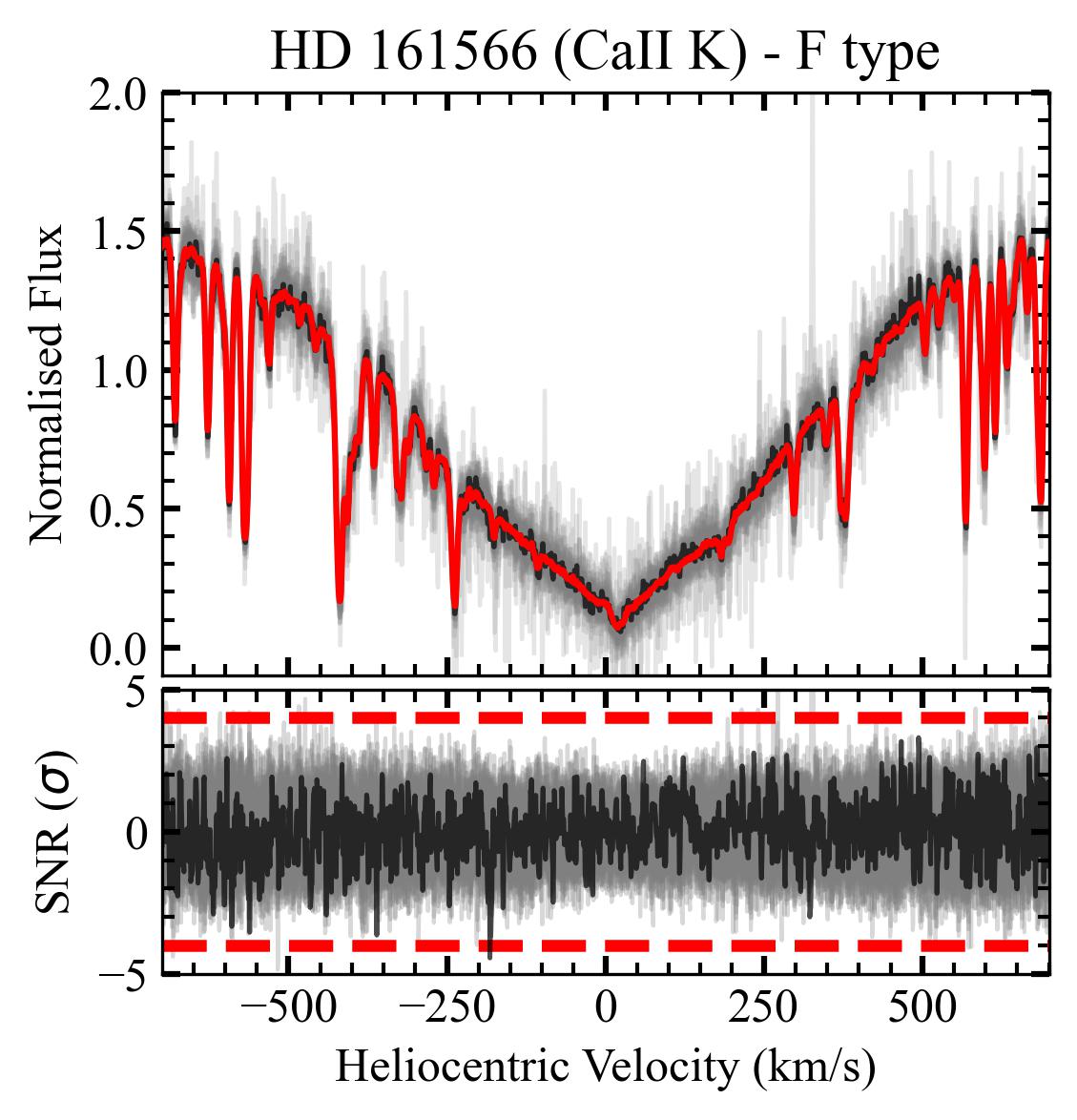}
      \includegraphics[width=0.49\linewidth]{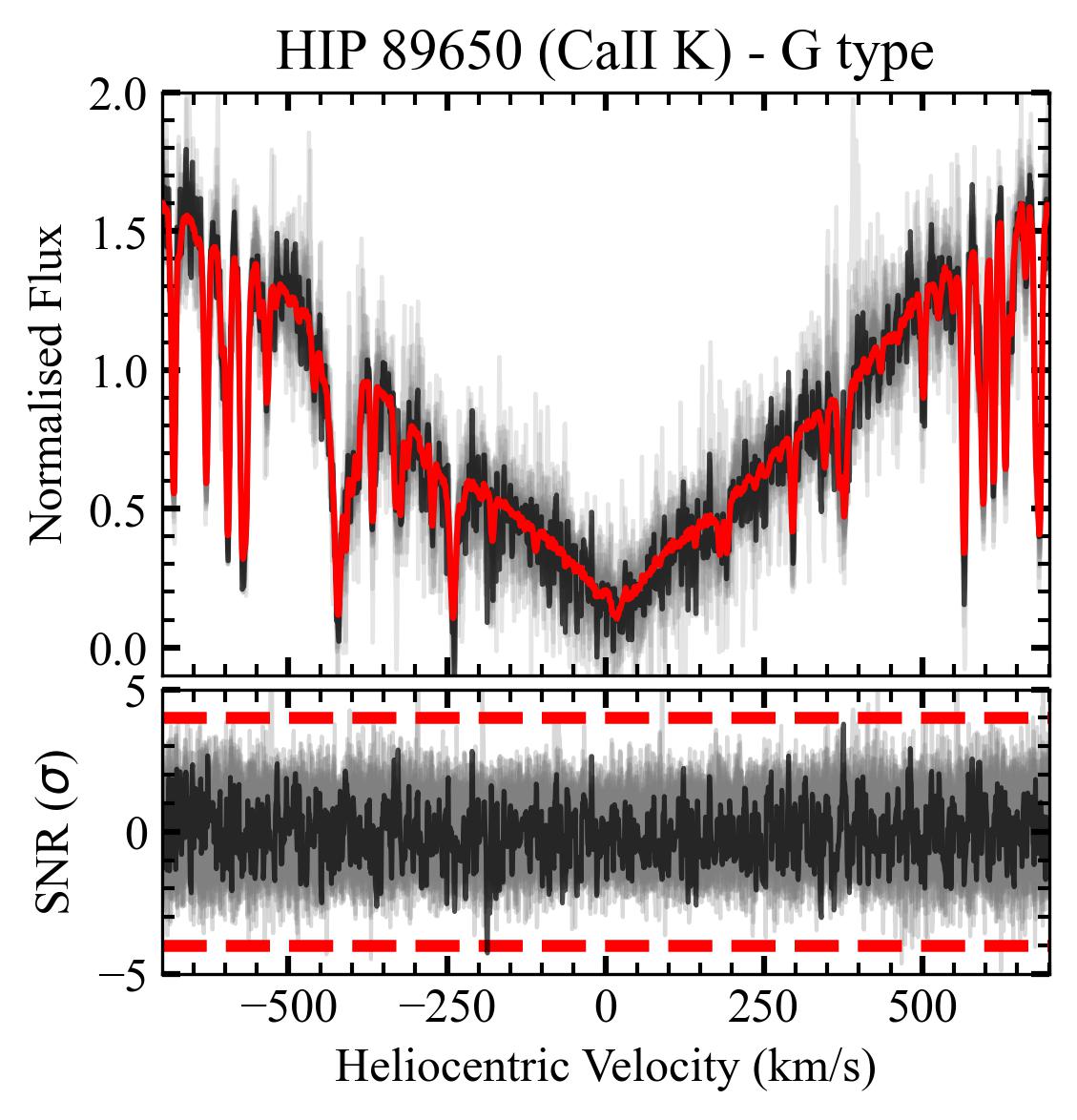} \hfill

      \includegraphics[width=0.49\linewidth]{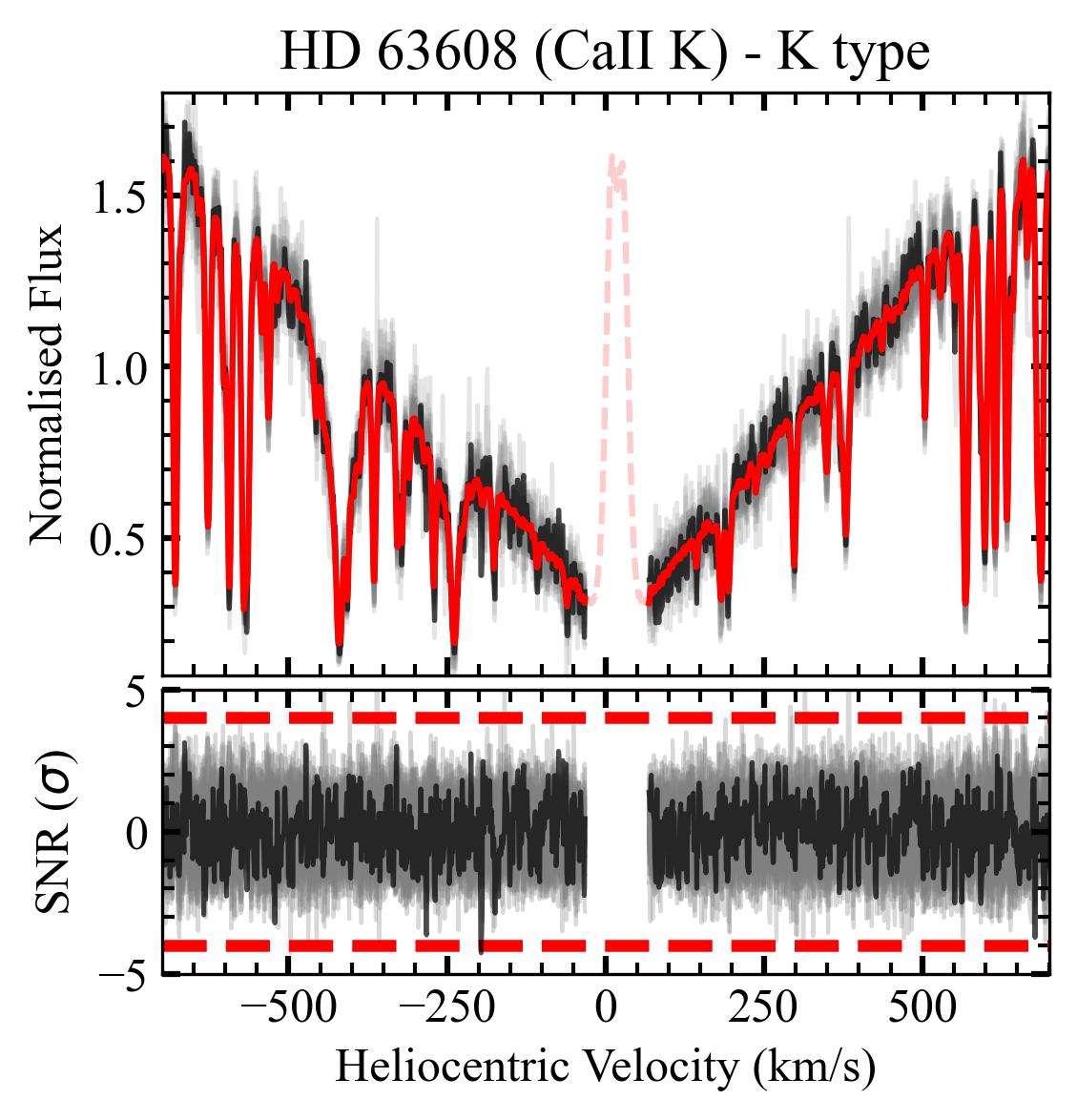}
      \includegraphics[width=0.49\linewidth]{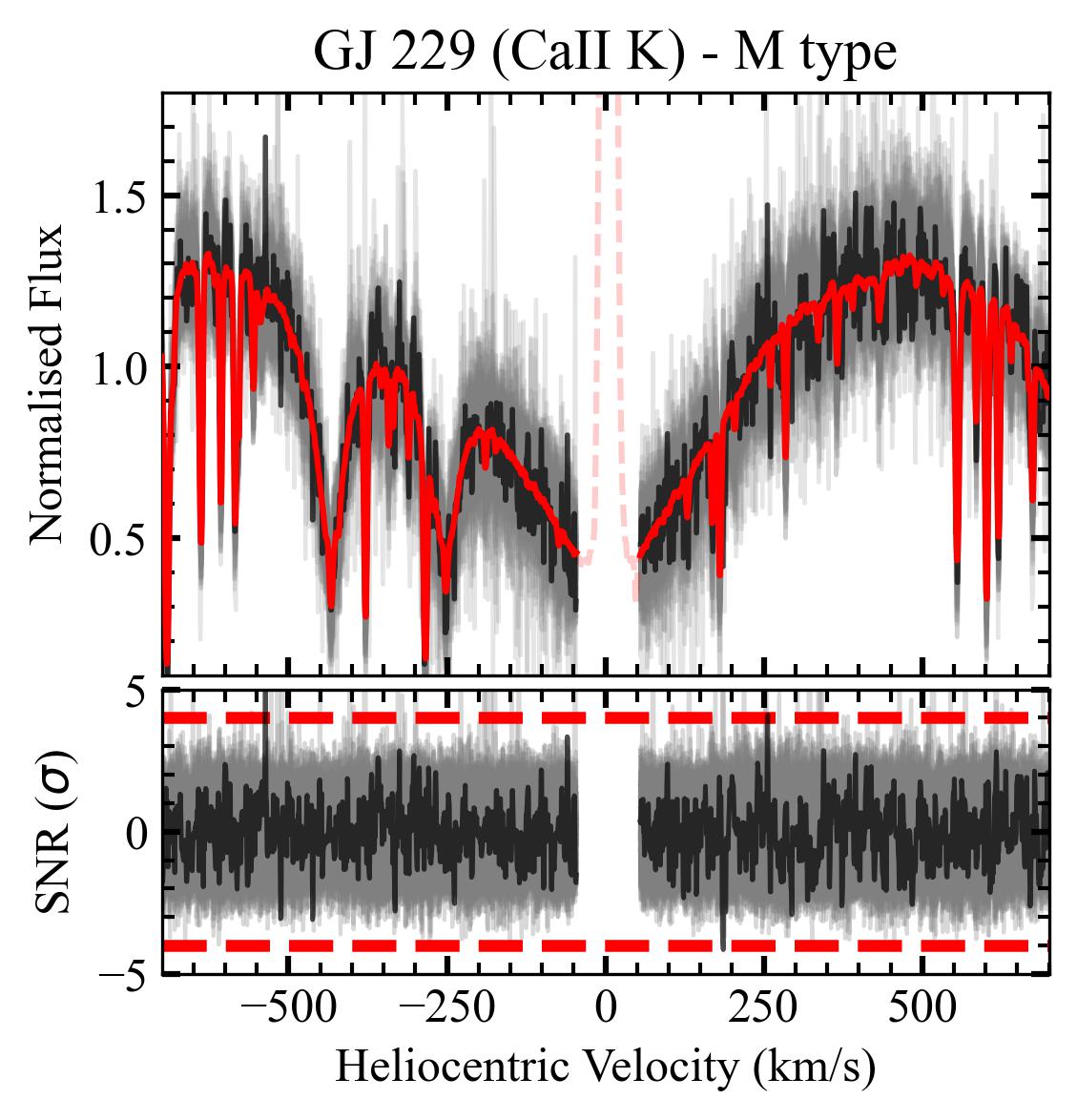} \hfill
    \caption{Normalisation as a function of spectral type. Each panel shows representative stars for each spectral types (B, A, F, G, K, M). Normalisation appears to be good for all spectral types, and there is no variation in lines that are within the normalisation region. For K and M stars, the active region near zero velocity has been masked (see Section~\ref{add-filters}).}
    \label{fig:normalisation} 
  \end{figure}

In Fig.~\ref{fig:normalisation}, we present the result of our spectrum normalisation process for the entire range of spectral types of our sample.

\section{Tier 2 Ca II H line}
\begin{figure*}
    \begin{subfigure}[t]{0.33\textwidth}
      \centering
      \includegraphics[width=\linewidth]{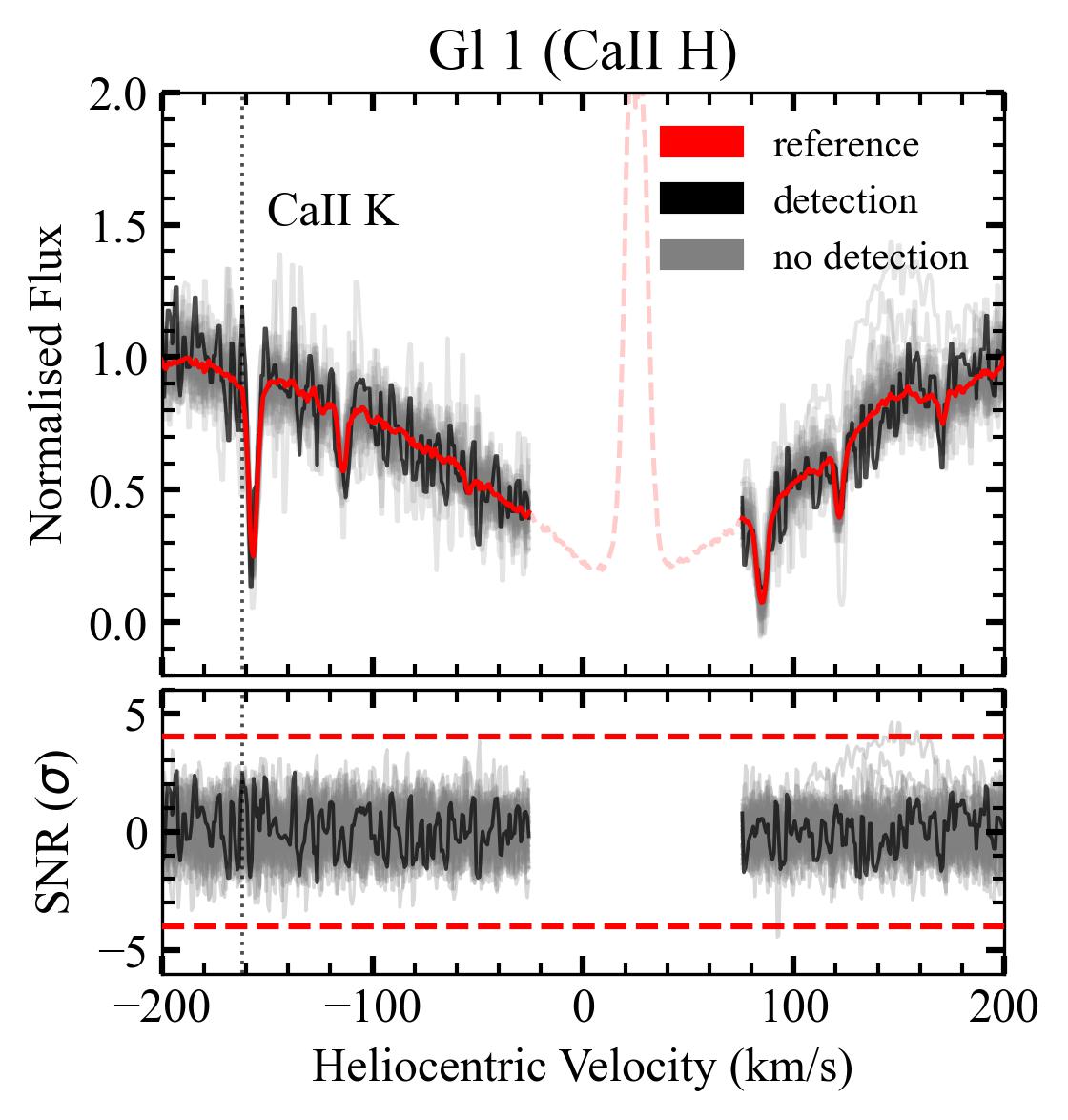} 
    \end{subfigure} \hfill
    \begin{subfigure}[t]{0.33\textwidth}
      \centering
      \includegraphics[width=\linewidth]{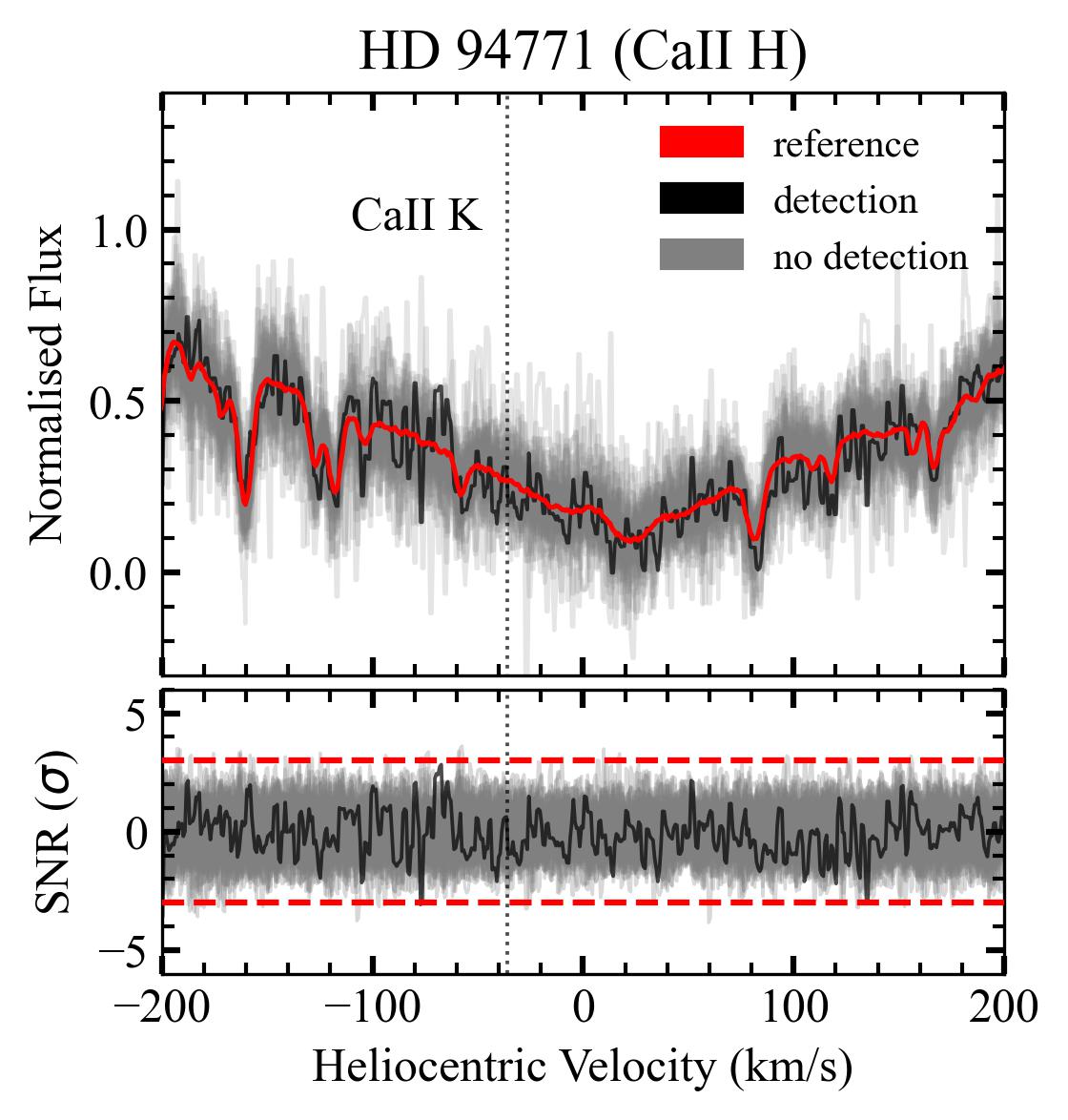} 
    \end{subfigure} \hfill
    \begin{subfigure}[t]{0.33\textwidth}
      \centering
      \includegraphics[width=\linewidth]{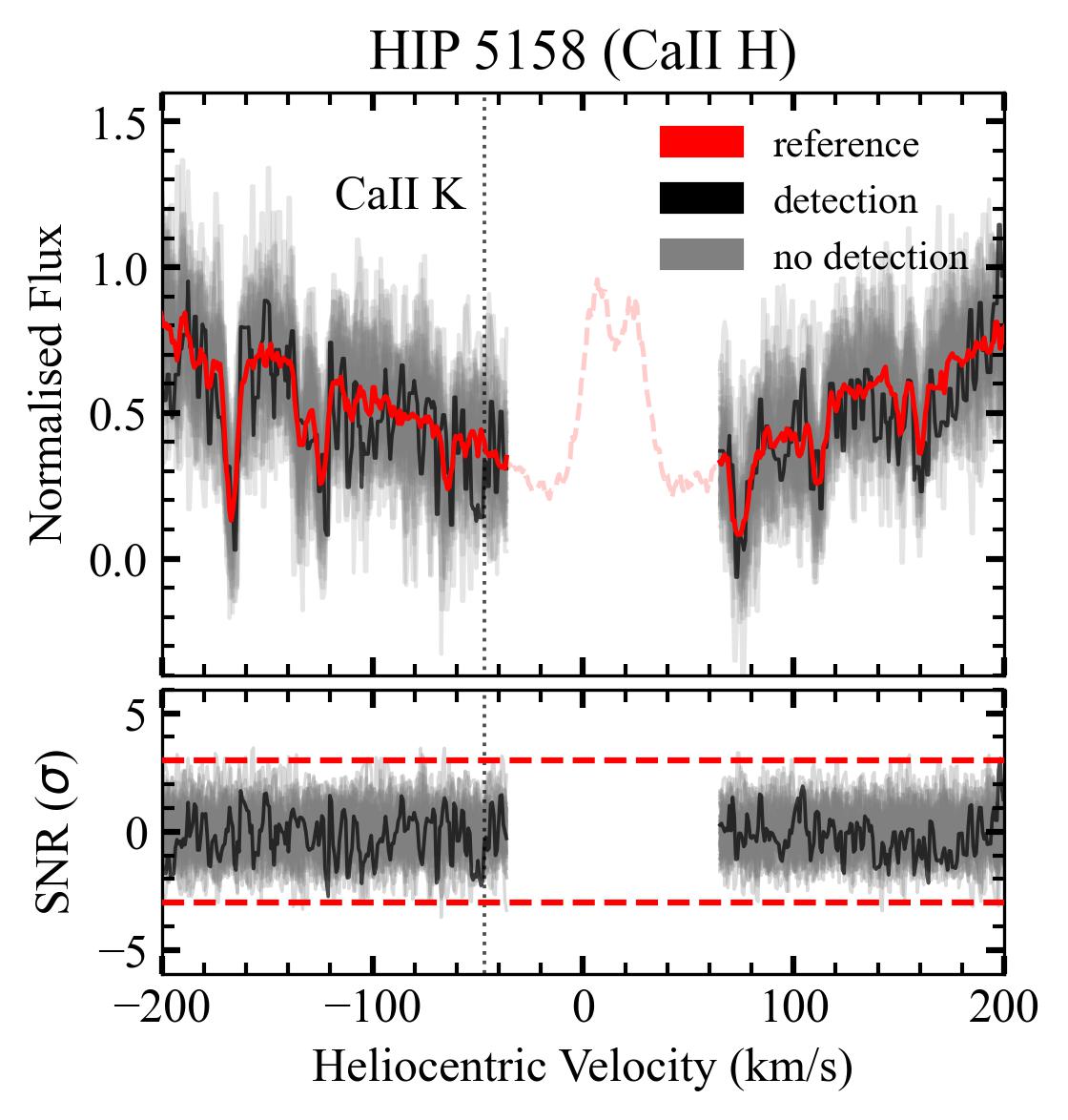} 
    \end{subfigure} \hfill
    \caption{Tier 2 candidates with no strong H counterparts to the detections in Ca\,{\sc ii} K. The format of the plots is the same as in Fig.~\ref{fig:T1-H}. The black spectra represent those with detections in the K line. The dotted vertical black lines denote the radial velocity of each detection in the Ca\,{\sc ii} K line.}
    \label{fig:tier2-H} 
\end{figure*}

In Fig.~\ref{fig:tier2-H}, we show the Ca\,{\sc ii} H line plots for the Tier 2 candidates with no strong detections in H: Gl~1, HD~94771 and HIP~5158. The spectrum shown in black in each panel is the one with a detected absorption feature in the Ca\,{\sc ii}~K line. In all three cases, the H line signal is too faint to be distinguishable from the noise level of the spectrum that is on average around -2$\sigma$ in SNR. Additionally, for Gl~1, the location of the detection in K seems to be adjacent to a stable absorption feature in the H line, which can make it even harder to distinguish.

This plot completes Fig.~\ref{fig:T1-H} which shows the two Tier 1 candidates (Beta Pictoris, HD~172555) as well as the only Tier 2 candidate with strong Ca\,{\sc ii} H detections (HR~1996).

\section{Vetting process of Tiered candidates}
\label{appendix-vetting}

This appendix presents the vetting process of all 22 Tiered candidates, which is also summarised in \autoref{tab:final-vetting}. Plots for each of the 22 candidates are shown, with the caption explaining the thought process linked to the Tiered classification. ISM cloud absorption is checked using the LISM Kinematic Calculator\footnote{\url{http://lism.wesleyan.edu/LISMdynamics.html}}. Each plot shows

\begin{itemize}
    \item (\textit{header}) names of the star used for the grouping process when building the HARPS dataset.
    
    \item (\textit{top}) Table showing interesting SIMBAD information: star name (MAIN ID), list of object type (OTYPES), star's radial velocity (RV VALUE), SIMBAD's preferred object type (OTYPE), parallax value (PLX VALUE), and spectral type (SP TYPE).
    
    \item (\textit{top left}) Superimposed spectra of the star with the reference spectrum in red, the spectra with detections in black, and the spectra with no detections in grey.
    
    \item (\textit{top right}) SNR distribution with the dashed red lines indicating the $\pm 4\sigma$ threshold, the SNR measurement with a detection in black, and with no detection in grey.

    \item (\textit{bottom left}) The radial velocity distribution of the most significant transient absorption feature per spectrum, where detections are in blue and non-detections in grey.

    \item (\textit{bottom right}) The position of the star (blue square) on the HR diagram of the entire sample of stars used in the search (grey dots).

    \item (\textit{bottom}) Timeline of all the observations, with a single dot being the time at which a spectrum is taken. Red dots are for spectra with detections and grey dots are for spectra with no detections.
    
\end{itemize}

\begin{table}
    \caption[Classification of all 32 Tiered candidates]{Manual classification of all 32 Tiered exocomet candidates.}
    \label{tab:final-vetting}
    \begin{adjustbox}{max width=0.98\linewidth}
    \begin{tabular}{l c c c c l}
        \hline
        \hline
        Star & SpT. & \# of det. & \# of spec. & Date (UT) & Tier \\
        \hline
        \hyperref[fig:app/vetting-bpic]{Beta Pictoris } & A6V & $>3000$ & $9100$ & many & 1\\
        \hyperref[fig:app/vetting-chioph]{Chi Oph } & B2Vne & $5$ & $41$ & 28/30-05-2015 & 3 (TC)\\
        \hyperref[fig:app/vetting-gl1]{Gl 1 } & M2V & $1$ & $44$ & 28-12-2008 01:05:41 & 2\\
        \hyperref[fig:app/vetting-gl229]{Gl 229 } & M1V & $1$ & $190$ & 14-03-2014 01:40:52 & 3 (other)\\
        \hyperref[fig:app/vetting-hd60753]{HD 60753 } & B3IV & $5$ & $25$ & various & 3 (TC)\\
        \hyperref[fig:app/vetting-hd63563]{HD 63563 } & B8/9II & $1$ & $27$ & 09-12-2005 03:57:07 & 3 (TC)\\
        \hyperref[fig:app/vetting-hd69830]{HD 69830 }& G8V& 1& 754& 06-11-2004 06:49:19&3 (other)\\
        \hyperref[fig:app/vetting-hd88955]{HD 88955 } & A2Va & $1$ & $40$ & 12-03-2006 04:14:21 & 3 (other)\\
        \hyperref[fig:app/vetting-hd91024]{HD 91024 } & B8Iab & $1$ & $26$ & 24-06-2012 22:53:29 & 3 (TC)\\
        \hyperref[fig:app/vetting-hd94771]{HD 94771 } & G3/5V & $1$ & $111$ & 12-12-2020 08:33:14 & 2\\
        \hyperref[fig:app/vetting-hd109200]{HD 109200 }& K1V& 9& 860& various&3 (TC)\\
        \hyperref[fig:app/vetting-hd126341]{HD 126341 } & B2IV & $3$ & $15$ & 23-08-2010 & 3 (TC)\\
        \hyperref[fig:app/vetting-hd135240]{HD 135240 } & O8V & $1$ & $95$ & 24-06-2011 00:11:49 & 3 (other)\\
        \hyperref[fig:app/vetting-hd141765]{HD 141765 } & B9IV & $1$ & $13$ & 22-08-2010 23:19:27 & 3 (TC)\\
        \hyperref[fig:app/vetting-hd161566]{HD 161566 } & F6V & $1$ & $57$ & 11-07-2010 01:45:29 & 3 (other)\\
        \hyperref[fig:app/vetting-hd172555]{HD 172555 } & A7V & $1$ & $537$ & 21-09-2004 23:54:43 & 1\\
        \hyperref[fig:app/vetting-hd205879]{HD 205879 } & B8V & $1$ & $6$ & 18-06-2010 09:52:12 & 3 (TC)\\
        \hyperref[fig:app/vetting-hip5158]{HIP 5158 } & K5V & $1$ & $20$ & 16-09-2009 07:11:31 & 2\\
        \hyperref[fig:app/vetting-hr1996]{HR 1996 } & O9.5V & $8$ & $120$ & various & 2\\
        \hyperref[fig:app/vetting-hr4468]{HR 4468 } & B9.5V & $5$ & $187$ & 28/29/30-05-2015 & 3 (TC)\\
        \hyperref[fig:app/vetting-hr6545]{HR 6545 } & Ap Si & $1$ & $18$ & 03-04-2005 09:02:21 & 3 (TC)\\
        \hyperref[fig:app/vetting-hycom]{HY Com } & A5 & $1$ & $11$ & 20-01-2021 07:50:05 & 3 (other)\\
    \end{tabular}
    \end{adjustbox}
    
    \vspace{1mm}
    \begin{itemize}
        \item [TC] Classified Tier 3 because the variability in the spectrum is too complex.
        \item [other] Classified Tier 3 because the absorption features are merged with other absorption lines or the absorption features seem too narrow to be exocometary.
    \end{itemize}
\end{table}

\begin{figure}
\centering
    \includegraphics[width=\linewidth]{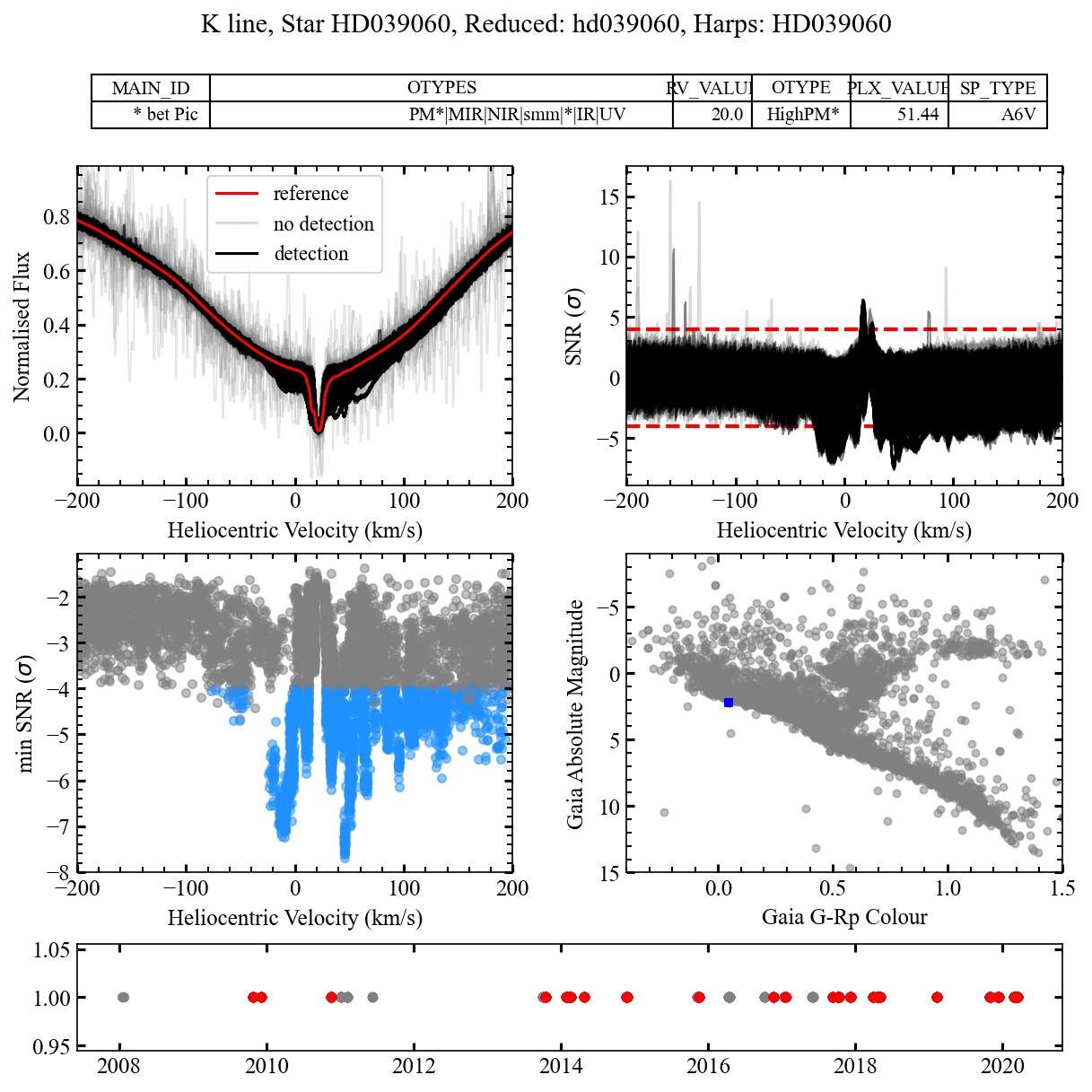}
    \caption[Beta Pictoris]{Beta Pictoris is an A6V star classified as a Tier 1 candidate, with more than 3000 detections in $\approx 9100$ spectra. Beta Pictoris is also known as the archetypal star in the exocomet field. More information in Section \ref{chap:exo sec:final-vet}.}
    \label{fig:app/vetting-bpic}
\end{figure} \hfill

\begin{figure}
\centering
    \includegraphics[width=\linewidth]{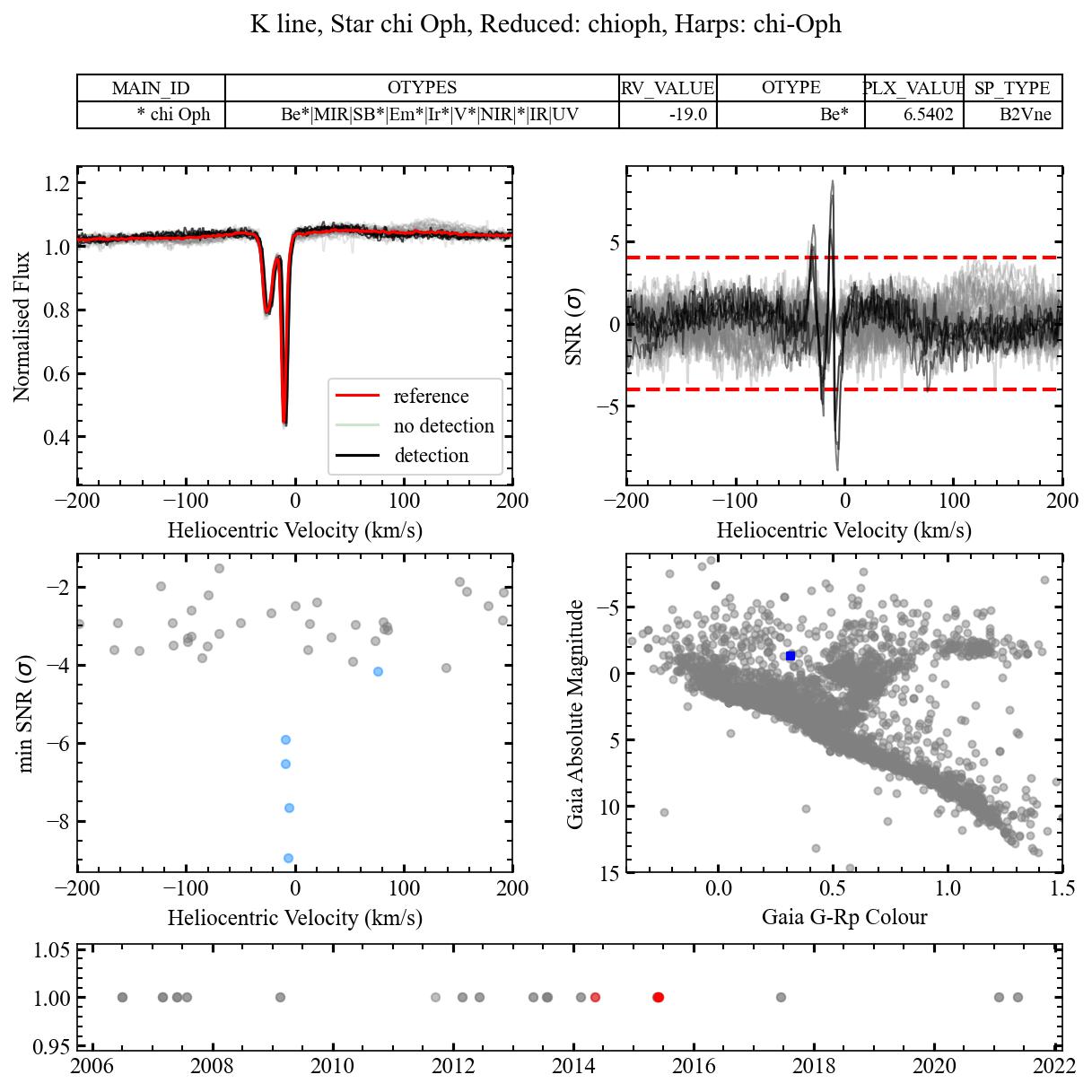}
    \caption[Chi Oph]{Chi Oph (HD 148184) is a B2Vne Be star classified as a Tier 3 candidate due to the multiple complex variability in the spectra. Be stars are fast rotators that can eject material from the star and form a decretion disc. This specificity of the star added to the apparent radial velocity shift at the position of the detections (hinted by the large peak followed by a dip in the SNR distribution) and the very variable region from 0 to 200\kms~(absorption + emission), complicates any exocomet host star classification.}
    \label{fig:app/vetting-chioph}
\end{figure} \hfill

\begin{figure}
\centering
    \includegraphics[width=\linewidth]{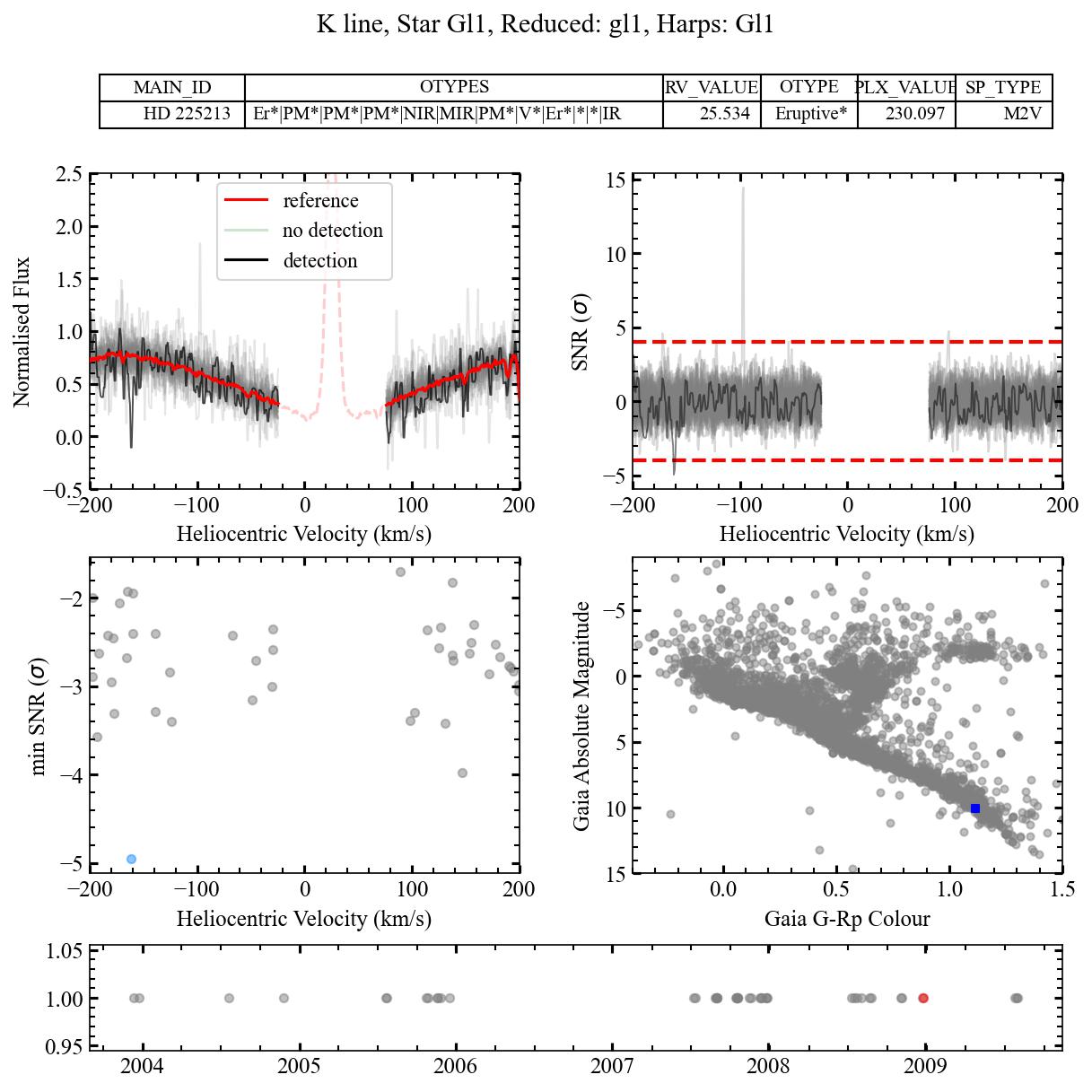}
    \caption[Gl 1]{Gl 1 (HD 225213) is a M2V star classified as a Tier 2 candidate due to the significant single detection in 44 spectra. More information in Section \ref{chap:exo sec:final-vet}.}
    \label{fig:app/vetting-gl1}
\end{figure} \hfill

\begin{figure}
\centering
    \includegraphics[width=\linewidth]{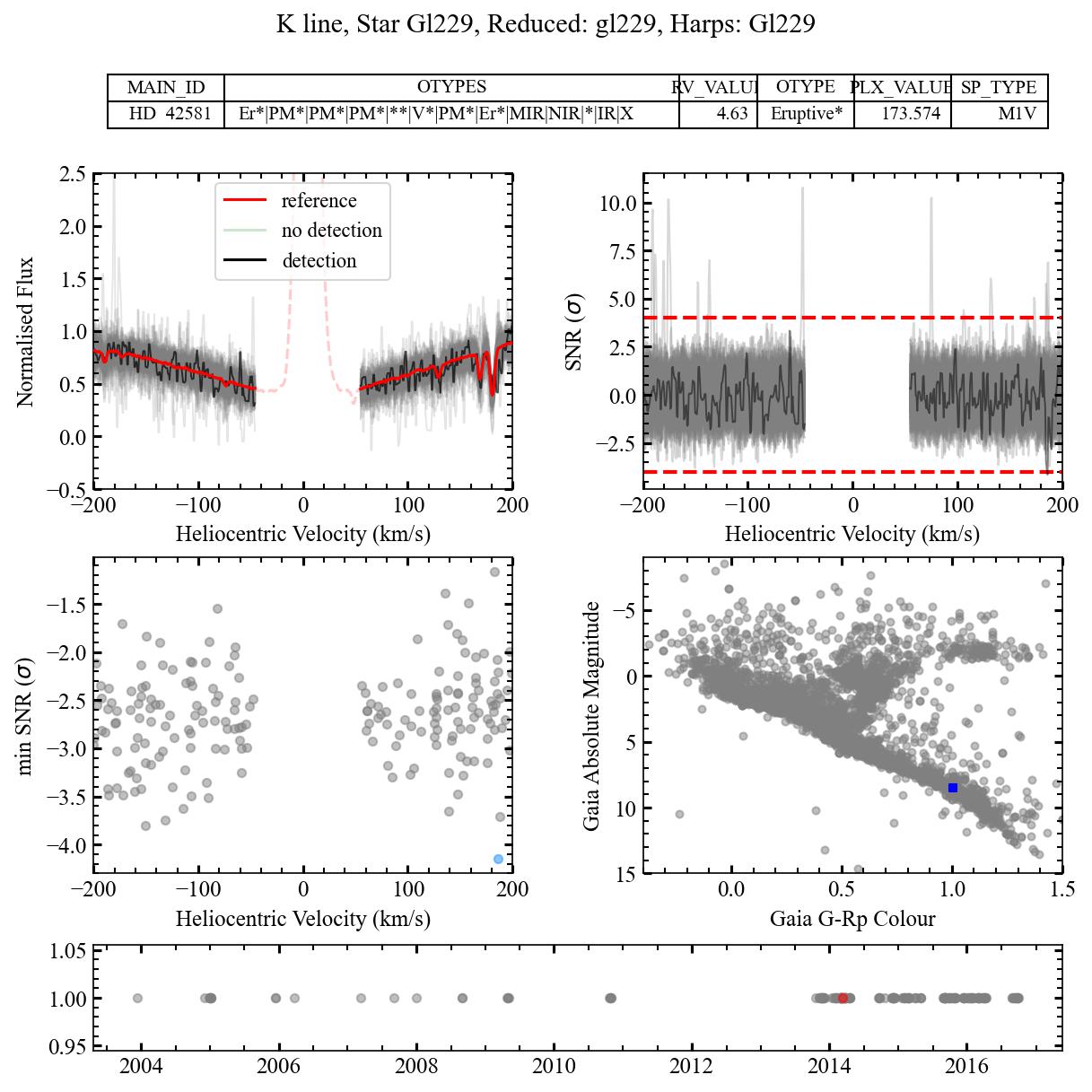}
    \caption[Gl 229]{Gl 229 (HD 42581) is a M1V star classified as a Tier 3 candidate, with 1 detection in 190 spectra. The detected absorption feature seems to be blended with some other stable absorption feature around $184$\kms, making the exocomet classification complicated.}
    \label{fig:app/vetting-gl229}
\end{figure}

\begin{figure}
\centering
    \includegraphics[width=\linewidth]{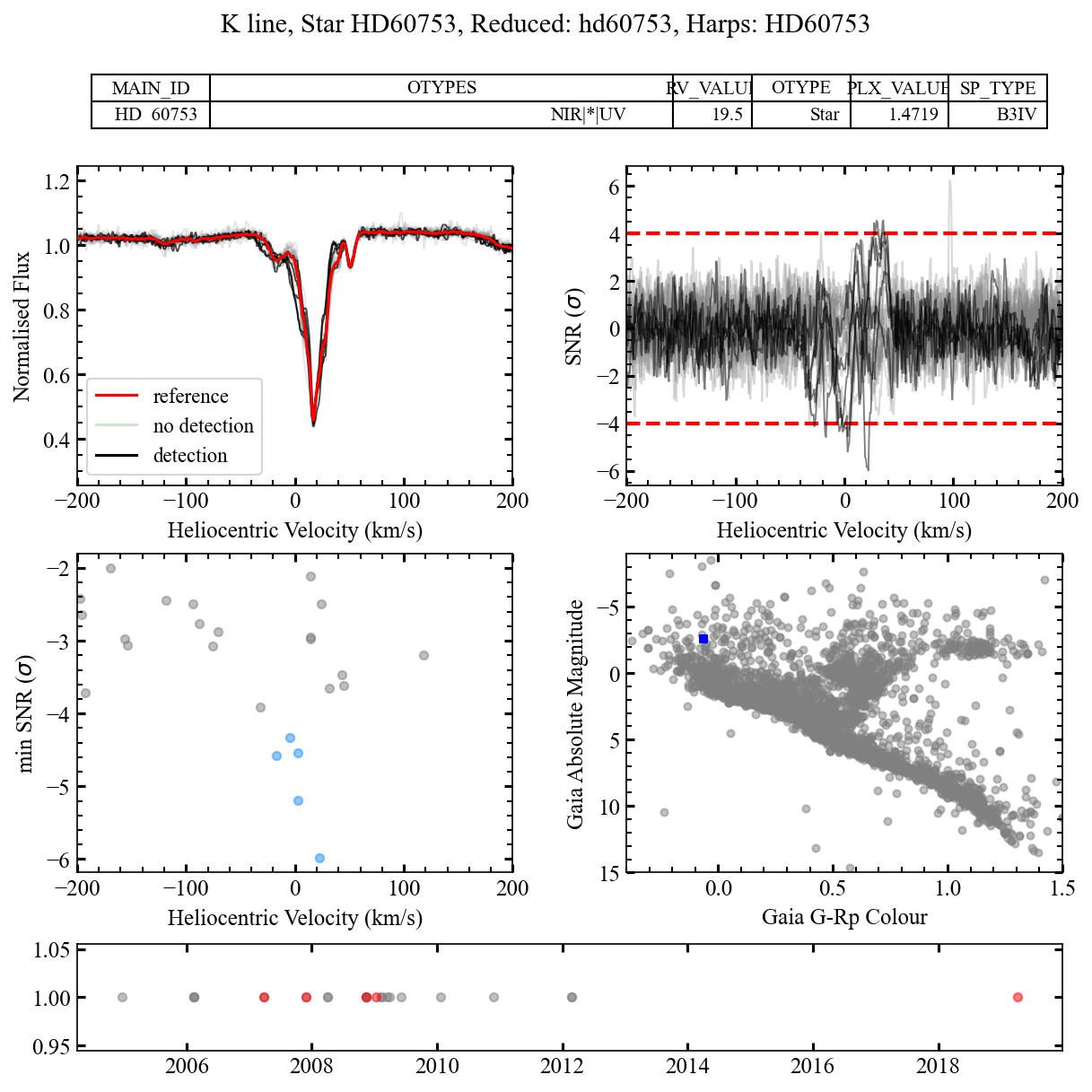}
    \caption[HD 60753]{HD 60753 is a B3IV star classified as a Tier 3 candidate, with 5 detections in 25 spectra. This Tier 3 classification can be explained by the fact that there is too much variability observed - some detections look like radial velocity shifts, some like a mix of absorption and emission, etc. }
    \label{fig:app/vetting-hd60753}
\end{figure}

\begin{figure}
\centering
    \includegraphics[width=\linewidth]{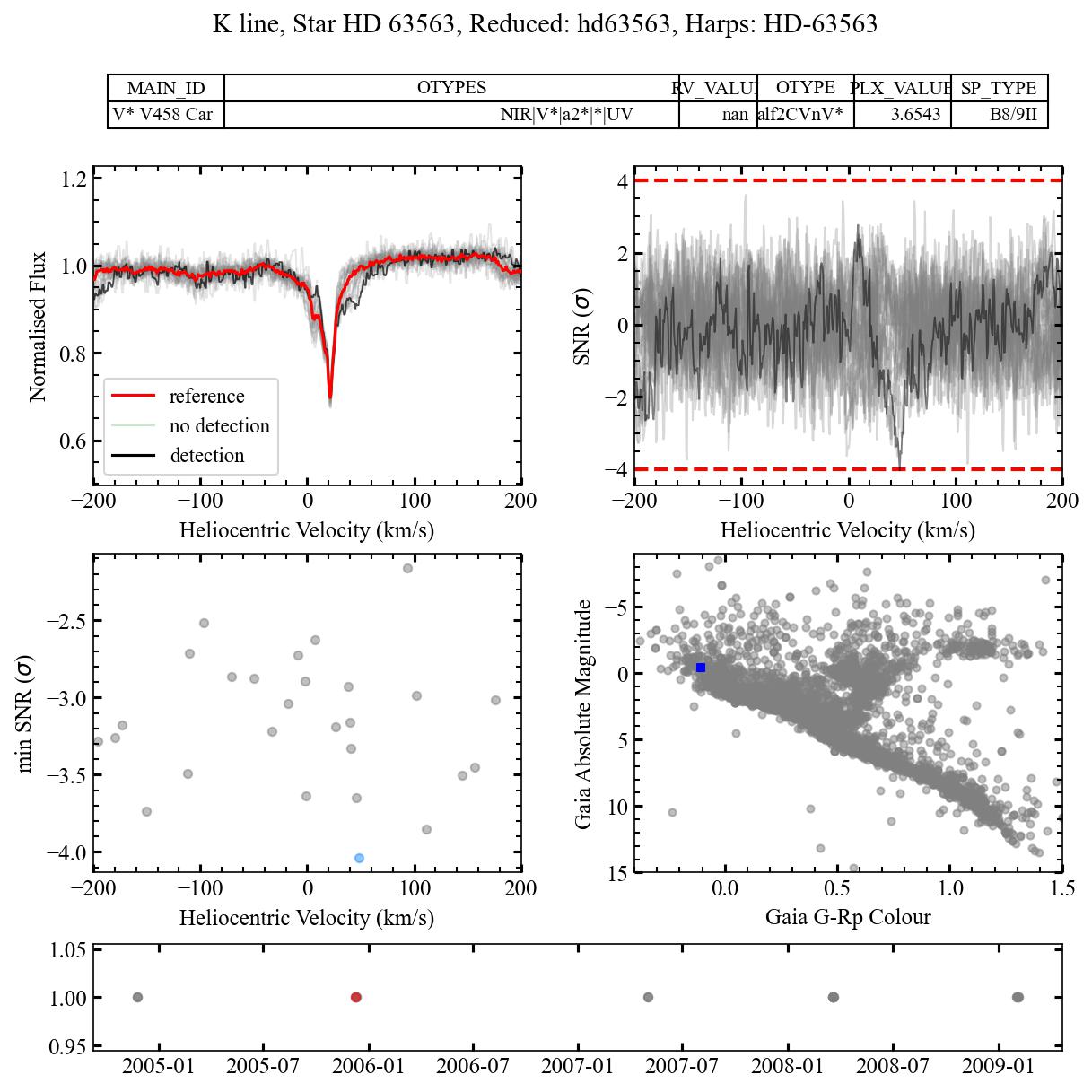}
    \caption[HD 63563]{HD 63563 is a B8/9II star classified as a Tier 3 candidate, with a single detection in 27 spectra. This is a Tier 3 candidate because the variability observed is too complex to be confident in an exocomet classification.}
    \label{fig:app/vetting-hd63563}
\end{figure}

\begin{figure}
\centering
    \includegraphics[width=\linewidth]{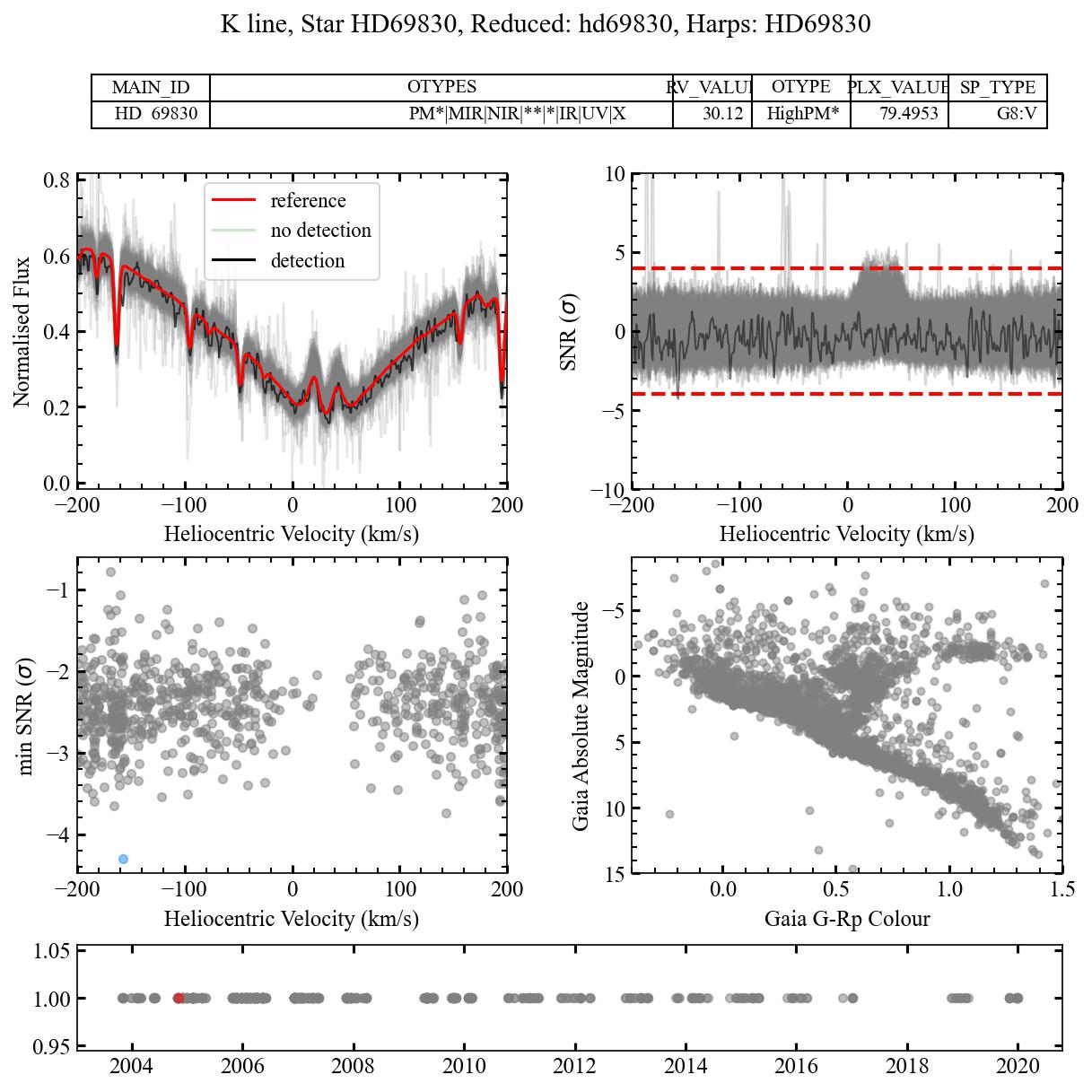}
    \caption[HD 69830]{HD 69830 is a G8V star classified as a Tier 3 candidate, with a single detection in 754 spectra. This is a Tier 3 candidate due to the difficulty in distinguishing the observed variability from another stable absorption line.}
    \label{fig:app/vetting-hd69830}
\end{figure}

\begin{figure}
\centering
    \includegraphics[width=\linewidth]{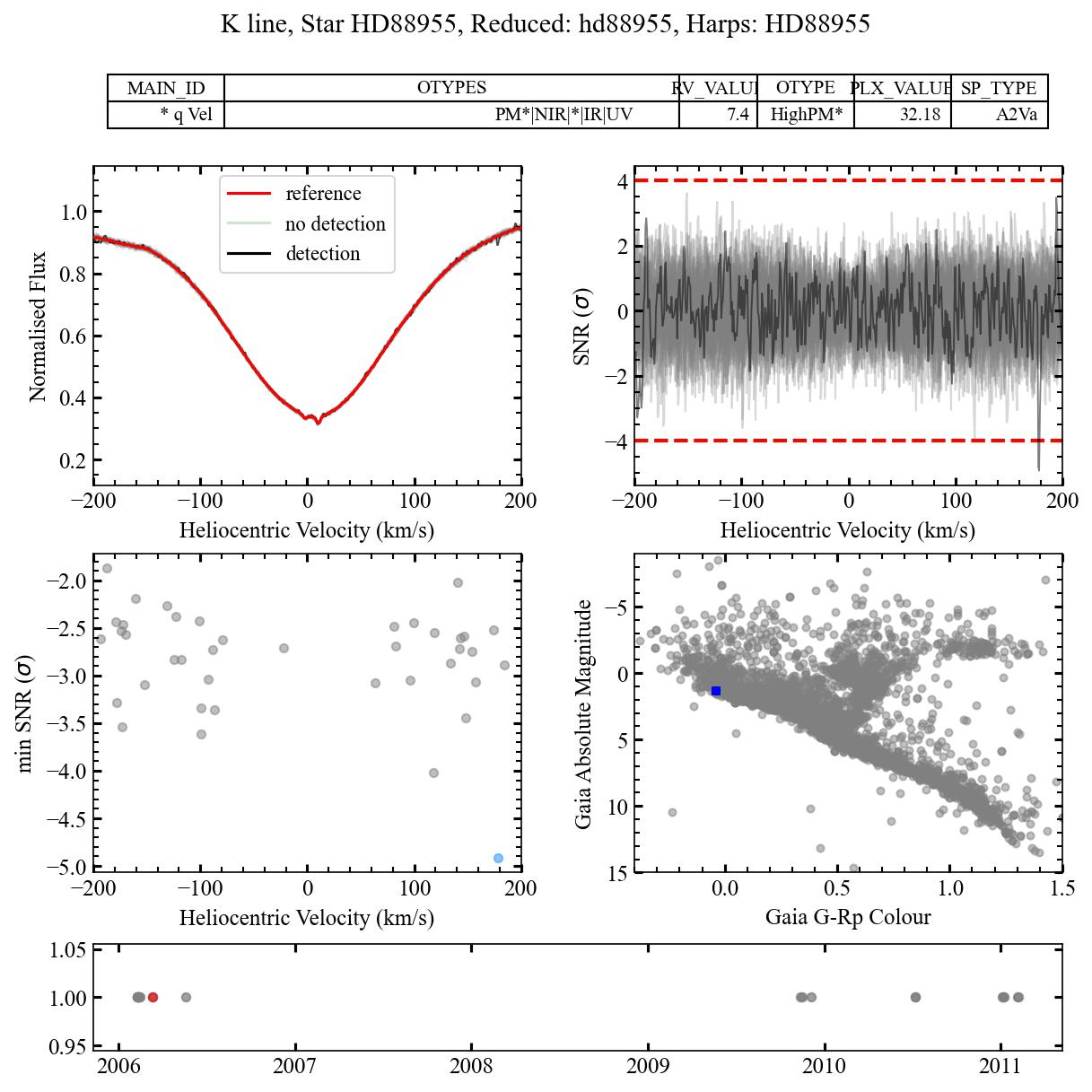}
    \caption[HD 88955]{HD 88955 is an A2Va star classified as a Tier 3 candidate, with a single detection in 40 spectra. The detected absorption feature is rather narrow for a high-velocity feature which would make this star a Tier 3 candidate. Note that there is also an emission feature just after the detected absorption ($\approx 200$\kms), which makes this detection even more suspicious.}
    \label{fig:app/vetting-hd88955}
\end{figure}

\begin{figure}
\centering
    \includegraphics[width=\linewidth]{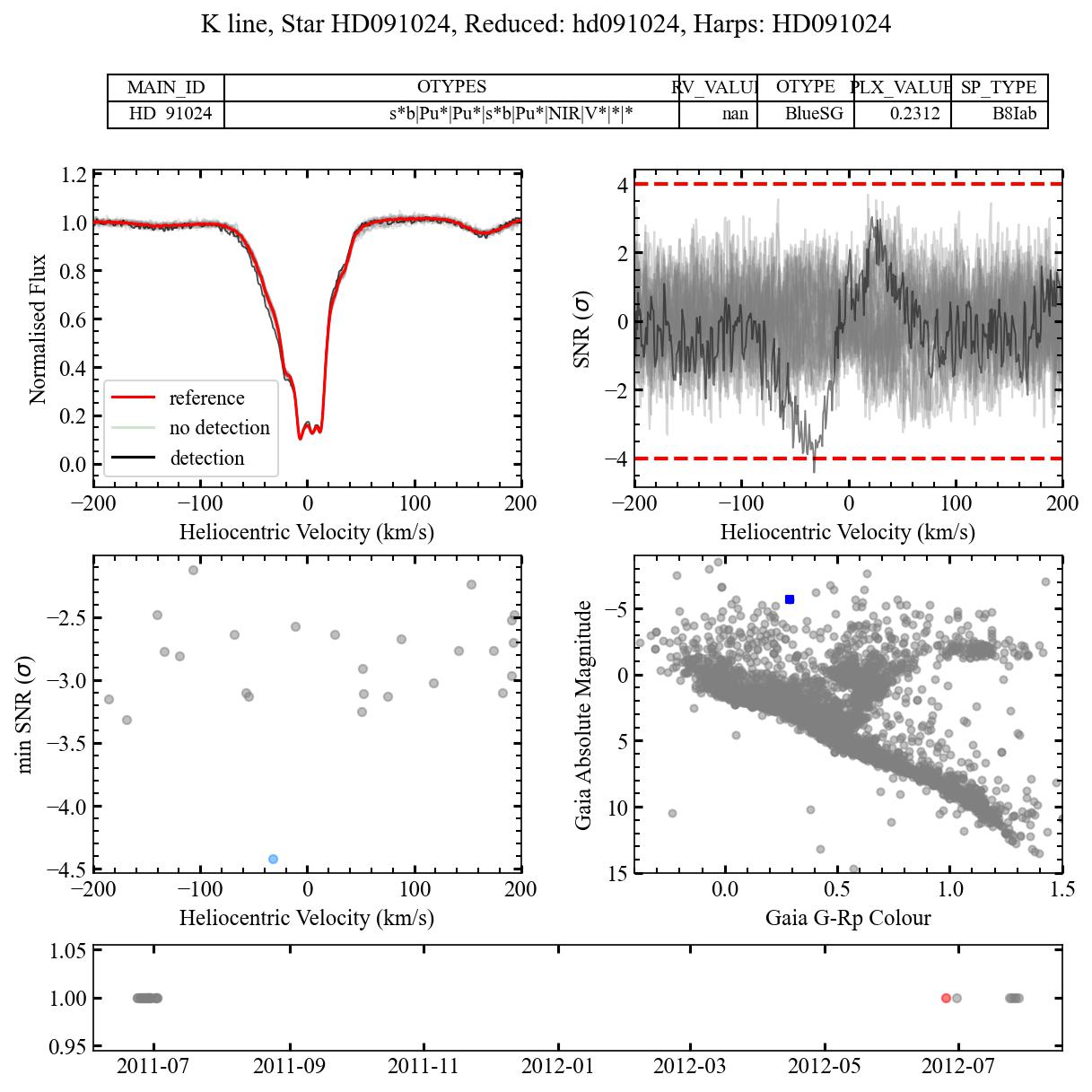}
    \caption[HD 91024]{HD 91024 is a B8Iab star classified as a Tier 3 candidate, with a single detection in 26 spectra. This star is classified as a Tier 3 candidate because of the spectra being too complicated for any exocomet classification; there seems to be a radial velocity shift but only at certain places, there is ISM cloud absorption at $-5$\kms~and $2$\kms, there is variability in the wings of the spectra as well, etc.}
    \label{fig:app/vetting-hd91024}
\end{figure}

\begin{figure}
\centering
    \includegraphics[width=\linewidth]{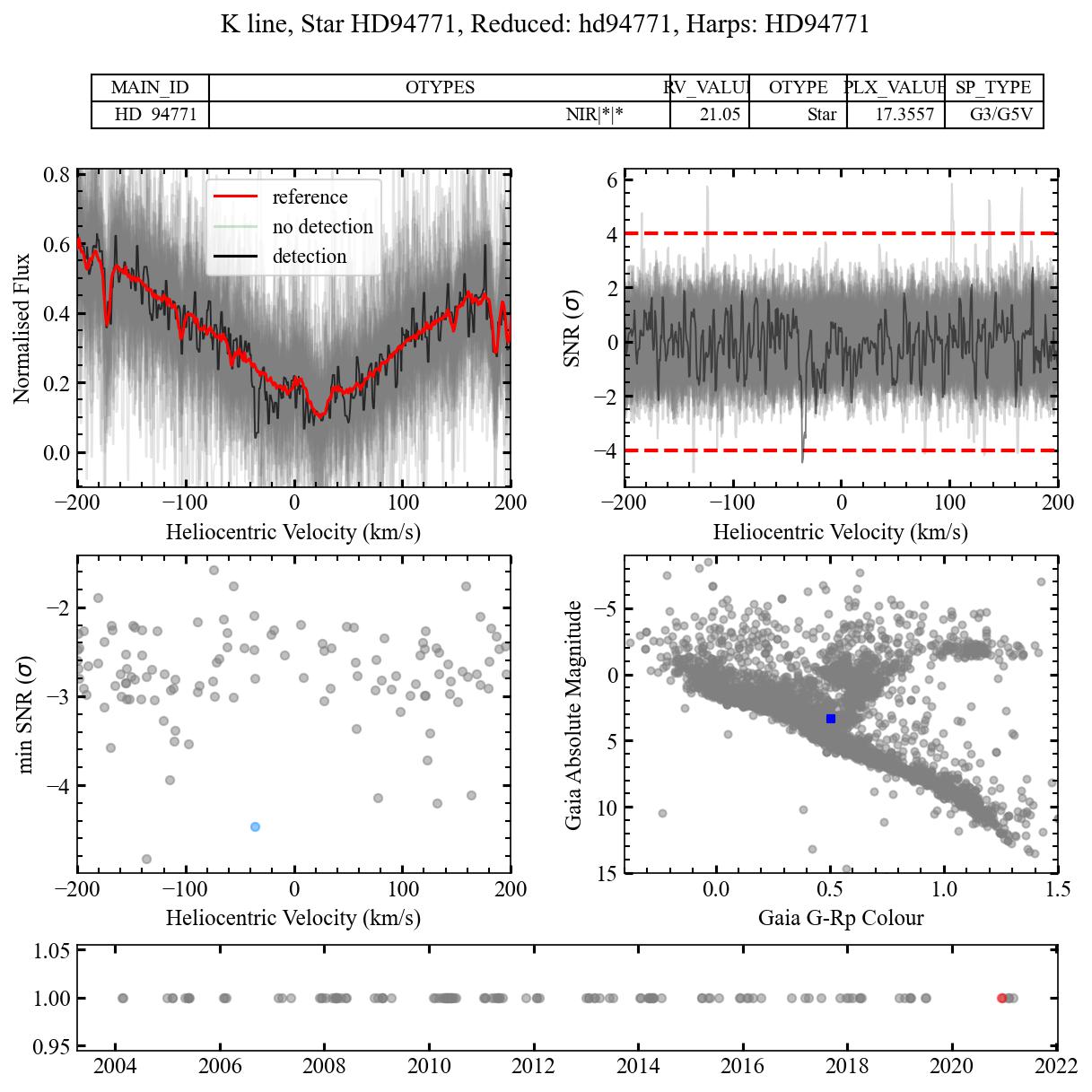}
    \caption[HD 94771]{HD 94771 is a G3/5V star classified as a Tier 2 candidate due to the significant single detection in 120 spectra. More information in Section \ref{chap:exo sec:final-vet}.}
    \label{fig:app/vetting-hd94771}
\end{figure}

\begin{figure}
\centering
    \includegraphics[width=\linewidth]{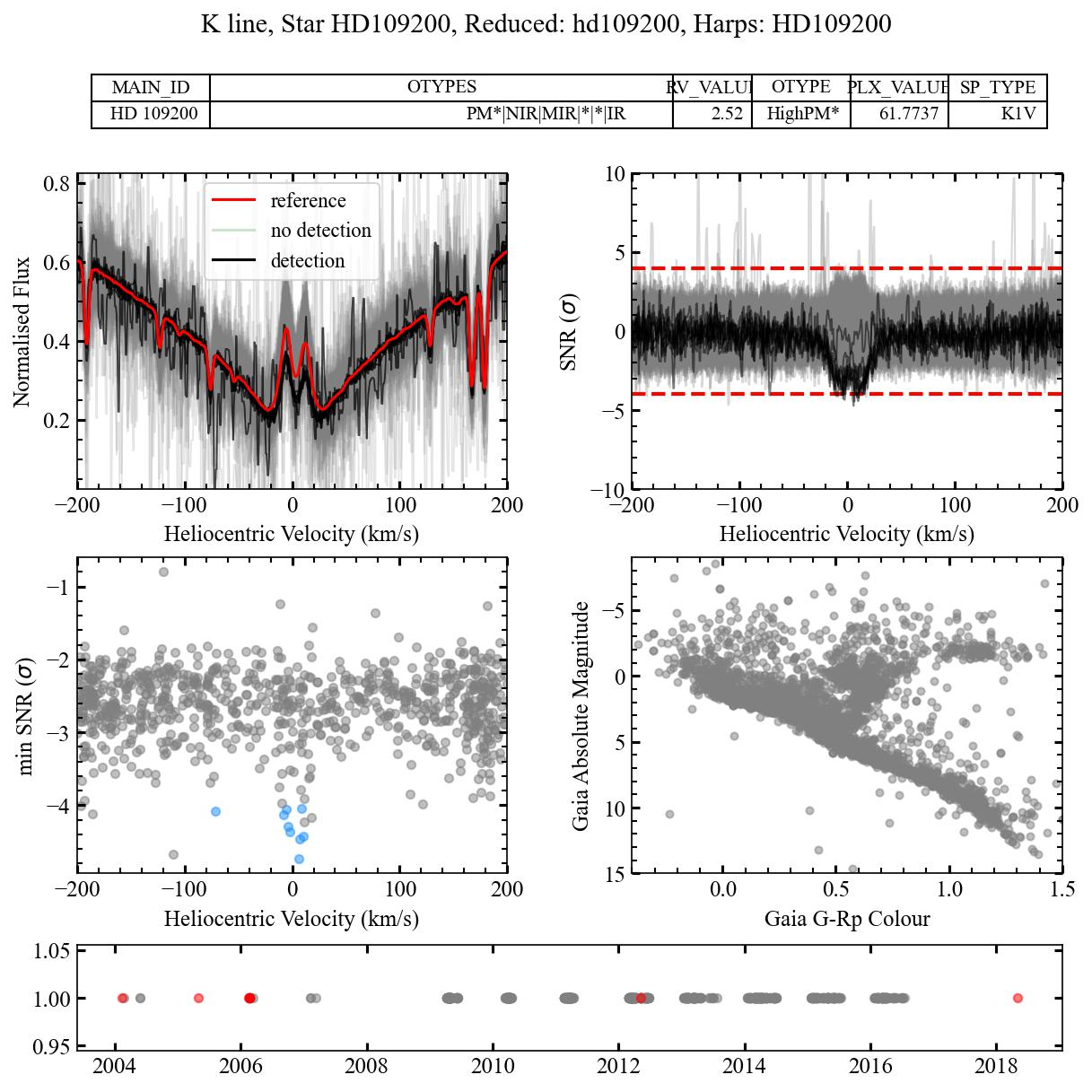}
    \caption[HD 109200]{HD 109200 is a K1V star classified as a Tier 3 candidate, with 9 detections in 860 spectra. This is a Tier 3 candidate because of the complex variability causing the detections. Some detections are results of stellar activity, others are absorption features merged with other stable absorption lines, but overall, it is hard to discern the scenarios from being exocometary.}
    \label{fig:app/vetting-hd109200}
\end{figure}

\begin{figure}
\centering
    \includegraphics[width=\linewidth]{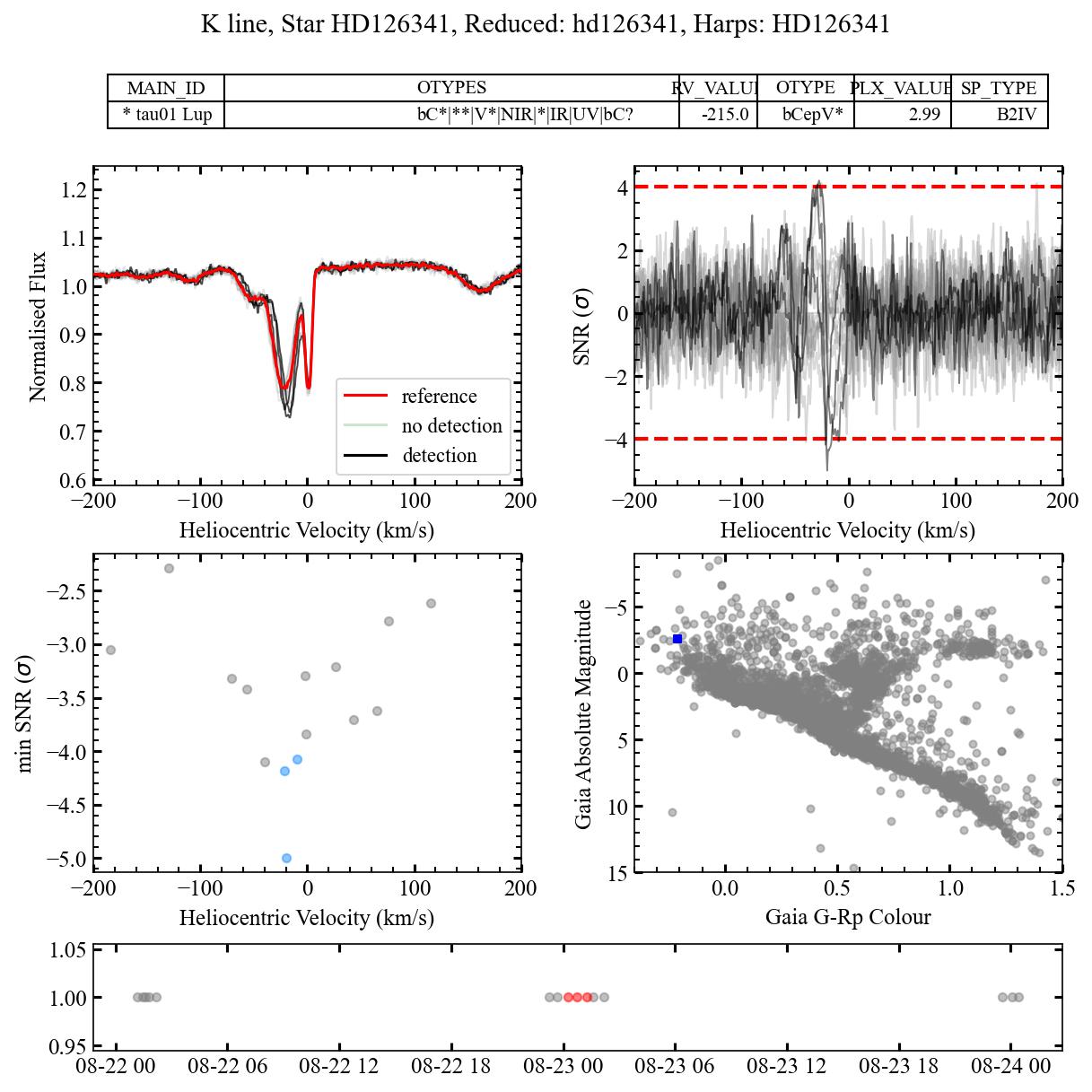}
    \caption[HD 126341]{HD 126341 is a B2IV classified as a Tier 3 candidate, with 3 detections in 15 spectra. This is a Tier 3 candidate because of complex variability, where hints of radial velocity shifts are seen at $-20$\kms~but none for others. To add to this complexity, the $-20$\kms~feature is caused by ISM absorption which brings the question of ISM variability. However, such variations for ISM absorption on a short 2 hour timescale are unlikely.}
    \label{fig:app/vetting-hd126341}
\end{figure}

\begin{figure} 
\centering
    \includegraphics[width=\linewidth]{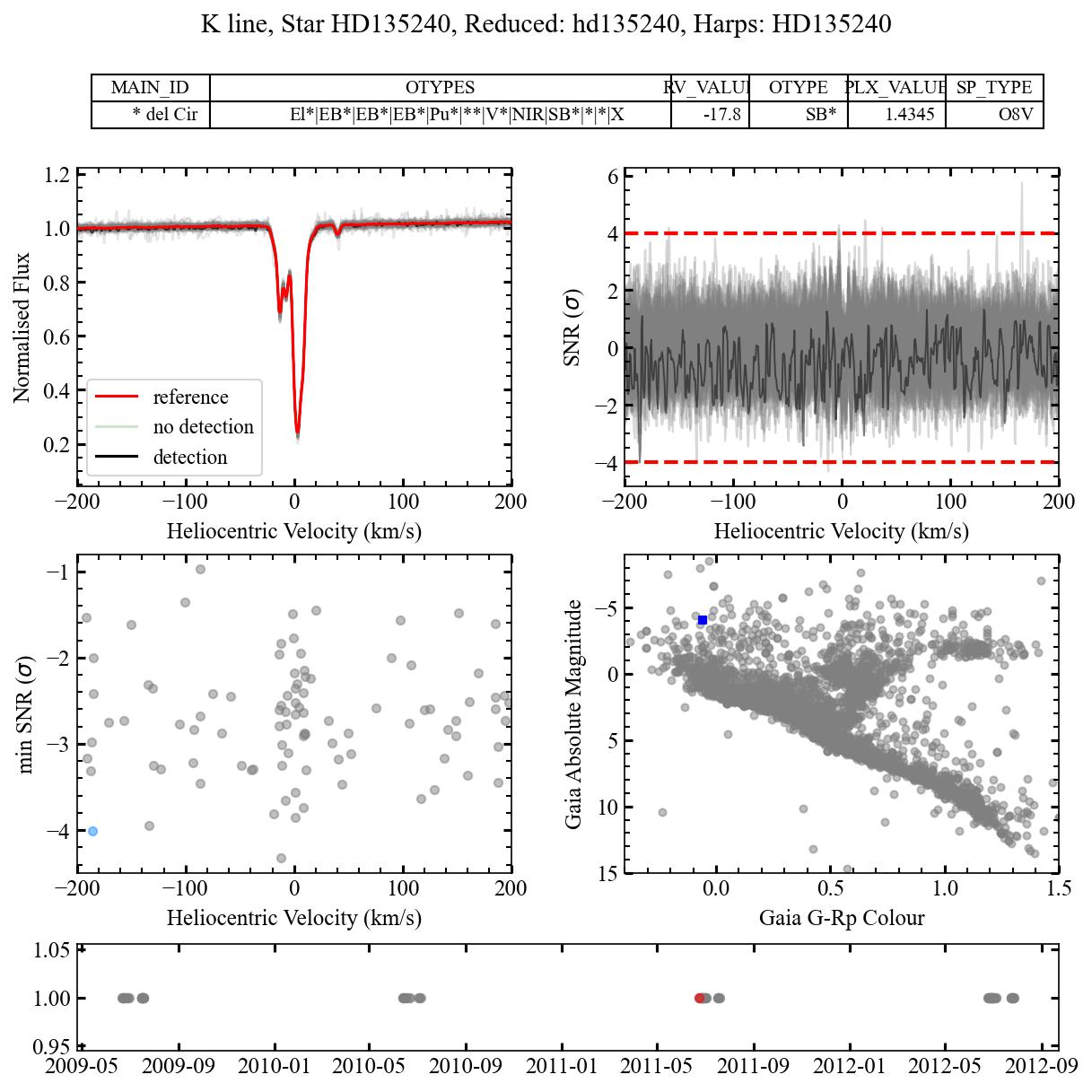}
    \caption{HD~135240 is an O8V star classified as a Tier 3 candidate, with a single detection in 95 spectra. The single transient absorption feature detected is narrow for a high-velocity feature and hardly makes the $-4\sigma$ threshold.}
    \label{fig:app/vetting-hd135240}
\end{figure}

\begin{figure}
\centering
    \includegraphics[width=\linewidth]{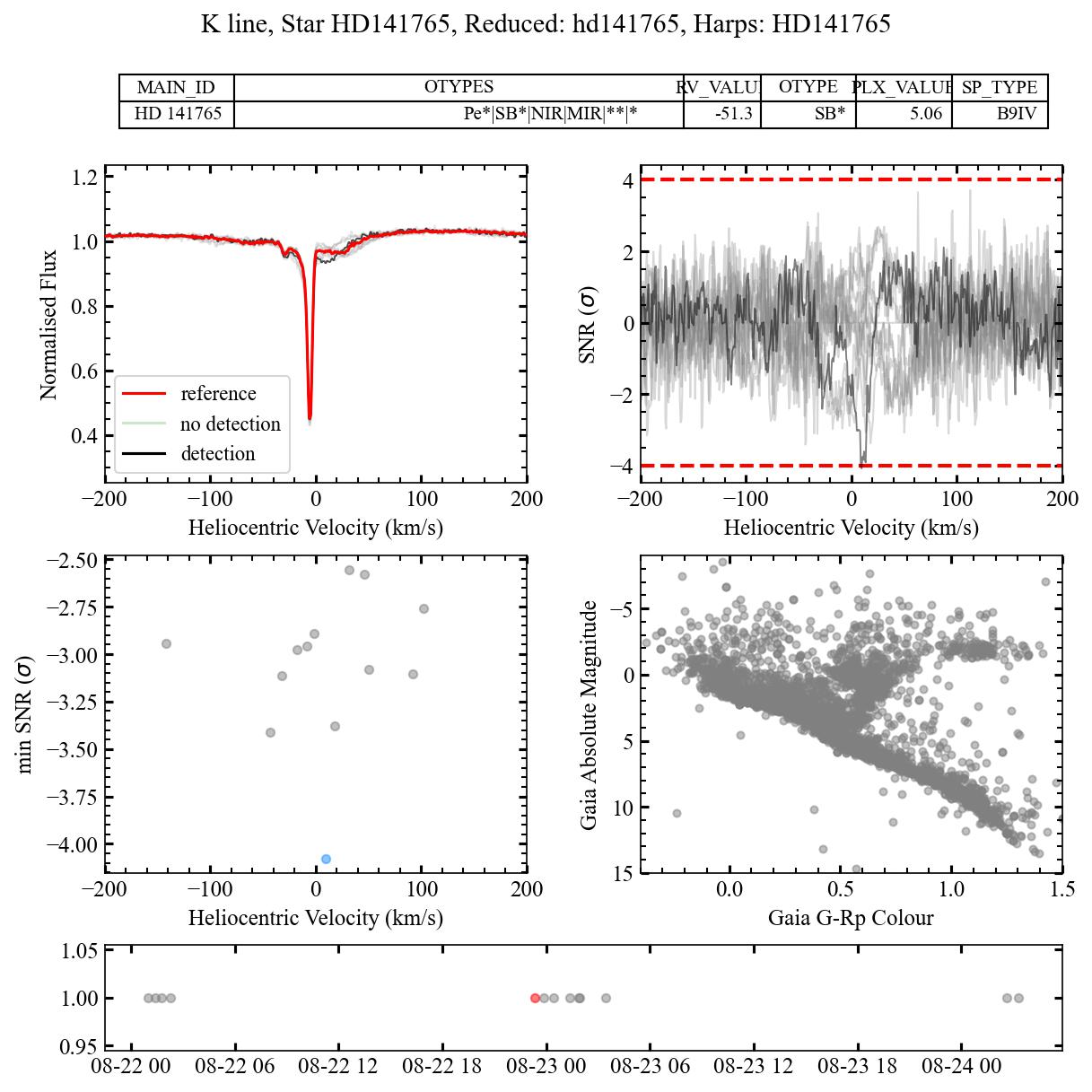}
    \caption[HD 141765]{HD 141765 is a B9IV star classified as a Tier 3 candidate, with a single detection in 13 spectra. The spectrum with the detection is too complicated to comment on possible exocomet transits, with multiple absorption features mixed with emission features outside the line observed at -5\kms. Also, note that there is no information in the HR diagram because there is no Gaia EDR3 data for this star.}
    \label{fig:app/vetting-hd141765}
\end{figure}

\begin{figure}
\centering
    \includegraphics[width=\linewidth]{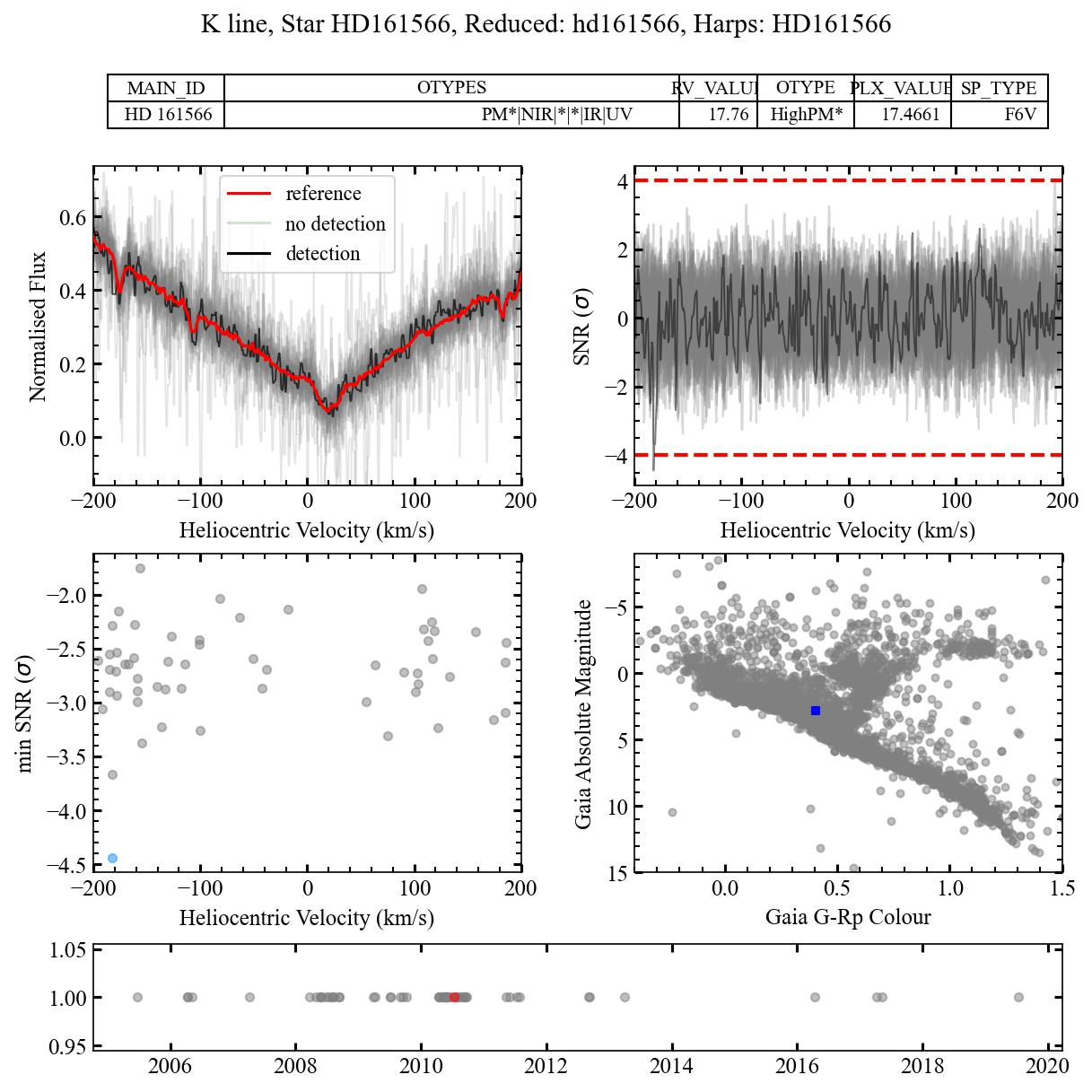}
    \caption[HD 161566]{HD 161566 is a F6V star classified as a Tier 3 candidate, with a single detection in 57 spectra. The detected absorption feature is found to be narrow and blended with some other stable absorption feature ($\approx 195$\kms~or $\approx 3931.1$\AA)}
    \label{fig:app/vetting-hd161566}
\end{figure}

\begin{figure}
\centering
    \includegraphics[width=\linewidth]{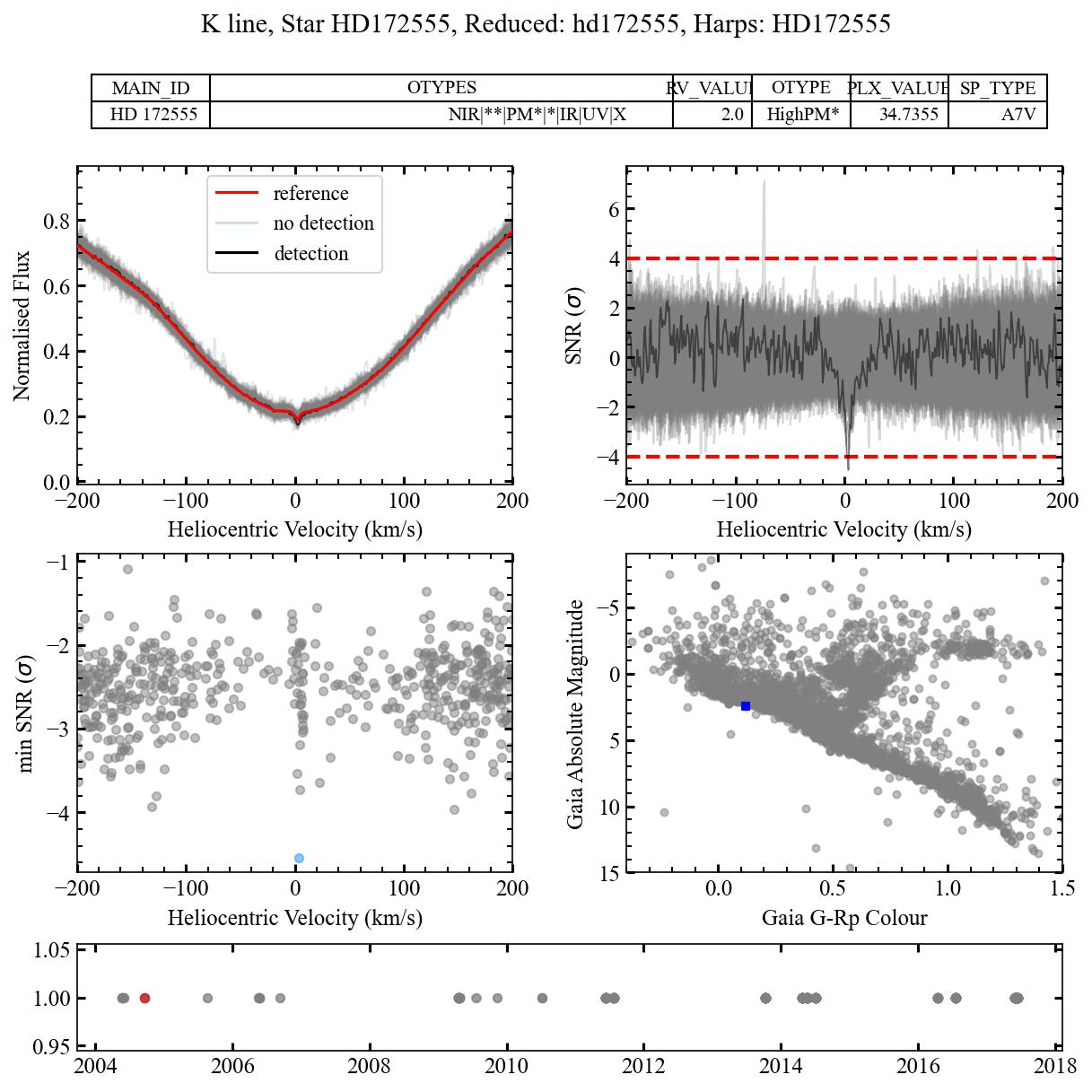}
    \caption[HD 172555]{HD 172555 is an A7V star classified as a Tier 1 candidate. More information about this detection can be seen in Section \ref{chap:exo sec:final-vet}.}
    \label{fig:app/vetting-hd172555}
\end{figure}

\begin{figure}
\centering
    \includegraphics[width=\linewidth]{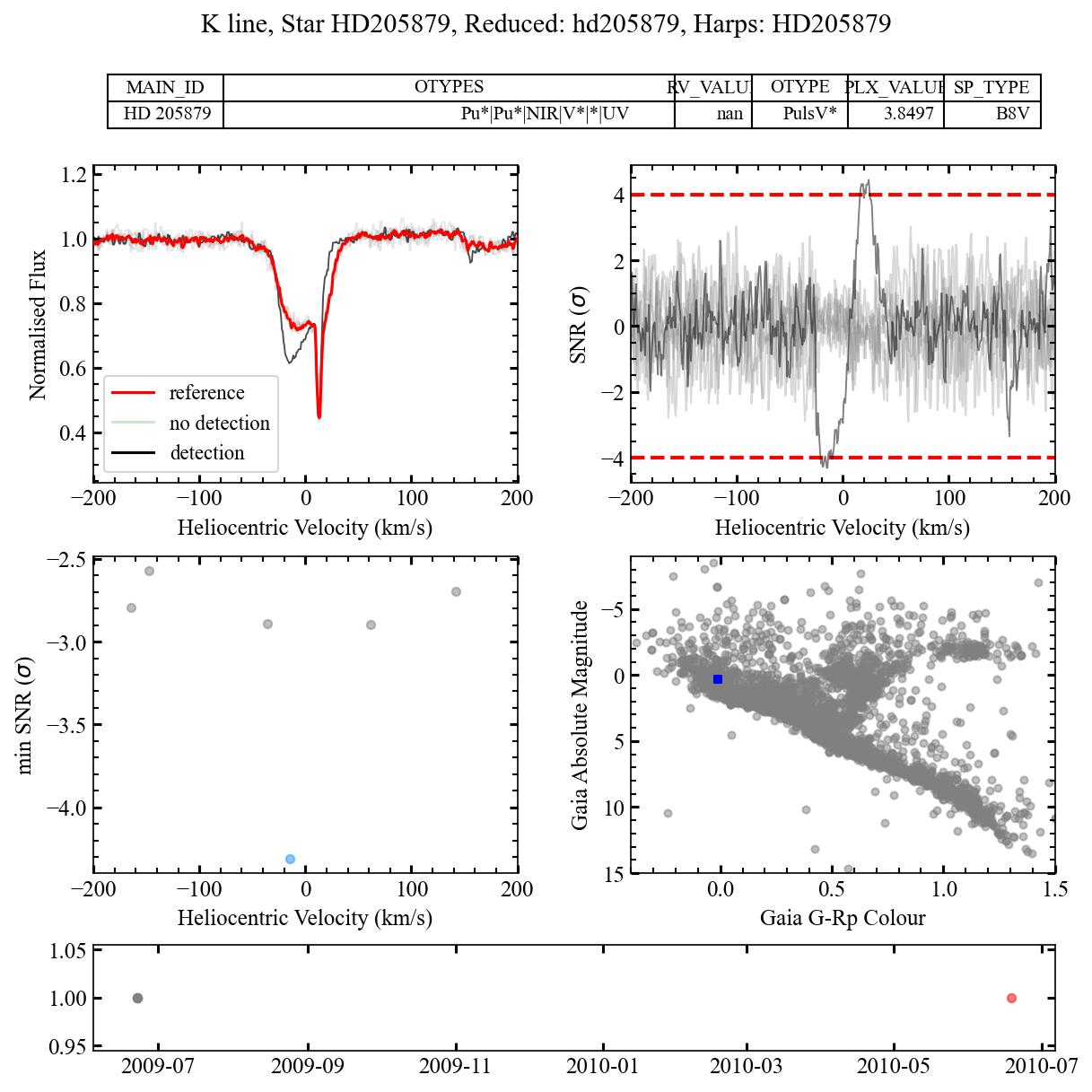}
    \caption[HD 205879]{HD 205879 is a B8V star classified as a Tier 3 candidate, with a single detection in 5 spectra. There are not many spectra for this star, but the only detection is observed in a spectrum showing too much complex variability - there is an absorption feature at $-20$\kms, then an emission around $10$\kms, followed by an absorption at $150$\kms. Note that most of the spectra are taken in 2009 but the detection is seen in the only spectrum from 2010, a spectrum that could have potentially varied in a year.}
    \label{fig:app/vetting-hd205879}
\end{figure}

\begin{figure} 
\centering
    \includegraphics[width=\linewidth]{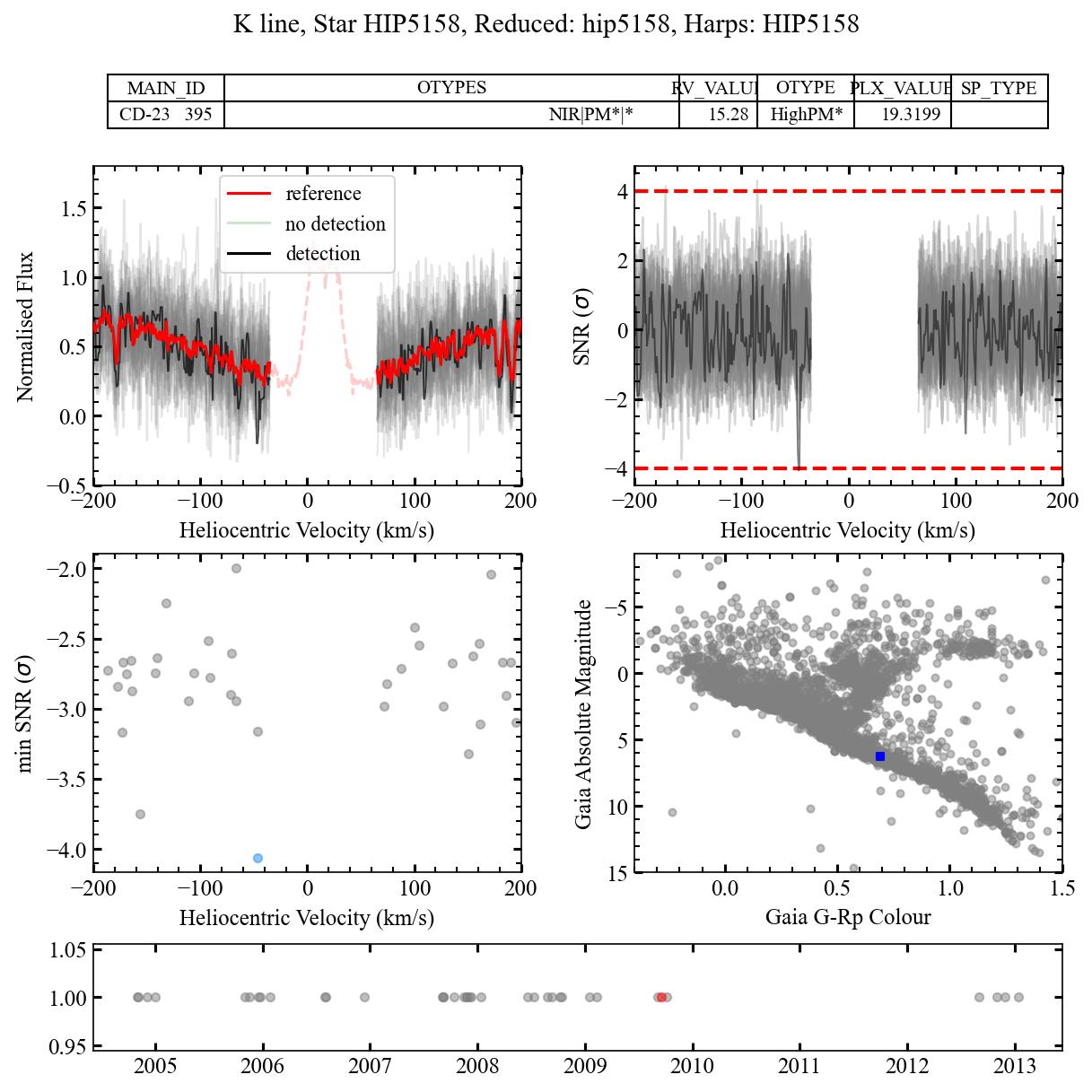}
    \caption[HIP 5158]{HIP 5158 is a K5V star classified as a Tier 2 candidate, with a single detection in 20 spectra. More information about this detection can be seen in Section \ref{chap:exo sec:final-vet}.}
    \label{fig:app/vetting-hip5158}
\end{figure}

\begin{figure} 
\centering
    \includegraphics[width=\linewidth]{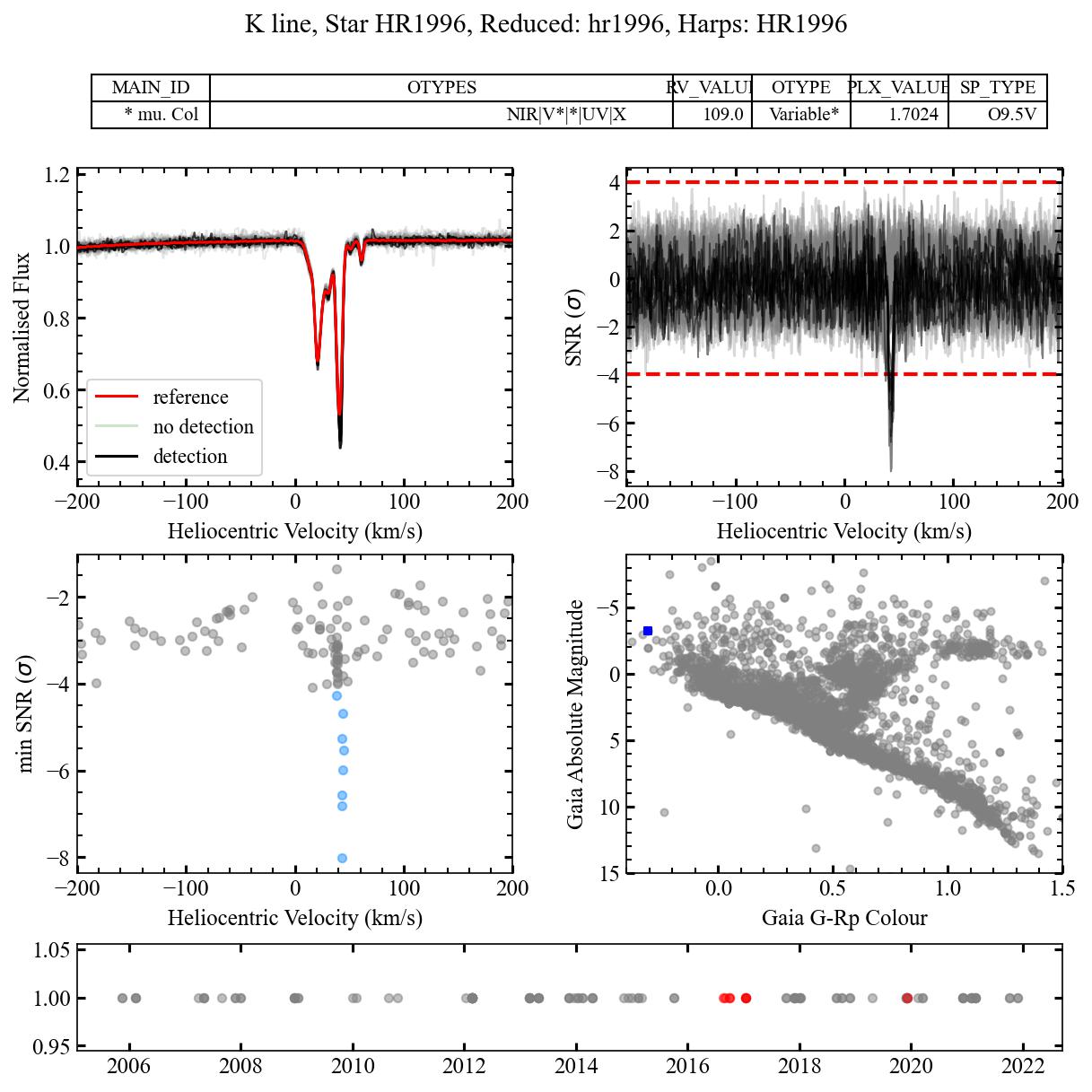}
    \caption[HR 1996]{HR 1996 ($\mu$ Col or HD 38666) is an O9.5V star classified as a Tier 2 candidate, with 8 detections in 120 spectra. More information about this detection can be seen in Section \ref{chap:exo sec:final-vet}.}
    \label{fig:app/vetting-hr1996}
\end{figure}

\begin{figure}
\centering
    \includegraphics[width=\linewidth]{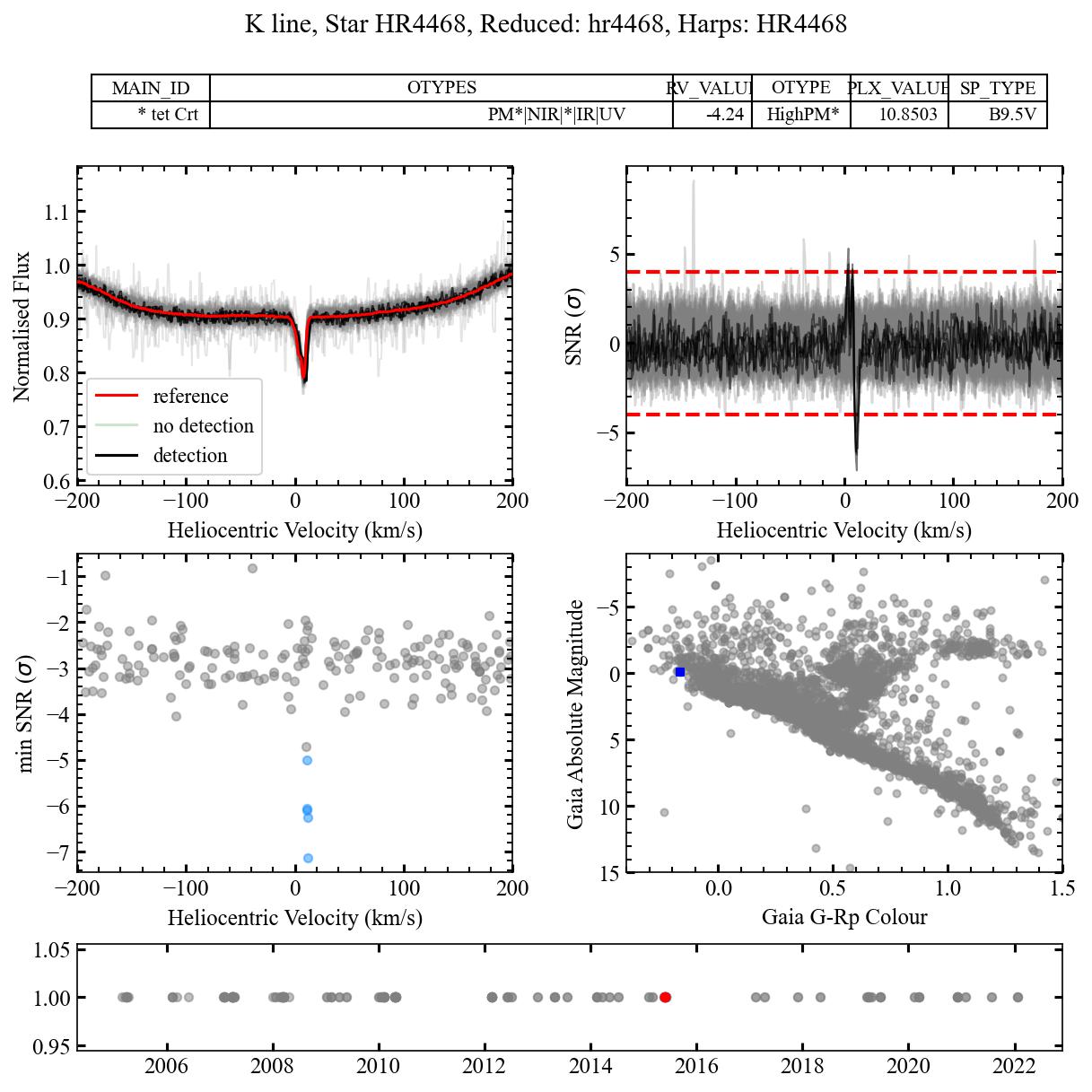}
    \caption[HR 4468]{HR 4468 (HD~100889) is a B9.5V star classified as a Tier 3 candidate, with 5 detections in 187 spectra. The extra absorption feature visible at the bottom of the broad main Ca\,{\sc ii}~K line is caused by ISM cloud absorption. There seems to be some shift at the location of this ISM causing the detected features, which adds complexity in determining whether this detection is exocometary-like.}
    \label{fig:app/vetting-hr4468}
\end{figure}

\begin{figure}
\centering
    \includegraphics[width=\linewidth]{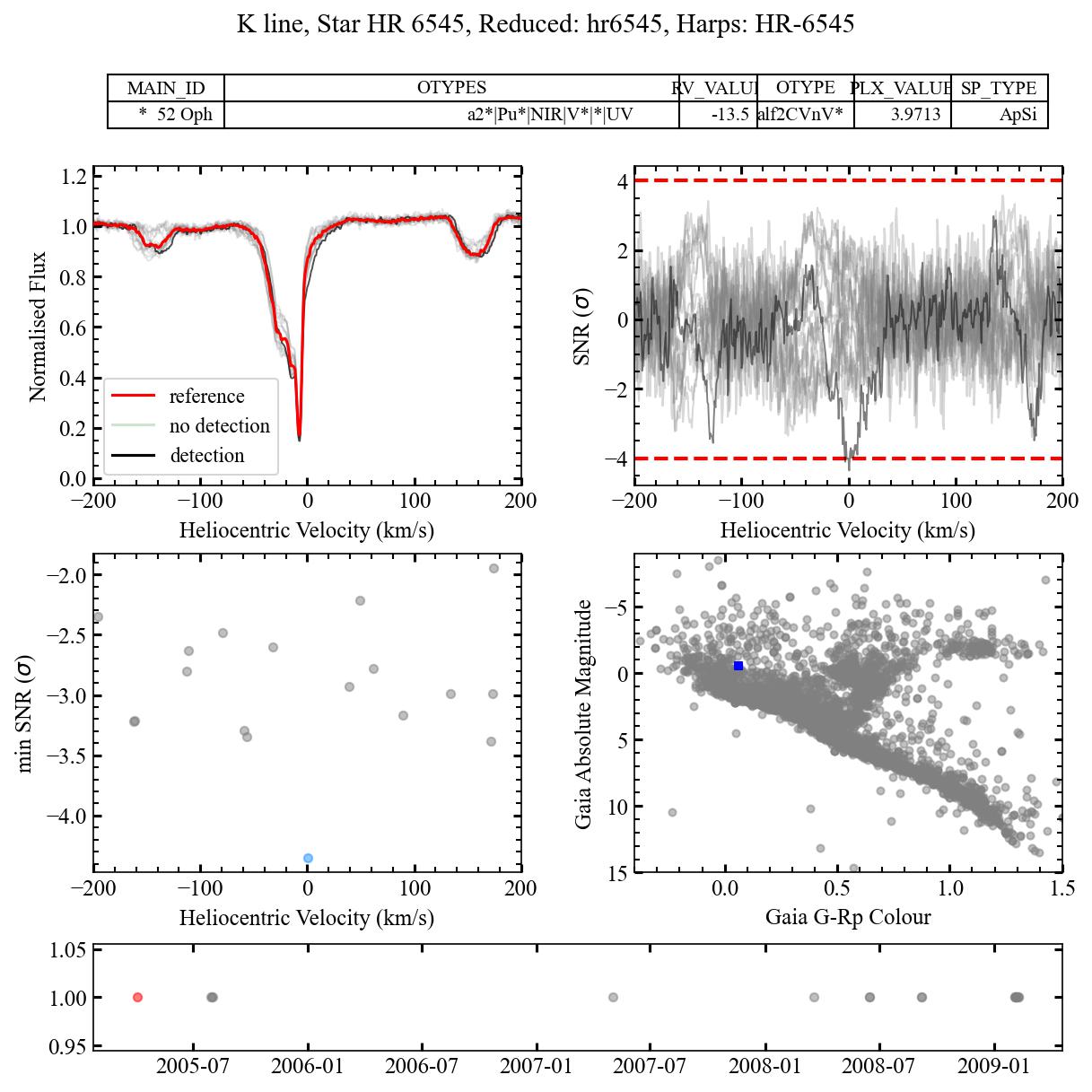}
    \caption[HR 6545]{HR 6545 (HD 159376) is a peculiar A star (Ap Si) classified as a Tier 3 candidate, with a single detection in 18 spectra. This is a Tier 3 candidate due to the complicated spectra with variability that cannot be explained by an exocomet transit and must have some other explanation (sometimes absorption features, sometimes emission features, etc).}
    \label{fig:app/vetting-hr6545}
\end{figure}

\begin{figure}
\centering
    \includegraphics[width=\linewidth]{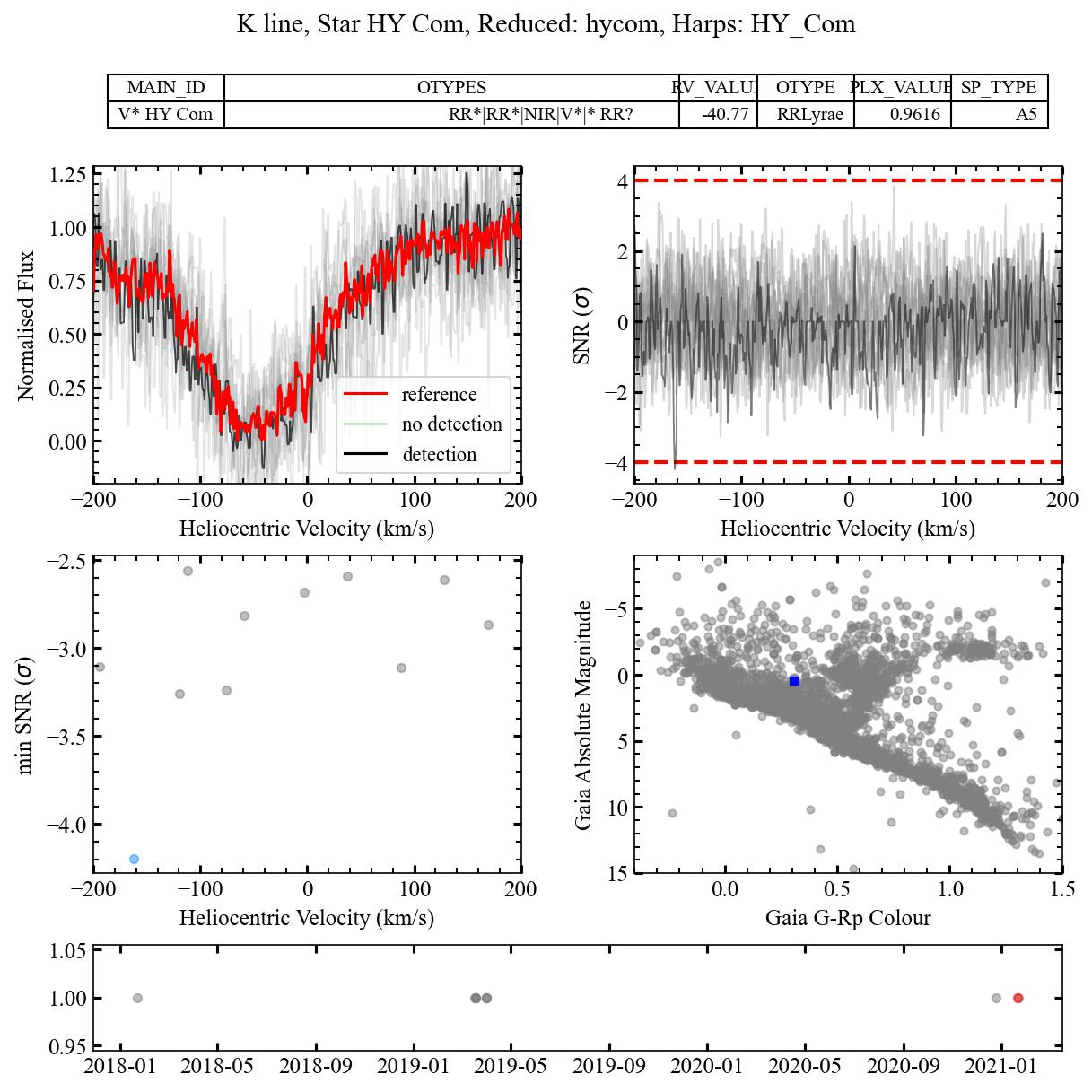}
    \caption[HY Com]{HY Com is an A5 star classified as a Tier 3 candidate, with a single detection in 11 spectra. The detected absorption feature seems to be a combination of multiple narrow dips that are merged with a stable absorption line. This detection hints towards spectra on the noisy end, hence the Tier 3 classification for this suspicious detection.}
    \label{fig:app/vetting-hycom}
\end{figure}


\bsp	
\label{lastpage}
\end{document}